\documentclass[onecolumn]{article}

\usepackage{geometry}
\usepackage{setspace}
\usepackage{textcomp}
\usepackage{diagbox}
\usepackage{titlesec}
\usepackage[titletoc,toc,page,header]{appendix}
\usepackage{tabularx}
\usepackage{authblk}
\usepackage{booktabs}
\usepackage{graphicx}
\usepackage{amsmath}
\usepackage{amssymb}
\usepackage{float}
\usepackage{makecell}
\usepackage{multirow}
\usepackage{tikz,pgf}
\usepackage{cite}
\usepackage{gensymb}
\usepackage{amstext}
\usepackage{bm}
\usepackage{bbm}
\usepackage{indentfirst}
\usepackage{latexsym}
\usepackage{color}
\usepackage{subfigure}
\usepackage{slashed}
\usepackage{hyperref}
\usepackage[misc]{ifsym}
\usepackage{url}
\usepackage{multicol}
\usepackage{lipsum}
\usepackage{cuted}
\usepackage{flushend}
\usepackage{threeparttable}
\usepackage{braket}
\allowdisplaybreaks[4]
\usetikzlibrary{trees}
\usetikzlibrary{decorations.pathmorphing}
\usetikzlibrary{decorations.markings}
\usetikzlibrary{patterns}
\usetikzlibrary{quotes,angles}
\usetikzlibrary{snakes}
\usepackage{mathrsfs}
\usepackage{amsbsy}
\usepackage[normalem]{ulem}
\usepackage{extarrows}

\usetikzlibrary{calc}
\usetikzlibrary{intersections}
\usetikzlibrary{trees}
\usetikzlibrary{decorations.pathmorphing}
\usetikzlibrary{decorations.markings}
\usetikzlibrary{patterns}
\tikzset{
   global scale/.style={
      scale=#1,
      every node/.append style={scale=#1}},
   photon/.style={decorate, decoration={snake}, draw=red},
   nucleon/.style={draw=black, postaction={decorate},
      decoration={markings,mark=at position .55 with{\arrow[draw=black]{>}}}},
   pion/.style={draw=blue, postaction={decorate},
      decoration={markings,mark=at position .55 with{\arrow[draw=blue]{}}}},
    }

\newcommand{\bk}{\boldsymbol{k}}
\newcommand{\bp}{\boldsymbol{p}}

\newcommand{\bq}{\boldsymbol{q}}

\newcommand{\sk}{\slashed{k}} 

\newcommand{\sep}{\slashed{\epsilon}}

\newcommand{\gmu}{\gamma^\mu}

\newcommand{\md}{\mathrm{d}}

\allowdisplaybreaks{}

\begin{document}
\title{\Large\textbf{Radiative Correction to Lepton Proton Scatterings in Manifestly Lorentz-Invariant Chiral Perturbation Theory}}
\author{Xiong-Hui Cao\(^{1}\)~\footnote{xionghuicao@pku.edu.cn}}
\author{Qu-Zhi Li\(^{1}\)~\footnote{2001110075@stu.edu.pku.cn}}
\author{Han-Qing Zheng\(^{2}\)~\footnote{zhenghq@scu.edu.cn (corresponding author)}}
\affil{\(^1\)School of Physics and State Key Laboratory of Nuclear Physics and Technology, \\Peking University, Beijing 100871, P. R. China}
\affil{\(^2\)College of Physics, Sichuan University, Chengdu, Sichuan 610065, P. R. China}

\maketitle

\begin{abstract}
Manifestly Lorentz-invariant baryon chiral perturbation theory is used to calculate the radiative correction of low energy elastic lepton proton scatterings.
Corrections of differential cross section and charge asymmetry are given at chiral next-to-leading order $(\mathcal{O}(p^2))$ with a nonzero lepton mass, which are infrared and ultraviolet finite.
The results are basically consistent with previous predictions based on hadron model calculation, but they are somewhat different from calculations based on heavy baryon chiral perturbation theory, especially in charge asymmetry.
\end{abstract}

\section{Introduction}

The lepton proton ($\ell \text{p}$) elastic scatterings, involving arbitrary number of real and virtual photons, have been proven to be an important process in the study of the electromagnetic structure of proton.
An accurate experimental determination of the proton's electromagnetic form factors (FFs) can clarify the proton's structure and internal dynamics.
The electric ($G^p_E$) and magnetic ($G^p_M$) form factors of protons can be extracted by conventional Rosenbluth separation technique.
These FFs describe charge and magnetization distribution inside a proton.

To improve the accuracy in the determination of proton's FFs, the idea to employ a polarization transfer method was suggested in Ref.~\cite{Akhiezer:1973xbf}.
Instead of measuring electric and magnetic FFs separately, the method is to access the ratio $G^p_E/G^p_M$  by detecting the polarization of the recoil proton in elastic scattering of polarized leptons off unpolarized proton targets.
An accurate measurement of the ratio $G^p_E/G^p_M$ by utilizing the novel experimental recoil polarization transfer technique\cite{PhysRevLett.88.092301, PhysRevLett.84.1398}, exposed a discrepancy compared with the Rosenbluth technique.
This discrepancy is referred as ``proton form factor puzzle,'' raising serious concerns regarding our basic understanding of the proton structure.
In order to solve these problems, an idea of two-photon exchange (TPE) correction was extensively discussed in papers\cite{Guichon:2003qm, Blunden:2003sp, Chen:2004tw, Arrington:2011dn}. 

Meanwhile, proton's root-mean-square (rms) charge radius obtained from high precision muonic hydrogen Lamb-shift measurements~\cite{Pohl:2010zza, Antognini:2013txn}, turned out to be about $5 \sigma$ discrepancy away from previous value extracted from $e \text{p}$ scattering data.
This is a so-called ``proton radius puzzle''.
Recent PRad result~\cite{Xiong:2019umf} supports a smaller value extracted from $e \text{p}$ scattering experiment.
This result is rather close to most muonium spectroscopy measurements, and therefore is inconsistent with previous $e \text{p}$ scattering data.
According to Ref.~\cite{Mohr:2012tt}, there is a large discrepancy between the electron- and muon-based charge radius of the proton.
The electron-based value is obtained from both hydrogen spectroscopy measurements and elastic $e \text{p}$ scattering data.
But the muon-based value is only obtained from muonic hydrogen spectroscopy.
Currently, there exists no precisely determined value for muon rms charge radius extracted from elastic $\mu \text{p}$ scatterings. 
Despite the efforts such discrepancies are yet to be conclusively resolved, and it requires further improved approaches on experimental verification of complete radiative correction of $\mu \text{p}$ scattering.

The $e \text{p}$ elastic scattering experiments, at BINP Novosibirsk, CERN, DESY, Fermilab, JLab, MAMI, SLAC, etc.,  have provided precision data about electromagnetic structure of the proton.
Several recent experimental proposals aim at carrying out high precision measurements of low energy $e^\pm \text{p}$ (and $\mu^\pm \text{p}$) scattering cross sections.
PRad\cite{Xiong:2019umf} at JLab, COMPASS++/AMBER at CERN~\cite{Adams:2018pwt, COMPASSAMBERworkinggroup:2019amp}, and MUSE\cite{MUSE:2013uhu, MUSE:2017dod} at PSI are three such experiments.
In particular, MUSE experiment plans to measure the elastic $\mu^\pm \text{p}$ scattering cross sections at momentum transfer as low as $|q^2| \thicksim 0.0016-0.08 \mathrm{GeV}^2$, where $q$ means the four-momentum transfer.
MUSE's goal is to measure the proton's rms charge radius at a better than $1\%$ precision, with incoming electron and muon beam momenta to be 115, 153, and 210 $\mathrm{MeV}$.
In this kinematical region, an extra theoretical complication comes out.
More precisely, a widely used ultrarelativistic (UR) approximation~\cite{Maximon:2000hm} cannot be employed in MUSE to describe the scattering of muons. 
In other words, the mass of the muon is going to be comparable to its energy and thus cannot be neglected. 
This means that previous radiative corrections codes naturally using the UR limit to describe the scattering of electrons have to be reconsidered.

In addition to the proton's rms charge radius, another meaningful observable is the lepton proton charge asymmetry (or $\ell^+ \text{p}/\ell^- \text{p}$ ratio), which describes the difference of elastic $\ell^{+} \text{p}$ and $\ell^{-} \text{p}$ cross sections.
Here, the charge asymmetry to order $\alpha^3$ is derived from interference between one- and two-photon exchange amplitudes,
along with the interference between bremsstrahlung off lepton and proton.
It provides a valuable input for our understanding of the radiative corrections~\cite{Arrington:2011dn, Koshchii:2017dzr}.
Recently, the real hard photon contribution to the charge
asymmetry in elastic lepton- and antilepton-proton scattering was estimated for the first time beyond the ultrarelativistic limit in Ref.~\cite{Afanasev:2020ejr}.

The most challenging aspect of radiative correction is TPE, in low energy regions, more or less approximate methods have been used to estimate the TPE contributions. 
The first is soft-photon approximation (SPA) used in Refs.~\cite{Tsai:1961zz, Maximon:2000hm}.
However, only the IR divergent part can be calculated in such a way. 
Instead, we are interested in the IR finite part that cannot to be calculated just in the IR region.
The second approach uses a hadron model to parametrize the on shell FFs.
Because of the explicit form, the results can be calculated easily by using computer program~\cite{Blunden:2003sp, Kondratyuk:2005kk, Blunden:2005ew, Chen:2013udl, Tomalak:2014dja, Zhou:2016psq}.
The approximation is reasonable numerically, but still contains some problems.
The most serious problem is that the physical region of $\ell \text{p}$ elastic scatterings is $q^2<0$ and $q^2>4 M^2$ (corresponding to crossing channel: $\ell^- +\ell^+ \to N+\bar{N}$, $M$ is nucleon physical mass), the unphysical region $0<q^2<4 M^2$ is completely inaccessible.
But the loop integral involves FFs in the whole timelike region ($q^2 > 0$) as well. 
The uncertainties caused by the above problem is not easily controlled.
Similarly, in the use of dispersion relations~\cite{GORCHTEIN2007322, Borisyuk:2008es, Tomalak:2014sva, Tomalak:2015aoa, Tomalak:2015hva, Tomalak:2016vbf, Tomalak:2017shs, Tomalak:2018jak}, the difficulty is that the uncertainties are mainly from the unknown subtractions, due to that we cannot estimate well the high energy contributions of dispersive integrals.

In order to estimate the QED radiative corrections at low energies ($q^2\thicksim m_\pi^2$), effective field theory provides a systematic formalism to study these processes.
Recently, heavy baryon chiral perturbation theory (HB$\chi$PT) has been used to estimate TPE\cite{Talukdar:2019dko} and complete radiative corrections\cite{Talukdar:2020aui}.
However, HB$\chi$PT has its own shortcomings~\cite{Bernard:1996cc, Becher:1999he}.
These disadvantages are related to the nonrelativistic expansion in this approach.
The scheme to be used in this paper is based on manifestly Lorentz-invariant baryon chiral perturbation theory (B$\chi$PT).
In this work, we only include elastic intermediate state (proton), and other possible contribution arising from $\Delta(1232)$ and high nucleon resonances,
\footnote{Here, we also ignore the contribution of the subthreshold resonance (pole) $N^*(890)$~\cite{Wang:2018nwi, Li:2021tnt} because the amplitudes in the physical region can be estimated by chiral low-order results.}
are not included.
As described in Ref.~\cite{Tomalak:2015hva}, except the nucleon intermediate state there are effects of nucleon resonance in the TPE diagrams. 
But nucleon intermediate state dominates in MUSE kinematical region, and the size of nucleon resonance contributions are within the anticipated error of the forthcoming data.
Recent papers of considering $\Delta(1232)$ in low energy scatterings can be found in Refs.\cite{GuerreroNavarro:2019fqb, GuerreroNavarro:2020kwb, Rijneveen:2021bfw}.

For definiteness, QED radiative corrections include all one-loop virtual contributions, i.e., TPE, vertex corrections, and vacuum polarizations to $\ell \text{p}$ elastic scatterings.
Single soft photon emission as the only real contribution are required in order to cancel the IR divergences from loop contribution.
In this work both chiral and QED divergences will be treated by employing dimensional regularization (DR).
The IR divergences, will systematically cancel at each order of chiral expansions.
In particular, we explicitly show that IR divergences of the TPE contribution are only from chiral leading-order (LO), no new IR divergence occurs at any chiral high order.

This paper is organized as the following.
In Sec.~\ref{sec:1}, the general lepton proton scattering formalism with explicit lepton mass is given.
In Sec.~\ref{sec:2}, we discuss how to construct the effective Lagrangian.
Based on the chiral power counting scheme, a self-consistent chiral expansion for observables is discussed.
From Secs.~\ref{sec:3} to \ref{sec:6}, The details of the radiative corrections, involving evaluations of the corresponding TPE, single soft photon emission, vertex correction and vacuum polarization are presented, in chiral LO and next-to-leading-order (NLO).
In Sec.~\ref{sec:7}, we provide the numerical estimation of various contributions and charge asymmetry in MUSE kinematical region. 
The major sources of theoretical uncertainties are also discussed.
Finally our conclusions are summarized in Sec~\ref{sec:8}.
Several technical details of the calculation are relegated to the appendices.

\section{Lepton proton scattering kinematics}\label{sec:1}

\subsection{Relativistic kinematics in MUSE experiment region}

According to the applications considered in this paper, we mainly choose the laboratory (lab) frame, where the target nucleon is at rest.
Elastic process is denoted by: $\ell^{\pm}\left(k_{1}\right)+p\left(p_{1}\right) \rightarrow \ell^{\pm}\left(k_{2}\right)+p\left(p_{2}\right)$, and $q=p_2-p_1$ is the (nucleon) momentum transfer.
Mandelstam variables are defined below:
\begin{align}
    \begin{aligned}
        &s=\left(k_{1}+p_{1}\right)^{2}, \quad t=\left(p_{2}-p_{1}\right)^{2}=q^{2}=-Q^{2} \\
        &u=\left(k_{1}-p_{2}\right)^{2}, \quad b_{i j}=2\left(k_{i} \cdot p_{j}\right)\quad  (i, j=1,2)\ ,
    \end{aligned}
\end{align} 
where $Q^2>0$ means virtuality of the exchanging particle. 
By means of four-momentum conservation in elastic scatterings, the following identities are satisfied: $b_{11}=b_{22}$ and $b_{12}=b_{21}$.
We also consider the bremsstrahlung process simultaneously: $\ell^{\pm}\left(k_{1}\right)+p\left(p_{1}\right) \rightarrow \ell^{\pm}\left(k_{2}\right)+p\left(p_{2}\right)+\gamma(k)$, reintroducing a lepton momentum transfer, $q_\ell=k_1-k_2$, and the four-momentum conservation implies $q_\ell=q+k$.
In this section, the elastic process is of primary consideration, so in the following, we do not distinguish $q$ from $q_\ell$ unless stated otherwise.

The square of momentum transfer, $q^2$, can be written as a function of the scattering angle $\theta_\ell$,
\begin{align}
    \begin{aligned}
        &q^{2}=\left(k_{1}-k_{2}\right)^{2}=2 m^{2}-2 E_{1} E_{2}\left(1-\beta_{1} \beta_{2} \cos \theta_{\ell}\right), \quad \beta_{i}=\sqrt{1-m^{2} / E_{i}^{2}}\ , \\
        &q^{2}=\left(p_{2}-p_{1}\right)^{2}=-2 M\left(E_{2}^{\prime}-M\right)=-2 M\left(E_{1}-E_{2}\right)\ ,
    \end{aligned}
\end{align} 
where the incoming (outgoing) lepton energy is given by $E_1 (E_2)$, and $E_2^\prime$ is the energy of the recoil nucleon of the lab frame; 
$m$ denotes the mass of the lepton, and $\beta_{1} (\beta_2)$ is the velocity of the incoming (outgoing) lepton.
There are several commonly used reference systems as follows: lab frame, Breit frame, and the center of mass (CM) frame.
The four-momentum conventions of the three reference frames are shown in Tab.~\ref{T1}. 
Bold symbols denote three-momentum throughout the paper.
\begin{table}[H]
    \centering
    \begin{tabular}{|c||c|c|c|}
    \hline \hline & Lab & CM & Breit \\
    \hline \hline$q$ & $(\omega, \mathbf{q})$ & $(\omega^*, \mathbf{q}^*)$ & $\left(\omega_{B}=0, \mathbf{q}_{B}\right)$ \\
    $k_{1}$ & $\left(E_{1}, \mathbf{k}_{1}\right)$ & $\left(E^*_{1}, \mathbf{p}_{i}^{*}\right)$ & $\left(E_{1 B}, \mathbf{k}_{1 \mathrm{B}}\right)$ \\
    $p_{1}$ & $(E_1^\prime =M, \bp_1=\boldsymbol{0})$ & $\left(E^{\prime*}_{1},-\mathbf{p}_i^{*}\right)$ & $\left(E_{1 B}^\prime, \mathbf{p}_{1 \mathrm{B}}\right)$ \\
    $k_{2}$ & $\left(E_{2}, \mathbf{k}_{2}\right)$ & $\left(E^*_{2}, \mathbf{p}_f^* \right)$ & $\left(E_{2 B}, \mathbf{k}_{2 \mathrm{B}}\right)$ \\
    $p_{2}$ & $\left(E_{2}^\prime, \mathbf{p}_{2}\right)$ & $\left(E^{\prime*}_{2},-\mathbf{p}_f^*\right)$ & $\left(E_{2 B}^\prime,-\mathbf{p}_{1 \mathrm{B}}\right)$ \\
    \hline \hline
    \end{tabular}
    \caption{Notations of four-momentum of leptons and protons in various reference systems.}\label{T1}
\end{table}
In the Breit system, the electric and magnetic parts of the proton's form factor can be completely separated, so it has crucial physical meaning and can also be used to derive some kinematic relations in a straightforward manner.
As for massless lepton (like electron at high energies), Ref.~\cite{Pacetti:2015iqa} is a pedagogical reference. 
In our approach, the mass of lepton is kept in any time.
Here, we summarize the kinematical relations without neglecting lepton mass.

$Q^2$ can be defined in terms of scattering angle $\theta^*$ in the CM frame,
\begin{align}\label{Sigmas}
    Q^{2}=-\left(k_{1}-k_{2}\right)^{2}=\frac{\Sigma\left(s, M^{2}, m^{2}\right)}{2 s}\left(1-\cos \theta^{*}\right)\ ,
\end{align}
with the kinematical triangle function $\Sigma_s\equiv \Sigma(s,M^2,m^2)=\left(s-(M+m)^{2}\right)\left(s-(M-m)^{2}\right)$\cite{Tomalak:2014dja}.
Scattering angle $\theta_B$ in Breit frame can be connected with lab's scattering angle $\theta_\ell$,
\begin{align}
    \cot ^{2} \frac{\theta_{B}}{2}=\frac{\left(q^{2}-2\left(m^{2}-E_{1} E_{2}\left(1-\beta_{1} \beta_{2}\right)\right)^{2}\right)^{2}}{q^{4}(1+\tau)} \cot ^{2} \frac{\theta_{\ell}}{2}\ ,
\end{align}
where $\tau=\frac{-q^2}{4M^2}>0$.
The outgoing lepton's energy in lab frame was also obtained\cite{GAKH201552},
\begin{align}
    E_{2}=\frac{\left(E_{1}+M\right)\left(M E_{1}+m^{2}\right)+\bk_{1}^{2} \cos \theta_{\ell} \sqrt{M^{2}-m^{2} \sin ^{2} \theta_{\ell}}}{\left(E_{1}+M\right)^{2}-\bk_{1}^{2} \cos ^{2} \theta_{\ell}}\ .
\end{align}  
with $\bk_1^2=E_1^2-m^2$.
The scattering angle can be written in terms of the four-momentum of the outgoing lepton,
\begin{align}
    \cos \theta_{\ell}=\frac{E_{1} E_{2}-m^{2}-M\left(E_{1}-E_{2}\right)}{\left|\boldsymbol{k}_{1}\right|\left|\boldsymbol{k}_{2}\right|}\ .
\end{align}
The relationship between $Q^2$ and incident lepton energy-momentum and scattering angle is\cite{Tomalak:2014dja},
\begin{align}\label{Q^2 theta}
    Q^{2}=2 M \frac{\boldsymbol{k}_{1}^{2}\left(M+E_{1} \sin ^{2} \theta_{\ell}-\sqrt{M^{2}-m^{2} \sin ^{2} \theta_{\ell}} \cos \theta_{\ell}\right)}{\left(E_{1}+M\right)^{2}-\boldsymbol{k}_{1}^{2} \cos ^{2} \theta_{\ell}}\ ,
\end{align}
and,
\begin{align}
    \cos \theta_{\ell}=\frac{2 M \bk_{1}^{2}-Q^{2}\left(E_{1}+M\right)}{\left|\bk_{1}\right| \sqrt{4 M^{2} \bk_{1}^{2}-4 E_{1} M Q^{2}+Q^{4}}}\ .
\end{align}
When the scattering angle $\theta_\ell$ is limited (as in the MUSE experiment), the range of values can be obtained by referring to Ref.~\cite{Tomalak:2014dja}, see Tab.~\ref{T2}.
\begin{table}[H]
    \centering
    \begin{tabular}{|c||c|c|c|}
\hline \hline $\text { Momentum} |\bk|\ \text{in}\ \mathrm{GeV}$ & $0.115$ & $0.153$ & $0.210$ \\
\hline \hline \multicolumn{4}{|c|} {$Q^{2} \text { in }\mathrm{GeV} \text { for Electron }$} \\
\hline $\text { Angle } \theta_\ell=20^{\circ}$ & $0.0016$ & $0.0028$ & $0.0052$ \\
\hline $\text { Angle } \theta_\ell=100^{\circ}$ & $0.027$ & $0.046$ & $0.082$ \\
\hline \multicolumn{4}{|c|} {$Q^{2} \text { in }\mathrm{GeV} \text { for Muon }$} \\
\hline $\text { Angle } \theta_\ell=20^{\circ}$ & $0.0016$ & $0.0028$ & $0.0052$ \\
\hline $\text { Angle } \theta_\ell=100^{\circ}$ & $0.026$ & $0.045$ & $0.080$ \\
\hline \hline
    \end{tabular}
   \caption{The range of $Q^{2}$ values for $e \text{p}$ and $\mu \text{p}$ scatterings in MUSE at the two limits of the lab frame scattering angles, $\theta_\ell=20^{\circ}$ and $100^{\circ}$, obtained from Eq.~(\ref{Q^2 theta}). 
   For convenience, we borrow the Tab.1 in Ref.~\cite{Talukdar:2019dko}.}\label{T2}
\end{table}

\subsection{Extended Rosenbluth formula of unpolarized cross section}

 The lab frame differential cross section of $\ell \text{p}$ elastic scattering in one-photon exchange (OPE) can be described by extended Resenbluth formula (with a nonzero lepton mass)\cite{PhysRevC.36.2466, Tomalak:2014dja, Koshchii:2016muj, Koshchii:2017dzr},
\begin{align}
    &\frac{\mathrm{d} \sigma_{1 \gamma}}{\mathrm{d} \Omega_{\ell}}=\frac{1}{\epsilon(1+\tau)}\left[\tau G_{M}^{2}\left(Q^{2}\right)+\epsilon G_{E}^{2}\left(Q^{2}\right)\right] \frac{\mathrm{d} \sigma_{M}}{\mathrm{~d} \Omega_{\ell}}\ , \\
    &\frac{\mathrm{d} \sigma_{M}}{\mathrm{~d} \Omega_{\ell}}=\frac{\alpha^{2}}{Q^{4}} \frac{\left(4 E_{1} E_{2}-Q^{2}\right) \boldsymbol{k}_{2}^{2}}{\left|\boldsymbol{k}_{1}\right|\left(\left|\boldsymbol{k}_{2}\right|+\frac{E_{1}}{M}\left|\boldsymbol{k}_{2}\right|-\frac{E_{2}}{M}\left|\boldsymbol{k}_{1}\right| \cos \theta_{\ell} \mid\right)}\ , \\
    &\frac{1}{\epsilon}=\frac{16 \nu^{2}+Q^{2}\left(4 M^{2}+Q^{2}\right)-4 m^{2}\left(4 M^{2}+Q^{2}\right)}{16 \nu^{2}-Q^{2}\left(4 M^{2}+Q^{2}\right)}\ .
\end{align}
The definition and characteristic of the Sachs FFs $G_E$ and $G_M$ are referred to \cite{Pacetti:2015iqa}.
Here we define $\nu$ as an $s-u$ crossing symmetric variable, $\nu=(s-u)/4=M(E_1+E_2)/2$; 
$\epsilon$ is the so-called photon polarization parameter;
it can be interpreted as a quantity that characterizes the degree of freedom of the longitudinal polarization of the virtual photon without considering the lepton mass\cite{Tomalak:2014dja}.
$\Omega_\ell$ here is the solid angle of outgoing muons in lab frame.
It is advantageous to study the relation between the photon polarization parameter $\epsilon$ and $Q^2$\cite{Tomalak:2014dja}.
For fixed $Q^2>2m^2$, $\epsilon$ is in the interval $(\epsilon_0,1)$, if $Q^2<2m^2$, then $\epsilon$ falls on $(1,\epsilon_0)$, with $\epsilon_0=2m^2/Q^2$.
The critical case, $\epsilon=1$, corresponds to $Q^2=2m^2 \simeq 0.022\mathrm{GeV}^2$ (muon beam).
Meanwhile, $s$ can also be written as a function of $\epsilon$, 
\begin{align}
    s=s_{1,2}=m^{2}+M^{2}+\frac{Q^{2}}{2}\pm\frac{\sqrt{(\epsilon-1)\left(4 M^{2}+Q^{2}\right)\left(4 m^{2} \epsilon-Q^{2}(\epsilon+1)\right)}}{2(\epsilon-1)}\ ,
\end{align}
if we require $s>(m+M)^2$, then we set $s=s_1$ when $Q^2<2m^2$, and then $s=s_2$ when $Q^2>2m^2$.

\section{B$\chi$PT: radiative corrections and chiral corrections}\label{sec:2}

The relevant parts of manifestly Lorentz-invariant chiral Lagrangian up to $\mathcal{O}(p^2)$ are given in Refs.~\cite{Fettes:2000gb, Scherer:2012xha} (the pion loops arise at $\mathcal{O}(p^3)$, which is beyond the accuracy of this work),
\begin{align}
    &\mathcal{L}_{\pi N}=\mathcal{L}_{\pi N}^{(1)}+\mathcal{L}_{\pi N}^{(2)}+\cdots\ , \\
    &\mathcal{L}_{\pi N}^{(1)}=\bar{N}\left(i \slashed{D}-M+\frac{g_{A}}{2} \gamma^{\mu} \gamma_{5} u_{\mu}\right) N\ , \\
    &\mathcal{L}_{\pi N}^{(2)}=\bar{N}\left\{\sigma^{\mu \nu}\left[\frac{c_{6}}{2} f_{\mu \nu}^{+}+\frac{c_{7}}{2} v_{\mu \nu}^{(s)}\right]+\cdots\right\} N\ ,
\end{align}
where $N=(p,n)^T$ is the nucleon doublet. 
The covariant derivatives $\slashed{{D}}=\gamma^\mu D_\mu$, the chiral connection $\Gamma_\mu$ and the chiral vielbein $u_\mu$ in the Lagrangian are
\begin{align}
\begin{aligned}
    D_{\mu} N &=\left(\partial_{\mu}+\Gamma_{\mu}-i v_{\mu}^{(s)}\right) N\ ,\\
    \Gamma_{\mu} &=\frac{1}{2}\left[u^{\dagger}\left(\partial_{\mu}-i r_{\mu}\right) u+u\left(\partial_{\mu}-i l_{\mu}\right) u^{\dagger}\right]\ ,\\
    u_{\mu} &= i\left[u^{\dagger}\left(\partial_{\mu}-i r_{\mu}\right) u-u\left(\partial_{\mu}-i l_{\mu}\right) u^{\dagger}\right]\ ,\\
    v_{\mu v}^{(s)} &=\partial_{\mu} v_{v}^{(s)}-\partial_{v} v_{\mu}^{(s)}\ , \\
    f_{\mu v}^{\pm} &=u f_{L \mu v} u^{\dagger} \pm u^{\dagger} f_{R \mu v} u\ , \\
    f_{L \mu v} &=\partial_{\mu} l_{v}-\partial_{v} l_{\mu}-i\left[l_{\mu}, l_{v}\right]\ , \\
    f_{R \mu v} &=\partial_{\mu} r_{v}-\partial_{v} r_{\mu}-i\left[r_{\mu}, r_{v}\right]\ ,  
\end{aligned}
\end{align}
where $g_A=1.267, c_6=3.706/(4 M), c_7=-0.120/(2 M)$ are chiral low energy constants (LECs)~\cite{Scherer:2012xha}.
Due to the absence of pions in our calculation, $u=\mathbbm{1}_{2\times 2}$ is the identity matrix in isospin space.
Here in our case the only external source field is the electromagnetic four-vector potential $A_\mu(x)$.
Relevant external isoscalar and isovector sources are obtained by $r_{\mu}=l_{\mu}=e \tau_{3} A_{\mu} / 2, v_{\mu}^{(s)}=e A_{\mu} / 2\ (e>0)$, where $\tau_3$ is the third Pauli matrix.
For more recent applications such as the interactions between photon, nucleon and $\pi$, refer to Refs.~\cite{Ma:2020hpe, Cao:2021kvs}.

It is worth noting that we have two independent power counting schemes here.
One is following $\alpha$ as QED power counting, and the other is chiral expansion of momentum $p$, within the energy $Q^2 \sim m_\pi^2$, which can be set as $\frac{p}{4 \pi F_{\pi}} \sim \frac{Q}{4 \pi F_{\pi}} \sim \frac{Q}{M}$, and $F_\pi=92.4 \mathrm{MeV}$ is physical pion's decay constant.
Since we are considering QED radiative correction, the leading order of $\ell \text{p}$ scattering amplitudes come from pure QED pointlike interaction which are of chiral $\mathcal{O}(p)$.
Next-to-leading order result is just chiral $\mathcal{O}(p^2)$, suppressed by $Q/M$ compared with LO. 
However, it is more convenient to rearrange the chiral power counting of a product $\left(\mathcal{M}^{(m)}\right)^*\mathcal{M}^{(n)}$ as $\mathcal{O}(p^{m+n-1})$.

One of the main purpose of this paper is to calculate charge asymmetry and complete radiative corrections in the framework of DR.
For instance, all the virtual corrections in lab frame can be defined by
\begin{align}\label{virtual}
    \left[\frac{\mathrm{d} \sigma_{\mathrm{el}}\left(Q^{2}\right)}{\mathrm{d} \Omega_{\ell}}\right]_{2\gamma,\mathrm{v}}=\left[\frac{\mathrm{d} \sigma_{\mathrm{el}}\left(Q^{2}\right)}{\mathrm{d} \Omega_{\ell}}\right]_{\gamma} \bar{\delta}_{2\gamma,\mathrm{v}}\left(Q^{2}\right)\ ,
\end{align}  
where
\begin{align}
    \bar{\delta}_{2\gamma, \mathrm{v}}\left(Q^{2}\right)=\underbrace{\frac{2 \mathcal{R} e \sum_{\text {spins }}\left(\mathcal{M}_{\gamma}^{*} \mathcal{M}_{\gamma \gamma}\right)}{\sum_{\text {spins }}\left|\mathcal{M}_{\gamma}\right|^{2}}}_{\delta_{2\gamma, \mathrm{v}}\left(Q^{2}\right)}-\delta_{\mathrm{IR}}\left(Q^{2}\right)\ .
\end{align}
where the subscript ``$\mathrm{v}$'' is an abbreviation of ``virtual.''
IR divergence $\delta_{\mathrm{IR}}\left(Q^{2}\right)$  would be canceled by real photon emission, and $\mathcal{M}_\gamma$ is the so-called OPE amplitude. 
In the framework of $\chi$PT, $\mathcal{M}_\gamma$ is not only the chiral LO, but in principle it should include any high order of chiral corrections.
Therefore, our definition (\ref{virtual}) is slightly different from Ref.\cite{Talukdar:2020aui}, in which it has a factorization structure of chiral LO OPE cross section $\left[\frac{\mathrm{d} \sigma_{\mathrm{el}}\left(Q^{2}\right)}{\mathrm{d} \Omega_{\ell}}\right]_{\gamma}^{(1)} \bar{\delta}\left(Q^{2}\right)$.
$\mathcal{M}_{\gamma \gamma}$ indicates virtual contributions of radiative corrections.
Based on this, the $\chi$PT corrections of $\delta_{2\gamma, \mathrm{v}}$ can be written as
\begin{align}
    \delta_{2\gamma, \mathrm{v}}=\frac{2 \mathcal{R} e \sum_{\text {spins }}\left[\left(\mathcal{M}_{\gamma}^{(1)}+\mathcal{M}_{\gamma}^{(2)}+\mathcal{O}\left(\alpha \cdot p^{3}\right)\right)^{*}\left(\mathcal{M}_{\gamma \gamma}^{(1)}+\mathcal{M}_{\gamma \gamma}^{(2)}+\mathcal{O}\left(\alpha \cdot \alpha p^{3}\right)\right)\right]}{\sum_{\text {spins }}\left|\mathcal{M}_{\gamma}^{(1)}+\mathcal{M}_{\gamma}^{(2)}+\mathcal{O}\left(\alpha \cdot p^{3}\right)\right|^{2}}\ ,
\end{align}
where $\mathcal{M}_{\gamma}^{(1,2)}$ are the OPE amplitudes of chiral $\mathcal{O}(p^{(1,2)})$, and all the two (virtual) photon amplitudes such as TPE $et\ al$. of chiral $\mathcal{O}(p^{(1,2)})$, are encoded in $\mathcal{M}_{\gamma \gamma}^{(1,2)}$.
Such a definition can be calculated order by order, namely,
\begin{align}\label{del chptdef}
    \delta_{2\gamma, \mathrm{v}}=&\frac{2 \mathcal{R}e \sum_{\text{spins}} \left[\left(\mathcal{M}_\gamma^{(1)}\right)^*\mathcal{M}_{\gamma\gamma}^{(1)}+\left(\mathcal{M}_\gamma^{(1)}\right)^*\mathcal{M}_{\gamma\gamma}^{(2)}+\left(\mathcal{M}_\gamma^{(2)}\right)^*\mathcal{M}_{\gamma\gamma}^{(1)}\right]}{\sum_{\text{spins}} \left|\mathcal{M}_\gamma^{(1)}\right|^2} \nonumber\\
    &-\frac{2 \mathcal{R}e \sum_{\text{spins}} \left[\left(\mathcal{M}_\gamma^{(1)}\right)^*\mathcal{M}_{\gamma\gamma}^{(1)}\right]}{\sum_{\text{spins}} \left|\mathcal{M}_\gamma^{(1)}\right|^2}\times \frac{2 \mathcal{R}e \sum_{\text{spins}} \left[\left(\mathcal{M}_\gamma^{(1)}\right)^*\mathcal{M}_{\gamma}^{(2)}\right]}{\sum_{\text{spins}} \left|\mathcal{M}_\gamma^{(1)}\right|^2}+\mathcal{O}(\alpha p^3)\ .
\end{align}
The complete radiative (virtual) contributions of QED$+$B$\chi$PT to $\mathcal{O}(\alpha p^2)$ should be given by the above formula.
Bremsstrahlung (real correction) corrections are similar as Eq.(\ref{del chptdef}).

\section{The calculation of the TPE diagrams}\label{sec:3}

In this section, we evaluate the TPE amplitudes of elastic $\ell \text{p}$ scatterings at low energy transfer up to chiral $\mathcal{O}(p^2)$.
The chiral LO and NLO amplitudes of the OPE required are as follows:
\begin{align}
    &\mathcal{M}_{\gamma}^{(1)}=z\frac{e^{2} \bar{u}\left(k_{2}\right) \gamma^{\mu} u\left(k_{1}\right) \bar{p}\left(p_{2}\right) \gamma_{\mu} p\left(p_{1}\right)}{Q^{2}}\ ,\\
    &\mathcal{M}_{\gamma}^{(2)}=z\frac{i e^{2}\left(c_{6}+\frac{c_{7}}{2}\right) \bar{u}\left(k_{2}\right) \gamma^{\mu} u\left(k_{1}\right) \bar{p}\left(p_{2}\right) \sigma_{\mu \nu} q^{\nu} p\left(p_{1}\right)}{Q^{2}}\ ,
\end{align}
$z=\pm$ corresponds to $\ell^\pm \text{p}$ elastic scatterings, respectively.\footnote{Using charge conjugate symmetry, one can prove the notations in Ref.~\cite{Afanasev:2020ejr} are equivalent to ours.}
Lepton (proton) spinor with momentum $k$ is abbreviated as $u(k) (p(k))$.
Here only the $\ell^- \text{p}$ scattering is considered unless stated otherwise.
\begin{figure}[H]
    \centering
    \subfigure[]{
    \begin{tikzpicture}[scale = 0.7]
        \draw[nucleon](-2,3)to(-1,2);
        \draw[nucleon](-1,2)to(1,2);
        \draw[nucleon](1,2)to(2,3);
        \draw[nucleon, very thick](-2,-1)to(-1,0);
        \draw[nucleon, very thick](-1,0)to(1,0);
        \draw[nucleon, very thick](1,0)to(2,-1);
        \draw[photon, very thick](-1,2)to(-1,0);
        \draw[photon, very thick](1,2)to(1,0);
        
        \draw[very thick](-1,0)circle(0.3);
        \fill[white](-1,0) circle(0.3);
        \node at (-1,0) {$1$};
        \draw[very thick](1,0)circle(0.3);
        \fill[white](1,0) circle(0.3);
        \node at (1,0) {$1$};
    \end{tikzpicture}
    } 
    \subfigure[]{
    \begin{tikzpicture}[scale = 0.7]
        \draw[nucleon](-2,3)to(-1,2);
        \draw[nucleon](-1,2)to(1,2);
        \draw[nucleon](1,2)to(2,3);
        \draw[nucleon, very thick](-2,-1)to(-1,0);
        \draw[nucleon, very thick](-1,0)to(1,0);
        \draw[nucleon, very thick](1,0)to(2,-1);
        \draw[photon, very thick](-1,2)to(1,0);
        \draw[photon, very thick](1,2)to(-1,0);
        
        \draw[very thick](-1,0)circle(0.3);
        \fill[white](-1,0) circle(0.3);
        \node at (-1,0) {$1$};
        \draw[very thick](1,0)circle(0.3);
        \fill[white](1,0) circle(0.3);
        \node at (1,0) {$1$};
    \end{tikzpicture}
    }
    \subfigure[]{
    \begin{tikzpicture}[scale = 0.7]
        \draw[nucleon](-2,3)to(-1,2);
        \draw[nucleon](-1,2)to(1,2);
        \draw[nucleon](1,2)to(2,3);
        \draw[nucleon, very thick](-2,-1)to(-1,0);
        \draw[nucleon, very thick](-1,0)to(1,0);
        \draw[nucleon, very thick](1,0)to(2,-1);
        \draw[photon, very thick](-1,2)to(-1,0);
        \draw[photon, very thick](1,2)to(1,0);
        
        \draw[very thick](-1,0)circle(0.3);
        \fill[white](-1,0) circle(0.3);
        \node at (-1,0) {$2$};
        \draw[very thick](1,0)circle(0.3);
        \fill[white](1,0) circle(0.3);
        \node at (1,0) {$1$};
    \end{tikzpicture}
    }
    
    \subfigure[]{
    \begin{tikzpicture}[scale = 0.7]
        \draw[nucleon](-2,3)to(-1,2);
        \draw[nucleon](-1,2)to(1,2);
        \draw[nucleon](1,2)to(2,3);
        \draw[nucleon, very thick](-2,-1)to(-1,0);
        \draw[nucleon, very thick](-1,0)to(1,0);
        \draw[nucleon, very thick](1,0)to(2,-1);
        \draw[photon, very thick](-1,2)to(1,0);
        \draw[photon, very thick](1,2)to(-1,0);
        
        \draw[very thick](-1,0)circle(0.3);
        \fill[white](-1,0) circle(0.3);
        \node at (-1,0) {$2$};
        \draw[very thick](1,0)circle(0.3);
        \fill[white](1,0) circle(0.3);
        \node at (1,0) {$1$};
    \end{tikzpicture}
    }
    \subfigure[]{
    \begin{tikzpicture}[scale = 0.7]
        \draw[nucleon](-2,3)to(-1,2);
        \draw[nucleon](-1,2)to(1,2);
        \draw[nucleon](1,2)to(2,3);
        \draw[nucleon, very thick](-2,-1)to(-1,0);
        \draw[nucleon, very thick](-1,0)to(1,0);
        \draw[nucleon, very thick](1,0)to(2,-1);
        \draw[photon, very thick](-1,2)to(-1,0);
        \draw[photon, very thick](1,2)to(1,0);
        
        \draw[very thick](-1,0)circle(0.3);
        \fill[white](-1,0) circle(0.3);
        \node at (-1,0) {$1$};
        \draw[very thick](1,0)circle(0.3);
        \fill[white](1,0) circle(0.3);
        \node at (1,0) {$2$};
    \end{tikzpicture}
    }
    \subfigure[]{
    \begin{tikzpicture}[scale = 0.7]
        \draw[nucleon](-2,3)to(-1,2);
        \draw[nucleon](-1,2)to(1,2);
        \draw[nucleon](1,2)to(2,3);
        \draw[nucleon, very thick](-2,-1)to(-1,0);
        \draw[nucleon, very thick](-1,0)to(1,0);
        \draw[nucleon, very thick](1,0)to(2,-1);
        \draw[photon, very thick](-1,2)to(1,0);
        \draw[photon, very thick](1,2)to(-1,0);
        
        \draw[very thick](-1,0)circle(0.3);
        \fill[white](-1,0) circle(0.3);
        \node at (-1,0) {$1$};
        \draw[very thick](1,0)circle(0.3);
        \fill[white](1,0) circle(0.3);
        \node at (1,0) {$2$};
    \end{tikzpicture}
    }
    \caption{The TPE diagrams, thin lines represent a lepton; black thick lines represent a proton; and red wiggly lines represent virtual photons.
    The solid circles represent chiral vertices [1,2 indicate $\mathcal{O}(p), \mathcal{O}(p^2)$ vertices, respectively].
    The power counting of all seagull diagrams is chiral $\mathcal{O}(p^3)$ or higher. 
    Thus during the calculation we neglect the seagull diagrams.}\label{TPE}
\end{figure}
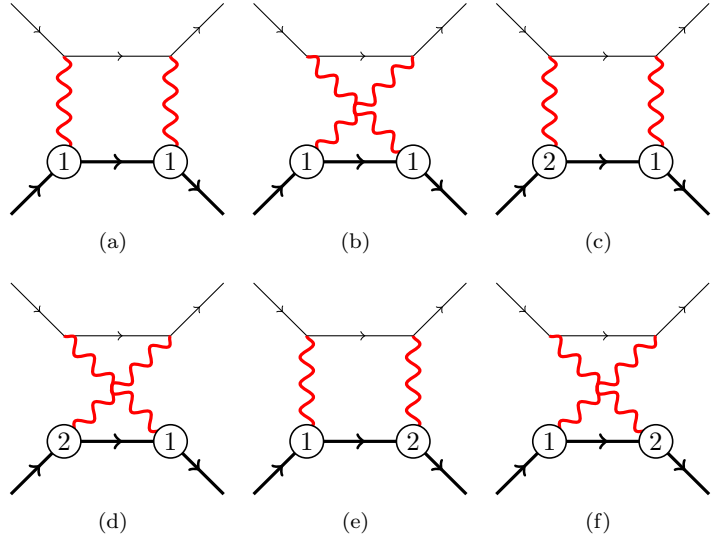

\begin{figure} 
    \centering
    \subfigure{
        \begin{minipage}[b]{0.99\linewidth}
          
        \begin{minipage}[b]{0.45\linewidth}
        \includegraphics[scale=0.5]{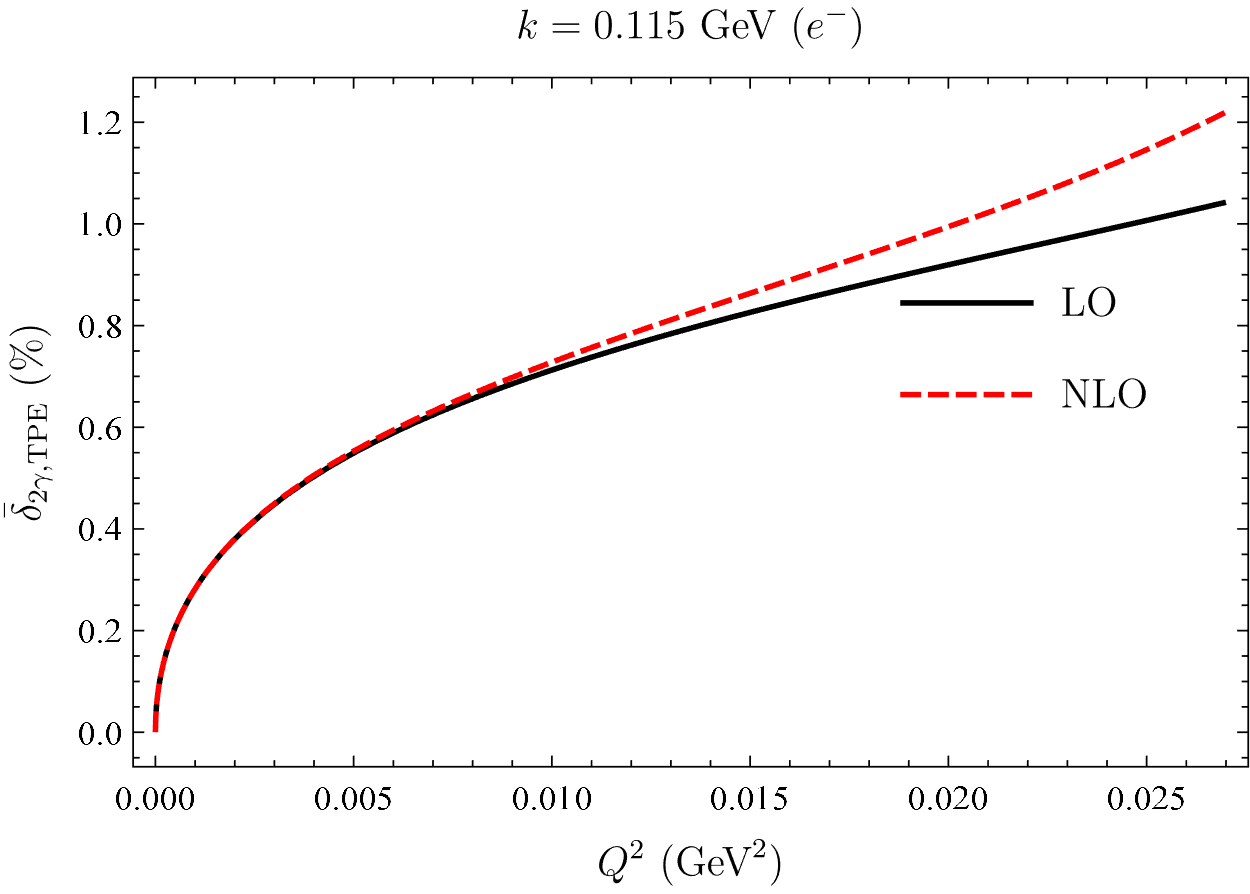}
        \vspace{0.1cm}
        \hspace{0.02cm}
        \end{minipage}
        \qquad
        \begin{minipage}[b]{0.45\linewidth}
        \includegraphics[scale=0.5]{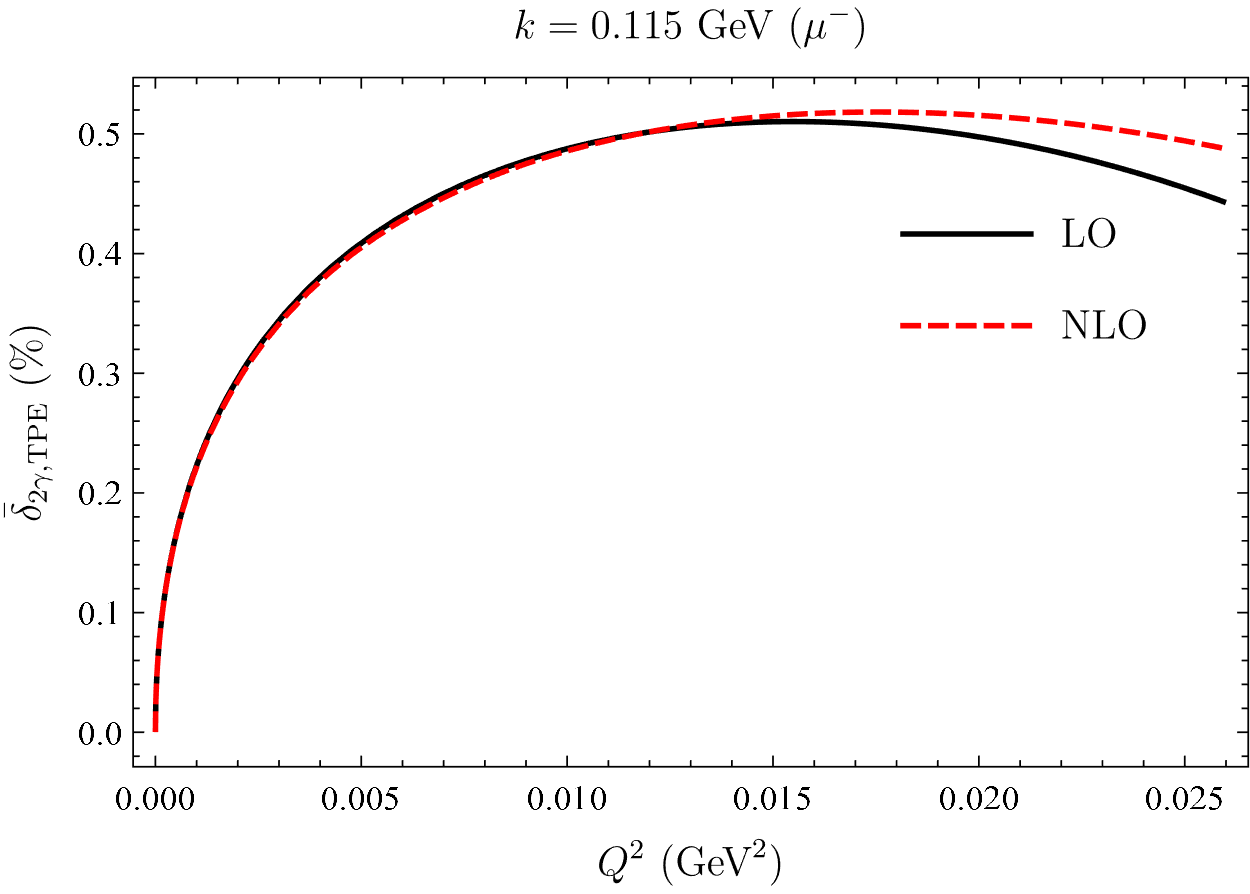}
        \vspace{0.1cm}
        \hspace{0.02cm}
        \end{minipage}
    
        \begin{minipage}[b]{0.45\linewidth}
        \includegraphics[scale=0.5]{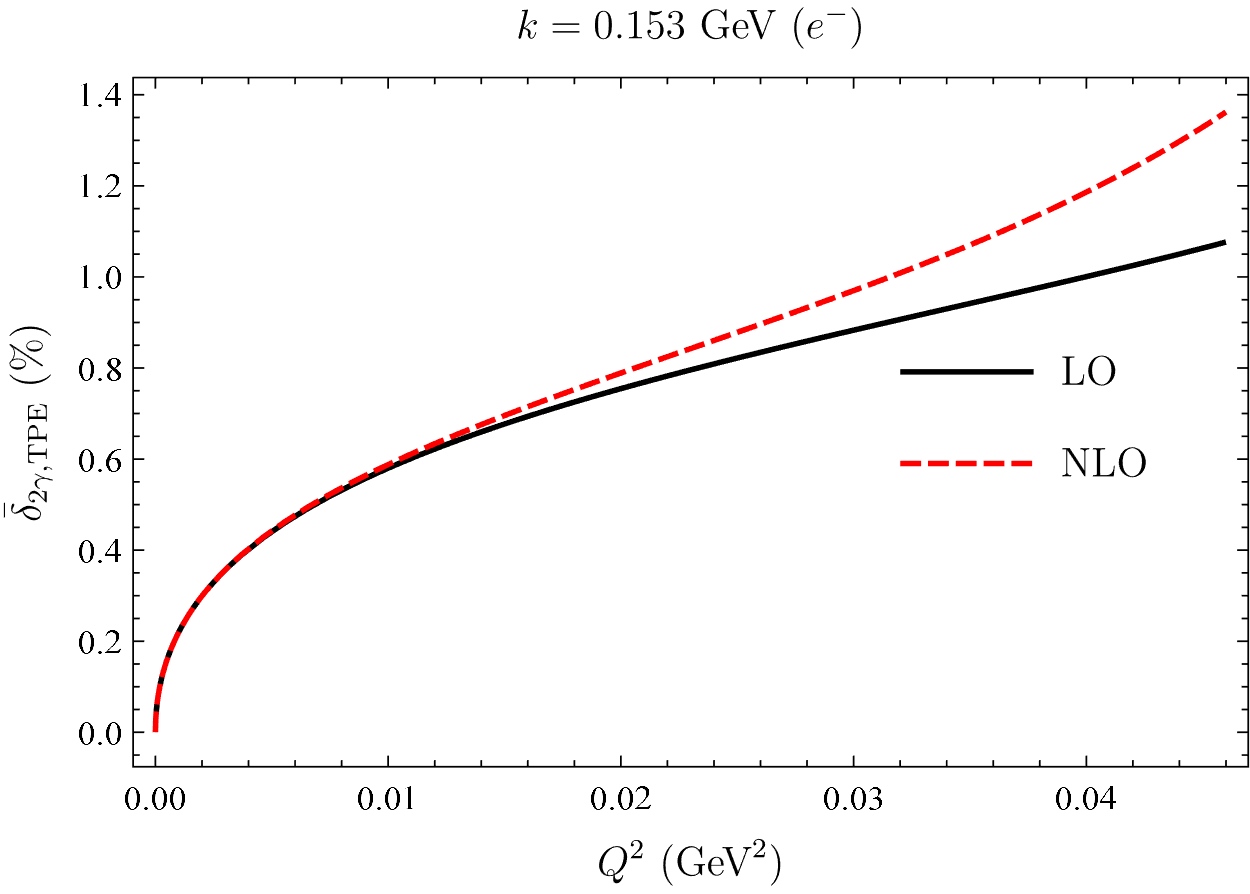}
        \vspace{0.1cm}
        \hspace{0.05cm}
        \end{minipage} 
        \qquad 
        \begin{minipage}[b]{0.45\linewidth}
        \includegraphics[scale=0.5]{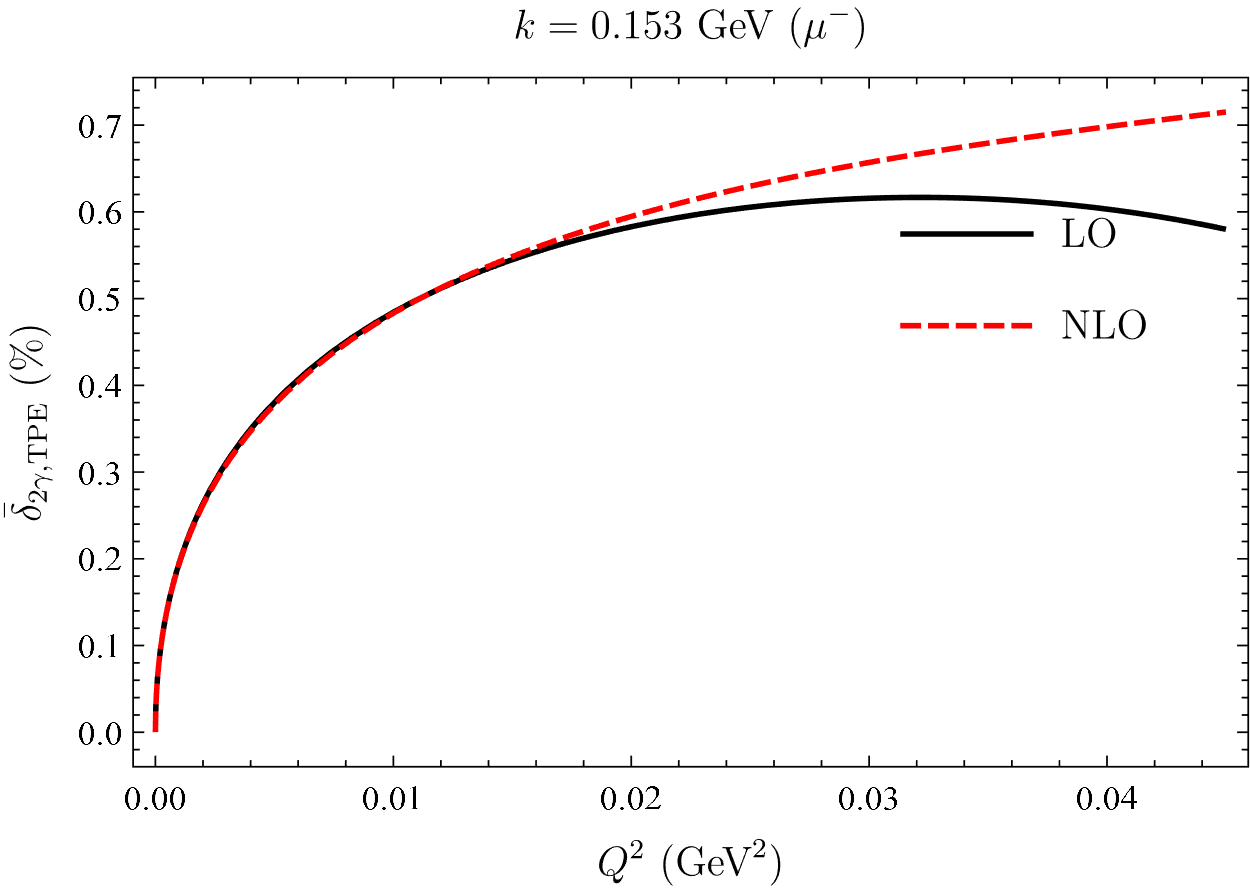}
        \vspace{0.1cm}
        \hspace{0.05cm}
        \end{minipage}

        \begin{minipage}[b]{0.45\linewidth}
        \includegraphics[scale=0.5]{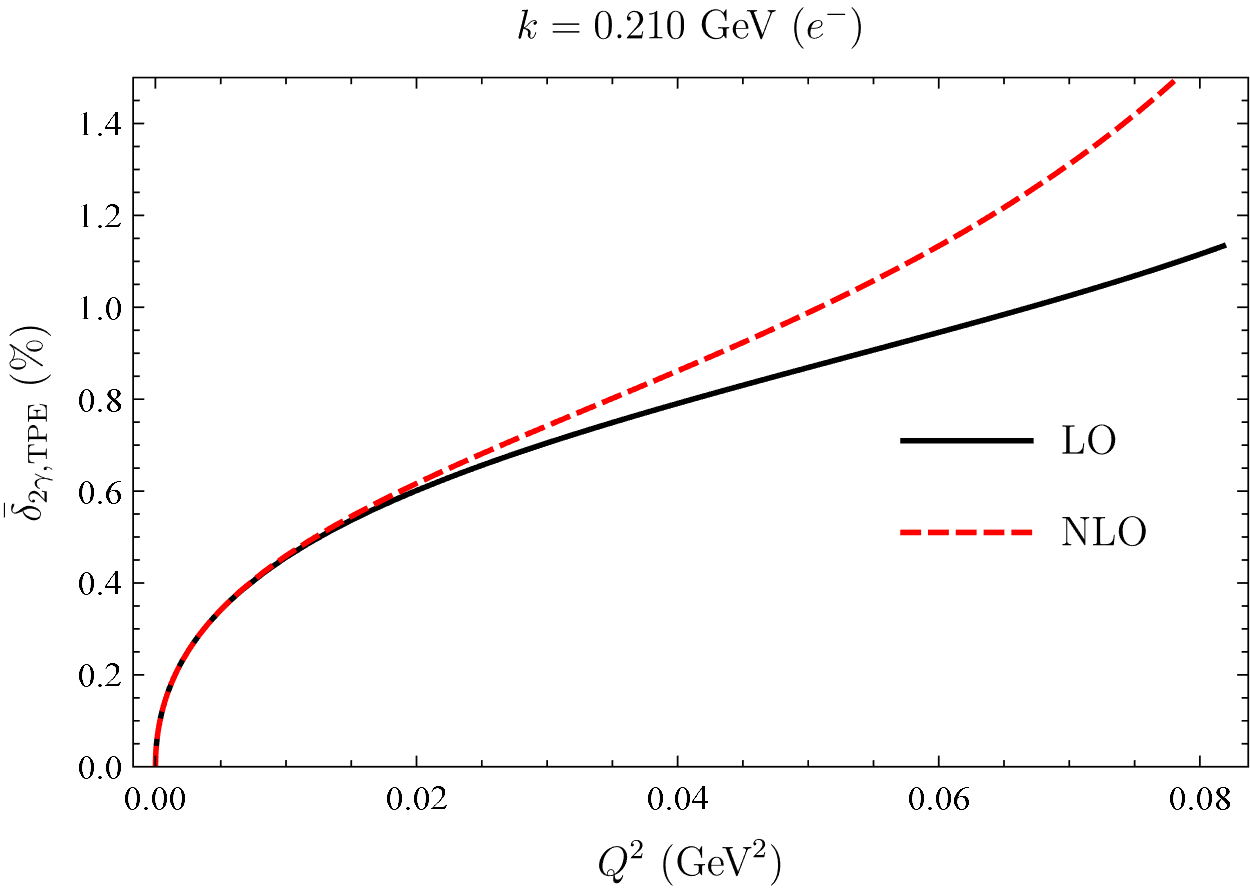}
        \vspace{0.1cm}
        \hspace{0.05cm}
        \end{minipage} 
        \qquad 
        \begin{minipage}[b]{0.45\linewidth}
        \includegraphics[scale=0.5]{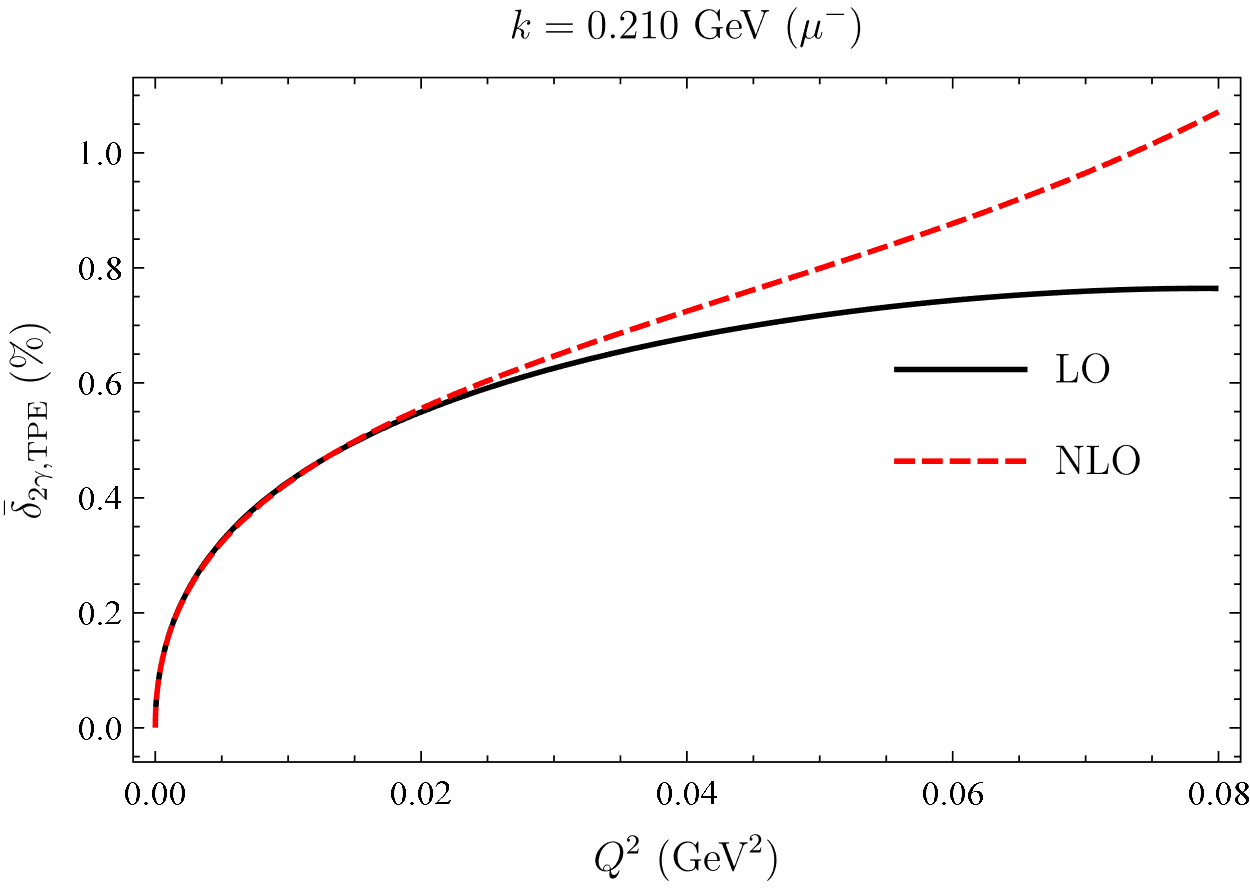}
        \vspace{0.1cm}
        \hspace{0.05cm}
        \end{minipage}
   
        \end{minipage}
    }
    \caption{Comparison of the TPE correction for $e^- \text{p}$ and $\mu^- \text{p}$ elastic scatterings as a function of $Q^2$ in MUSE kinematical region.
    Here, $k$ is the incoming lepton three-momentum in a lab frame.}\label{TPE1}
    \end{figure}

\begin{figure} 
        \centering

        \subfigure{
            \begin{minipage}[b]{0.99\linewidth}
              
            \begin{minipage}[b]{0.45\linewidth}
            \includegraphics[scale=0.45]{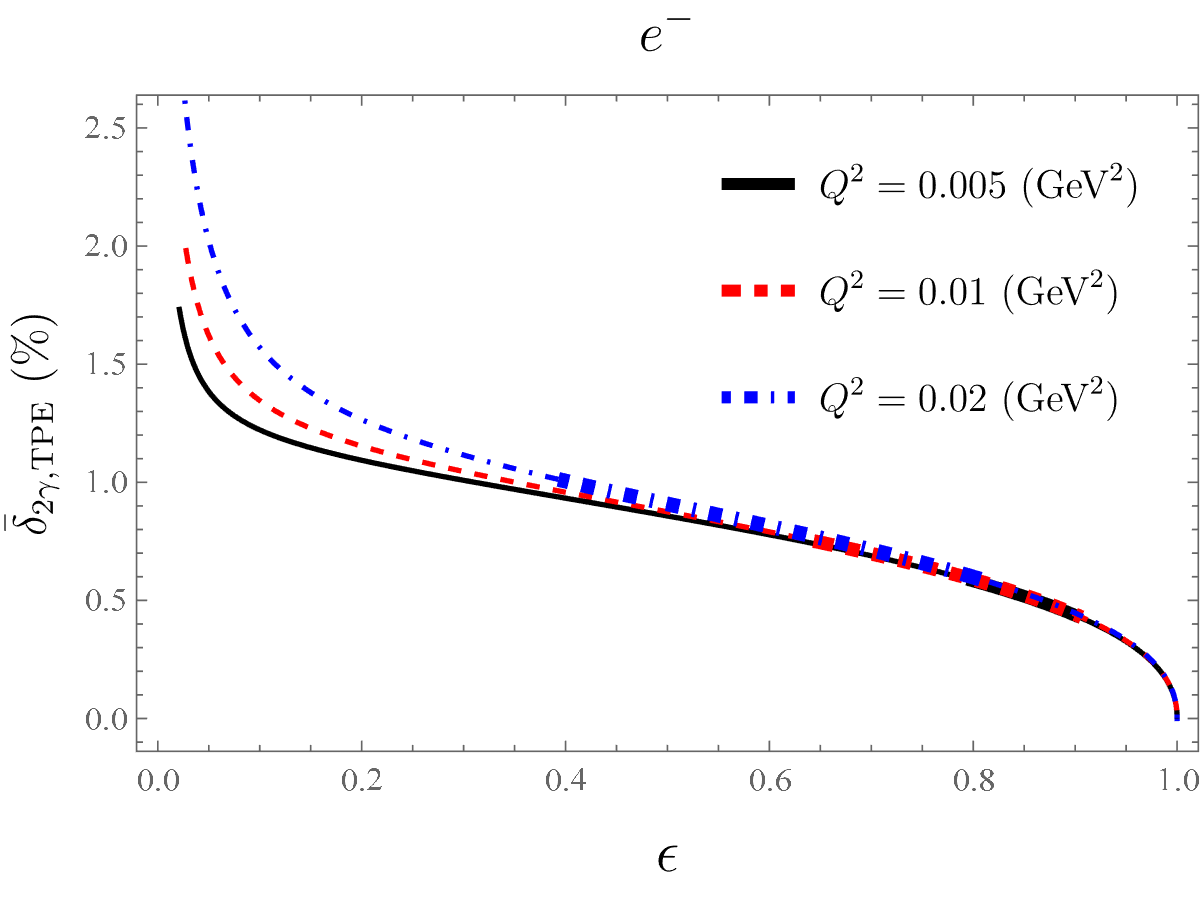}
            \vspace{0.1cm}
            \hspace{0.02cm}
            \end{minipage}
            \qquad
            \begin{minipage}[b]{0.45\linewidth}
            \includegraphics[scale=0.45]{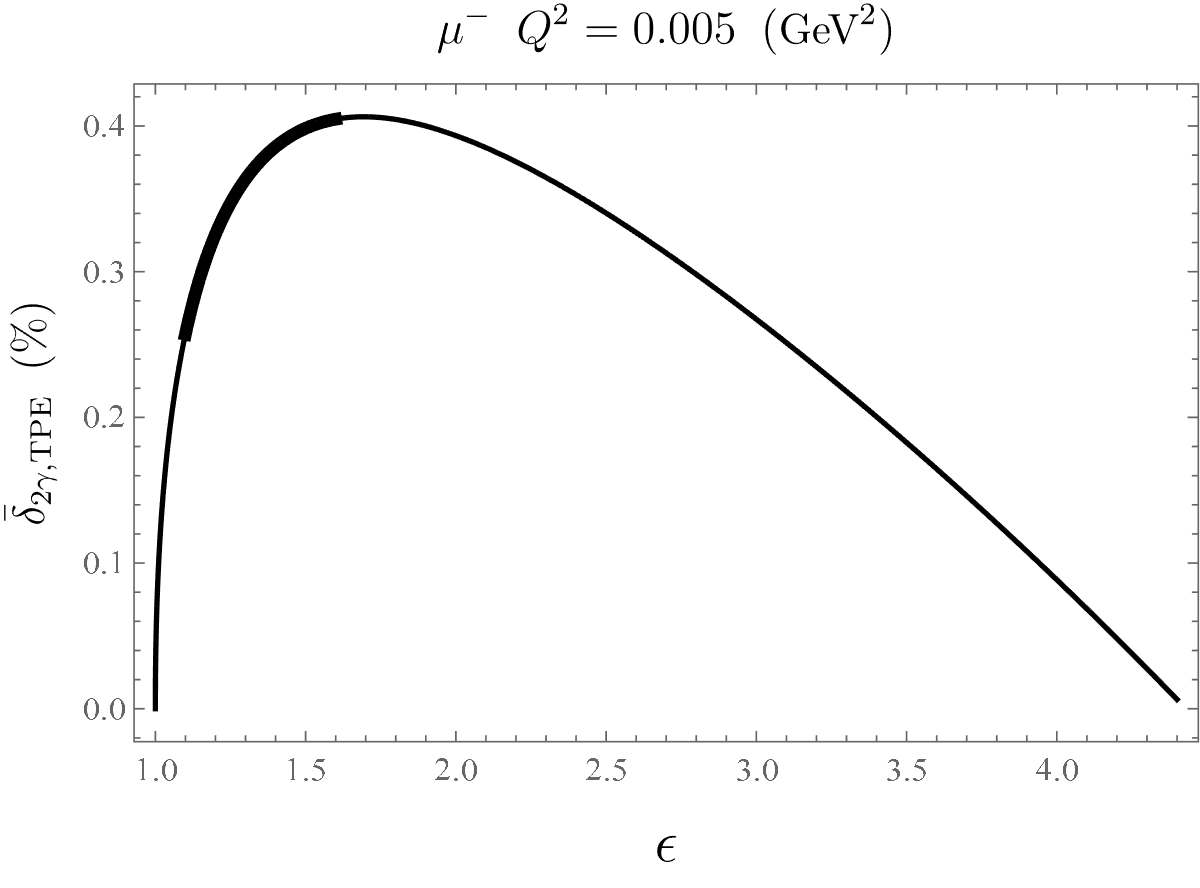}
            \vspace{0.1cm}
            \hspace{0.02cm}
            \end{minipage}
        
            \begin{minipage}[b]{0.45\linewidth}
            \includegraphics[scale=0.45]{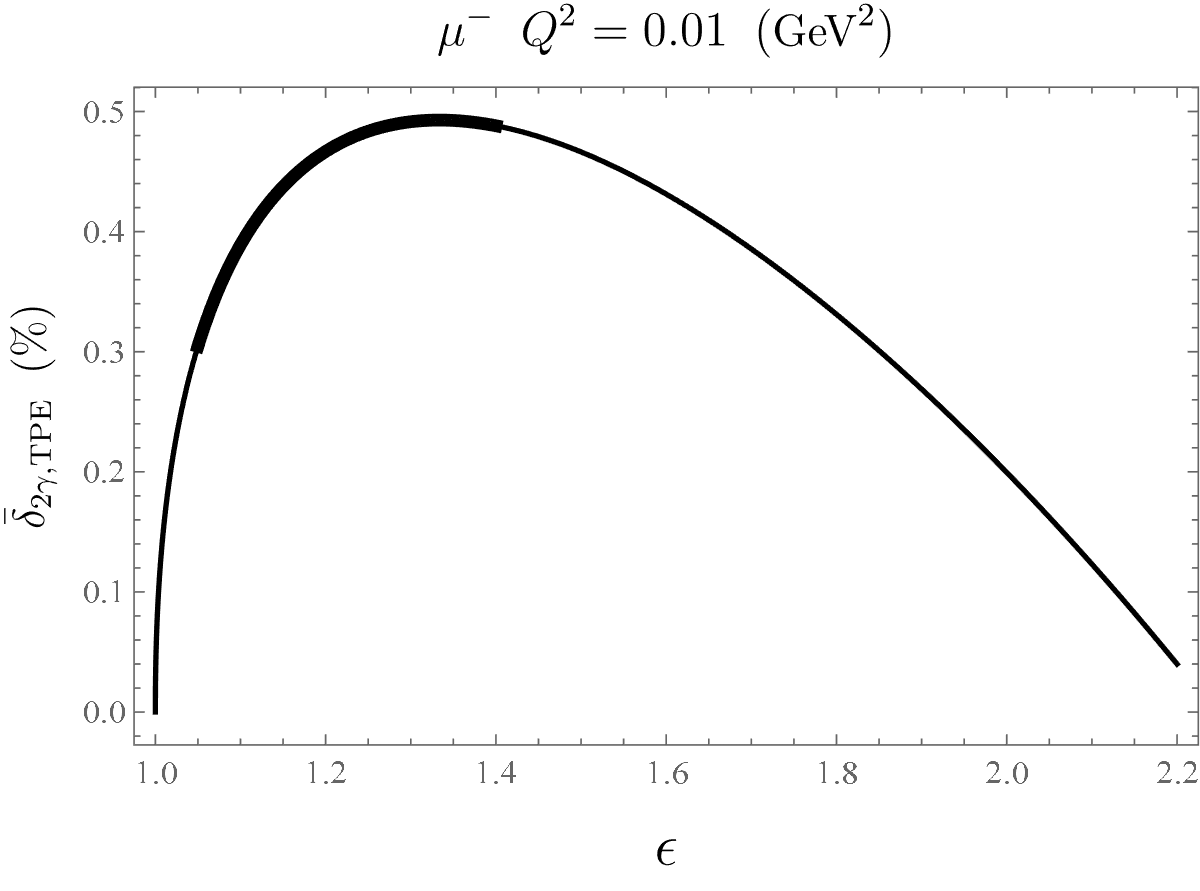}
            \vspace{0.1cm}
            \hspace{0.05cm}
            \end{minipage} 
            \qquad 
            \begin{minipage}[b]{0.45\linewidth}
            \includegraphics[scale=0.45]{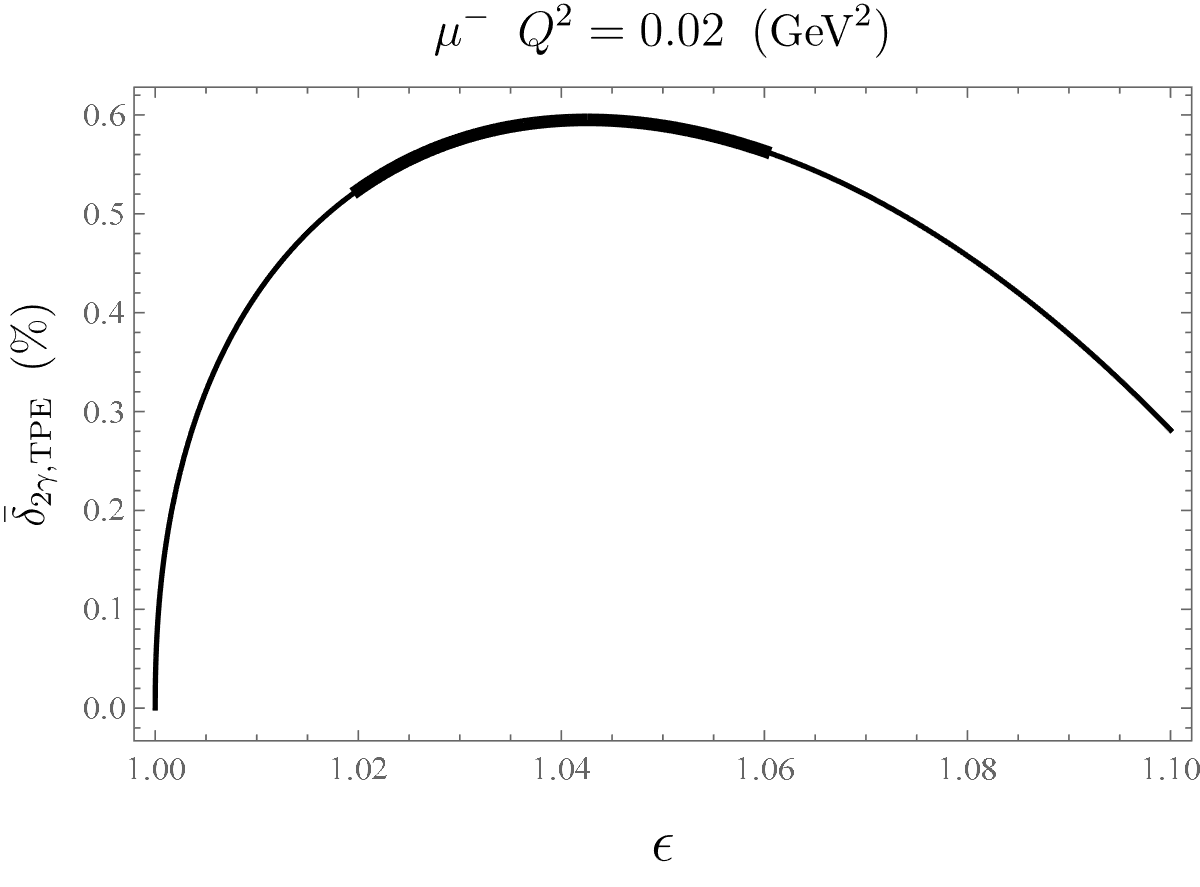}
            \vspace{0.1cm}
            \hspace{0.05cm}
            \end{minipage}
       
            \end{minipage}
        }
        \caption{The $\epsilon$ dependence of the NLO TPE corrections for $e^- \text{p}$ and $\mu^- \text{p}$ elastic scatterings, for different momentum transfers $Q^2$ in MUSE kinematical region.
        Hereafter the thickened segment of each curve corresponds to MUSE kinematical region derived from Tab.~\ref{T2}.
        We do not list the chiral LO results, because the differences are very small compared with the NLO results.}\label{TPE2}
    \end{figure}

As shown in Fig.~\ref{TPE}, the amplitudes from box and crossed box TPE diagrams can be expressed as,
\begin{align}
    i \mathcal{M}_{\text {box }}^{(a)} &=e^{4} \int \frac{\mathrm{d}^{4} k}{(2 \pi)^{4}} \frac{\left[\bar{u}\left(k_2\right) \gamma^{\mu}\left(\sk_1-\sk+m\right) \gamma^{\nu} u(k_1)\right]\left[\bar{p}\left(p_{2}\right) \gamma_\mu \left(\slashed{p_1}+\sk+M\right) \gamma_\nu p\left(p_{1}\right)\right]}{D_1}\ ,  \\
    i \mathcal{M}_{\text {xbox }}^{(b)} &=e^{4} \int \frac{\mathrm{d}^{4} k}{(2 \pi)^{4}} \frac{\left[\bar{u}\left(k_2\right) \gamma^{\mu}\left(\sk_1-\sk+m\right) \gamma^{\nu} u(k_1)\right]\left[\bar{p}\left(p_{2}\right) \gamma_\nu \left(\slashed{p_2}-\sk+M\right) \gamma_\mu p\left(p_{1}\right)\right]}{D_2}\ ,  \\
    i \mathcal{M}_{\text {box }}^{(c)} &=e^{4} \left(c_6+\frac{c_7}{2}\right) \int \frac{\mathrm{d}^{4} k}{(2 \pi)^{4}} \frac{\left[\bar{u}\left(k_2\right) \gamma^{\mu}\left(\sk_1-\sk+m\right) \gamma^{\nu} u(k_1)\right]\left[\bar{p}\left(p_{2}\right) \gamma_\mu \left(\slashed{p_1}+\sk+M\right) \sigma_{\nu\alpha}k^\alpha p\left(p_{1}\right)\right]}{D_1}\ ,  \\
    i \mathcal{M}_{\text {xbox }}^{(d)} &=e^{4} \left(c_6+\frac{c_7}{2}\right) \int \frac{\mathrm{d}^{4} k}{(2 \pi)^{4}} \frac{\left[\bar{u}\left(k_2\right) \gamma^{\mu}\left(\sk_1-\sk+m\right) \gamma^{\nu} u(k_1)\right]\left[\bar{p}\left(p_{2}\right) \gamma_\nu \left(\slashed{p_2}-\sk+M\right) \sigma_{\mu\alpha}(q-k)^\alpha p\left(p_{1}\right)\right]}{D_2}\ ,  \\
    i \mathcal{M}_{\text {box }}^{(e)} &=e^{4} \left(c_6+\frac{c_7}{2}\right) \int \frac{\mathrm{d}^{4} k}{(2 \pi)^{4}} \frac{\left[\bar{u}\left(k_2\right) \gamma^{\mu}\left(\sk_1-\sk+m\right) \gamma^{\nu} u(k_1)\right]\left[\bar{p}\left(p_{2}\right) \sigma_{\mu\alpha}(q-k)^\alpha \left(\slashed{p_1}+\sk+M\right)\gamma_\nu p\left(p_{1}\right)\right]}{D_1}\ ,  \\
    i \mathcal{M}_{\text {xbox }}^{(f)} &=e^{4} \left(c_6+\frac{c_7}{2}\right) \int \frac{\mathrm{d}^{4} k}{(2 \pi)^{4}} \frac{\left[\bar{u}\left(k_2\right) \gamma^{\mu}\left(\sk_1-\sk+m\right) \gamma^{\nu} u(k_1)\right]\left[\bar{p}\left(p_{2}\right) \sigma_{\nu\alpha}k^\alpha \left(\slashed{p_2}-\sk+M\right)\gamma_\mu p\left(p_{1}\right)\right]}{D_2}\ .
\end{align}
where the subscript ``xbox'' is an abbreviation of ``crossed box,'' and 
\begin{align*}
    D_1 &=\left(k^{2}+i 0_+\right)\left[(q-k)^{2}+i 0_+\right]\left[(k_1-k)^2-m^2+i 0_+\right]\left[(p_1+k)^2-M^2+i 0_+\right]\ ,\\
    D_2 &=\left(k^{2}+i 0_+\right)\left[(q-k)^{2}+i 0_+\right]\left[(k_1-k)^2-m^2+i 0_+\right]\left[(p_2-k)^2-M^2+i 0_+\right]\ .
\end{align*}
For brevity, in particular, as for crossed box diagram, one can make use of crossing symmetry.
This requires that the TPE amplitudes obey the relation \cite{Arrington:2011dn}
\begin{align}
    \mathcal{M}_{\mathrm{xbox}}(u, t)=+\left.\mathcal{M}_{\text {box }}(s, t)\right|_{s \rightarrow u}\ .
\end{align} 
So the problems are reduced to how to analytically calculate the box contributions.
By means of the PV reduction~\cite{Passarino:1978jh}, which transforms the complicated calculation of the box integral into the calculation of standard $n$-point integrals.
For those Lorentz-invariant Feynman integrals with massless propagators, the analytical results have been given in Refs.~\cite{Beenakker:1988jr, Ellis:2007qk}.
We use different tools such as \emph{FeynCalc}\cite{Shtabovenko:2016sxi, Shtabovenko:2020gxv}, \emph{PackageX}\cite{Patel:2016fam} and \emph{FeynHelpers}\cite{Shtabovenko:2016whf} to evaluate above integrals and find an unique result.
The final results are lengthy, hence complete analytical expressions are not listed here, but can be obtained from the authors upon request.
The IR divergence of the Feynman diagrams in Fig.~\ref{TPE} may occur in each order of $\chi$PT.
But for differential cross sections, i.e., $\delta_{2\gamma, \text{TPE}}$, no more IR divergences occur except the one from QED of pointlike particles~\cite{Arrington:2011dn}.
In $\chi$PT scheme, the IR divergent term of $\delta_{2\gamma, \text{TPE}}$, $\delta_{2\gamma, \text{TPE}}^{\mathrm{IR}}$, is only from chiral $\mathcal{O}(p)$ contributions~\cite{Tomalak:2014dja},
\begin{align}\label{TPE IR}
    & \delta_{2\gamma, \text{ TPE }}^{\mathrm{IR}} \nonumber\\
    =&-\frac{2 \alpha}{\pi}\left(\frac{1}{\epsilon_{\mathrm{IR}}}-\gamma_{\mathrm{E}}+\ln \left(\frac{4 \pi \nu^{2}}{Q^{2}}\right)\right) \nonumber\\
    & \times\left(\frac{\left(s-m^{2}-M^{2}\right) \ln \left(-\frac{\sqrt{\Sigma_{s}}+m^{2}+M^{2}-s}{2 m M}\right)}{\sqrt{\Sigma_{s}}}-\frac{\left(u-m^{2}-M^{2}\right) \ln \left(\frac{\sqrt{\Sigma_{u}}+m^{2}+M^{2}-u}{2 m M}\right)}{\sqrt{\Sigma_{u}}}\right)\ ,
\end{align} 
with
\begin{align}
    \Sigma_{s} \equiv\left(s-(m+M)^{2}\right)\left(s-(m-M)^{2}\right), \quad \Sigma_{u} \equiv\left(u-(m+M)^{2}\right)\left(u-(m-M)^{2}\right)\ ,
\end{align}
where $\nu$ corresponds to the subtraction scale in DR; $\gamma_\mathrm{E}$ is Euler constant and $\epsilon_{\mathrm{IR}}=(4-D)/2$.
In literature, a nonzero photon mass $\lambda$ was used to renormalize IR divergence, in which the IR part is represented by $\ln\left(\lambda^2/Q^2\right)$, and the simplest comparison can be taken by a substitution:
\begin{align}
    \frac{1}{\epsilon_{\mathrm{IR}}}-\gamma_{E}+\ln \left(\frac{4 \pi \nu^{2}}{Q^{2}}\right) \leftrightarrow \ln \left(\frac{\lambda^{2}}{Q^{2}}\right)\ .
\end{align}
In this way, IR divergence obtained by $\chi$PT is completely consistent with the previous results obtained by SPA calculation\cite{Maximon:2000hm}.
The numerical TPE corrections of the analytical expressions for 
$\overline{\delta}_{2\gamma, \text{ TPE }}=\delta_{2\gamma, \text{ TPE }}-\delta_{2\gamma, \text{ TPE }}^{\mathrm{IR}}$ of $e^- \text{p}$ and $\mu^- \text{p}$ scatterings up to $\mathcal{O}(\alpha p^2)$, 
are shown in Figs.~\ref{TPE1} and \ref{TPE2}.

\begin{figure}[H] 
    \centering
    \subfigure{
        \begin{minipage}[b]{0.98\linewidth}
          
        \begin{minipage}[b]{0.45\linewidth}
        \includegraphics[scale=0.5]{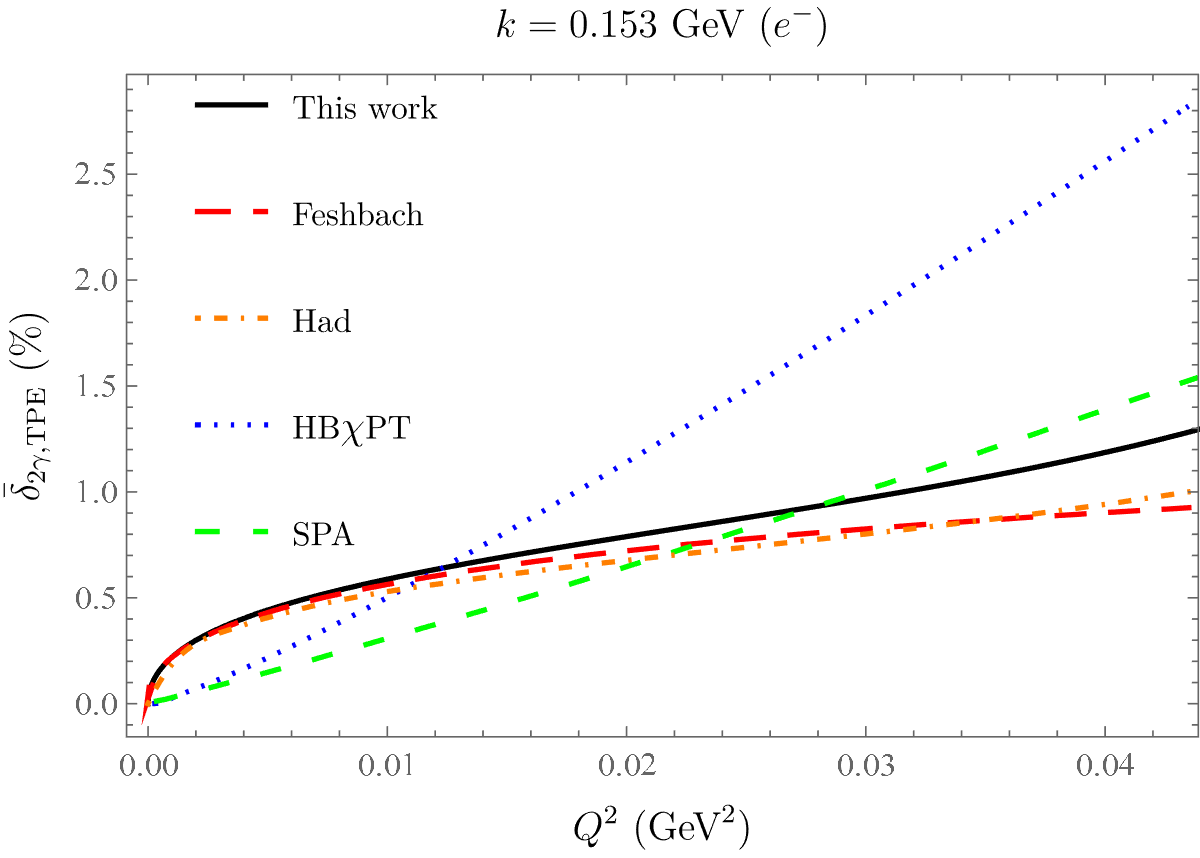}
        \vspace{0.1cm}
        \hspace{0.02cm}
        \end{minipage}
        \qquad
        \begin{minipage}[b]{0.45\linewidth}
        \includegraphics[scale=0.5]{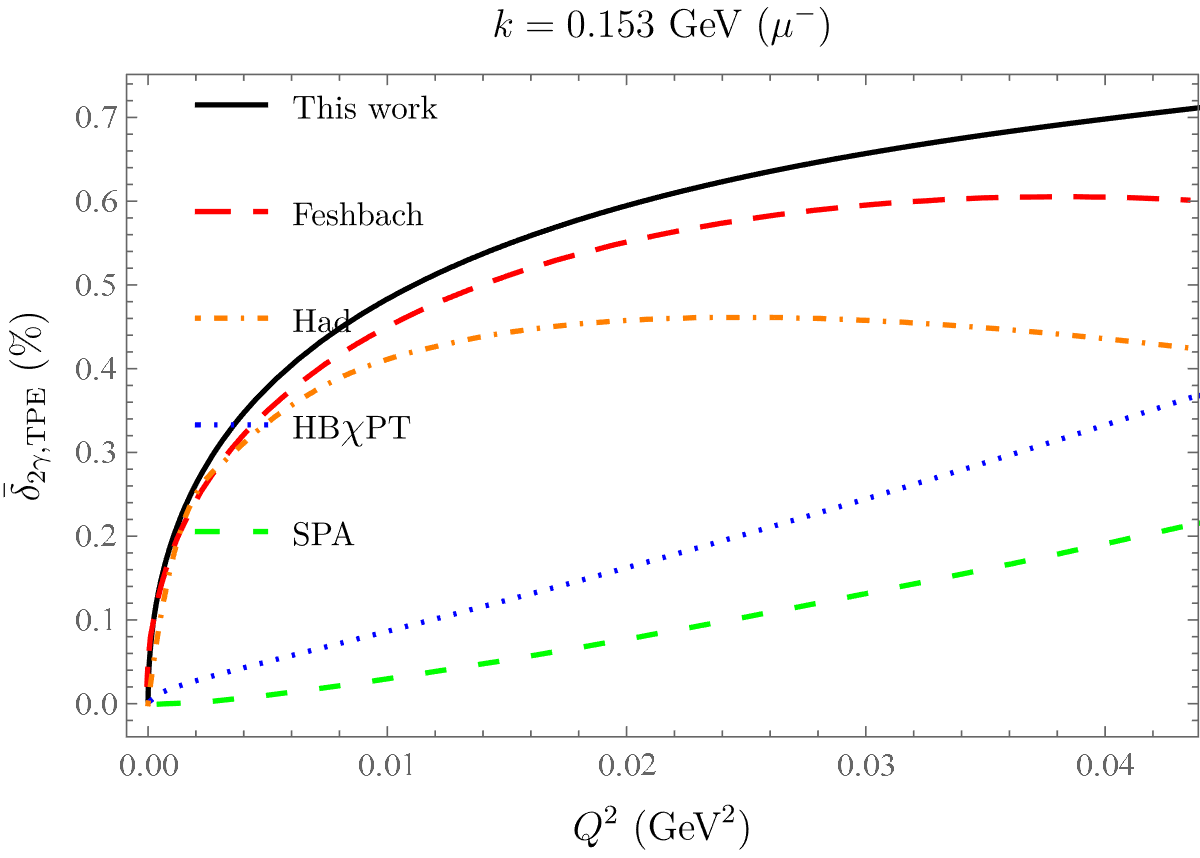}
        \vspace{0.1cm}
        \hspace{0.02cm}
        \end{minipage}

        \end{minipage}
    }
    \caption{Comparison of TPE finite contribution to $e^- p$ and $\mu^- p$ elastic scatterings.
    The contributions of the Feshbach result~\cite{McKinley:1948zz}(labeled as ``Feshbach''), 
    hadron model calculation~\cite{Tomalak:2014dja}(labeled as ``Had''), 
    model-independent calculation based on SPA~\cite{Koshchii:2017dzr}(labeled as ``SPA''), and recent HB$\chi$PT calculation also based on SPA~\cite{Talukdar:2019dko}(labeled as ``HB$\chi$PT'')
    are displayed.
    }\label{TPE comparison fig}
\end{figure}

The estimation of the results displayed in Fig.~\ref{TPE1}, indicates that the TPE corrections in $e \text{p}$ elastic scatterings vary between $1\%$ and $1.5\%$ in MUSE kinematical region, and between $0.5\%$ and $1\%$ for $\mu \text{p}$ scatterings.
In Fig.~\ref{TPE comparison fig}, comparing with conventional Feshbach's result~\cite{McKinley:1948zz} and recent papers, e.g., Refs.~\cite{Tomalak:2014dja ,Zhou:2016psq,Talukdar:2019dko}, 
the contributions are close to the results of hadron model\cite{Tomalak:2014dja ,Zhou:2016psq} without using SPA. 
There are significant differences comparing with the estimation of HB$\chi$PT\cite{Talukdar:2019dko} when using SPA, whatever in $e \text{p}$ or $\mu \text{p}$ scatterings.
At the same time, it can be seen that the so-called model independent results obtained from SPA in Ref.~\cite{Koshchii:2017dzr} also underestimates TPE effects due to ignoring the contribution of hard momentum region of box diagrams.
Interestingly, using SPA naively may result in an unphysical consequence that $\delta_{2\gamma, \text{TPE}} \nrightarrow 0$ in the forward limit $Q^2 \to 0$.
The authors of Ref.\cite{Koshchii:2017dzr} therefore forced $\delta_{2\gamma, \text{TPE}} \to 0$ by virtue of shifting a constant factor when $Q^2 \to 0$ (similar to applying an additional on shell renormalization procedure).
Such manipulation also corresponds to the subtracted dispersion relation evaluation with a $Q^2$ dependent subtraction function in the forward limit~\cite{Tomalak:2018jak}.
The subtraction function renormalizes the effects of some momentum dependent couplings in a proper way.

\section{The calculation of the bremsstrahlung diagrams}\label{sec:4}
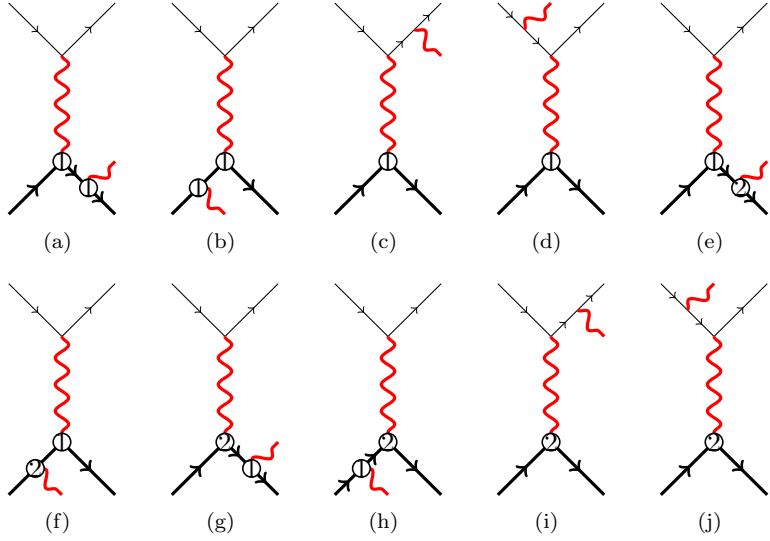
\begin{figure}
    \centering
    \subfigure[]{
    \begin{tikzpicture}[scale = 0.7]
        \draw[nucleon](-1,3)to(0,2);
        \draw[nucleon](0,2)to(1,3);
        \draw[nucleon, very thick](-1,-1)to(0,0);
        \draw[nucleon, very thick](0,0)to(0.5,-0.5);
        \draw[nucleon, very thick](0.5,-0.5)to(1,-1);
        \draw[photon, very thick](0,2)to(0,0);
        \draw[photon, very thick](0.5,-0.5)to(1,0);
        
        \draw[very thick](0,0)circle(0.15);
        \fill[white](0,0) circle(0.15);
        \node at (0,0) {$1$};
        \draw[very thick](0.5,-0.5)circle(0.15);
        \fill[white](0.5,-0.5) circle(0.15);
        \node at (0.5,-0.5) {$1$};
    \end{tikzpicture}
    } 
    \quad
    \subfigure[]{
    \begin{tikzpicture}[scale = 0.7]
        \draw[nucleon](-1,3)to(0,2);
        \draw[nucleon](0,2)to(1,3);
        \draw[nucleon, very thick](-1,-1)to(0,0);
        \draw[nucleon, very thick](0,0)to(1,-1);
        \draw[photon, very thick](0,2)to(0,0);
        \draw[photon, very thick](-0.5,-0.5)to(0,-1);
        
        \draw[very thick](0,0)circle(0.15);
        \fill[white](0,0) circle(0.15);
        \node at (0,0) {$1$};
        \draw[very thick](-0.5,-0.5)circle(0.15);
        \fill[white](-0.5,-0.5) circle(0.15);
        \node at (-0.5,-0.5) {$1$};
    \end{tikzpicture}
    }
    \quad
    \subfigure[]{
    \begin{tikzpicture}[scale = 0.7]
        \draw[nucleon](-1,3)to(0,2);
        \draw[nucleon](0,2)to(0.5,2.5);
        \draw[nucleon](0.5,2.5)to(1,3);
        \draw[nucleon, very thick](-1,-1)to(0,0);
        \draw[nucleon, very thick](0,0)to(1,-1);
        \draw[photon, very thick](0,2)to(0,0);
        \draw[photon, very thick](0.5,2.5)to(1,2);
        
        \draw[very thick](0,0)circle(0.15);
        \fill[white](0,0) circle(0.15);
        \node at (0,0) {$1$};
    \end{tikzpicture}
    } 
    \quad
    \subfigure[]{
    \begin{tikzpicture}[scale = 0.7]
        \draw[nucleon](-1,3)to(-0.5,2.5);
        \draw[nucleon](-0.5,2.5)to(0,2);
        \draw[nucleon](0,2)to(1,3);
        \draw[nucleon, very thick](-1,-1)to(0,0);
        \draw[nucleon, very thick](0,0)to(1,-1);
        \draw[photon, very thick](0,2)to(0,0);
        \draw[photon, very thick](-0.5,2.5)to(0,3);
        
        \draw[very thick](0,0)circle(0.15);
        \fill[white](0,0) circle(0.15);
        \node at (0,0) {$1$};
    \end{tikzpicture}
    }
    \quad
    \subfigure[]{
        \begin{tikzpicture}[scale = 0.7]
            \draw[nucleon](-1,3)to(0,2);
            \draw[nucleon](0,2)to(1,3);
            \draw[nucleon, very thick](-1,-1)to(0,0);
            \draw[nucleon, very thick](0,0)to(0.5,-0.5);
            \draw[nucleon, very thick](0.5,-0.5)to(1,-1);
            \draw[photon, very thick](0,2)to(0,0);
            \draw[photon, very thick](0.5,-0.5)to(1,0);
            
            \draw[very thick](0,0)circle(0.15);
            \fill[white](0,0) circle(0.15);
            \node at (0,0) {$1$};
            \draw[very thick](0.5,-0.5)circle(0.15);
            \fill[white](0.5,-0.5) circle(0.15);
            \node at (0.5,-0.5) {$2$};
        \end{tikzpicture}
    }

    \subfigure[]{
        \begin{tikzpicture}[scale = 0.7]
            \draw[nucleon](-1,3)to(0,2);
            \draw[nucleon](0,2)to(1,3);
            \draw[nucleon, very thick](-1,-1)to(0,0);
            \draw[nucleon, very thick](0,0)to(1,-1);
            \draw[photon, very thick](0,2)to(0,0);
            \draw[photon, very thick](-0.5,-0.5)to(0,-1);
            
            \draw[very thick](0,0)circle(0.15);
            \fill[white](0,0) circle(0.15);
            \node at (0,0) {$1$};
            \draw[very thick](-0.5,-0.5)circle(0.15);
            \fill[white](-0.5,-0.5) circle(0.15);
            \node at (-0.5,-0.5) {$2$};
        \end{tikzpicture}
    }
    \quad
        \subfigure[]{
        \begin{tikzpicture}[scale = 0.7]
            \draw[nucleon](-1,3)to(0,2);
            \draw[nucleon](0,2)to(1,3);
            \draw[nucleon, very thick](-1,-1)to(0,0);
            \draw[nucleon, very thick](0,0)to(0.5,-0.5);
            \draw[nucleon, very thick](0.5,-0.5)to(1,-1);
            \draw[photon, very thick](0,2)to(0,0);
            \draw[photon, very thick](0.5,-0.5)to(1,0);
            
            \draw[very thick](0,0)circle(0.15);
            \fill[white](0,0) circle(0.15);
            \node at (0,0) {$2$};
            \draw[very thick](0.5,-0.5)circle(0.15);
            \fill[white](0.5,-0.5) circle(0.15);
            \node at (0.5,-0.5) {$1$};
        \end{tikzpicture}
    } 
    \quad
        \subfigure[]{
        \begin{tikzpicture}[scale = 0.7]
            \draw[nucleon](-1,3)to(0,2);
            \draw[nucleon](0,2)to(1,3);
            \draw[nucleon, very thick](-1,-1)to(-0.5,-0.5);
            \draw[nucleon, very thick](-0.5,-0.5)to(0,0);
            \draw[nucleon, very thick](0,0)to(1,-1);
            \draw[photon, very thick](0,2)to(0,0);
            \draw[photon, very thick](-0.5,-0.5)to(0,-1);
            
            \draw[very thick](0,0)circle(0.15);
            \fill[white](0,0) circle(0.15);
            \node at (0,0) {$2$};
            \draw[very thick](-0.5,-0.5)circle(0.15);
            \fill[white](-0.5,-0.5) circle(0.15);
            \node at (-0.5,-0.5) {$1$};
        \end{tikzpicture}
    }
    \quad
        \subfigure[]{
        \begin{tikzpicture}[scale = 0.7]
            \draw[nucleon](-1,3)to(0,2);
            \draw[nucleon](0,2)to(0.5,2.5);
            \draw[nucleon](0.5,2.5)to(1,3);
            \draw[nucleon, very thick](-1,-1)to(0,0);
            \draw[nucleon, very thick](0,0)to(1,-1);
            \draw[photon, very thick](0,2)to(0,0);
            \draw[photon, very thick](0.5,2.5)to(1,2);
            
            \draw[very thick](0,0)circle(0.15);
            \fill[white](0,0) circle(0.15);
            \node at (0,0) {$2$};
        \end{tikzpicture}
    } 
    \quad
        \subfigure[]{
        \begin{tikzpicture}[scale = 0.7]
            \draw[nucleon](-1,3)to(-0.5,2.5);
            \draw[nucleon](-0.5,2.5)to(0,2);
            \draw[nucleon](0,2)to(1,3);
            \draw[nucleon, very thick](-1,-1)to(0,0);
            \draw[nucleon, very thick](0,0)to(1,-1);
            \draw[photon, very thick](0,2)to(0,0);
            \draw[photon, very thick](-0.5,2.5)to(0,3);
            
            \draw[very thick](0,0)circle(0.15);
            \fill[white](0,0) circle(0.15);
            \node at (0,0) {$2$};
        \end{tikzpicture}
    }
    \caption{Bremsstrahlung diagrams of chiral LO and NLO.}\label{BR}
\end{figure}

Bremsstrahlung diagrams are shown in Fig.~\ref{BR}.
For the convenience of comparison, we artificially distinguish between the bremsstrahlung caused by the interference of real photon emission from lepton and that of proton (crossed bremsstrahlung contributions), and other contributions (direct bremsstrahlung contributions). 
The former cancels the IR divergence derived from the TPE corrections, and the latter cancels the IR divergence of the vertex corrections.

Soft real photon bremsstrahlung where the emission energy below the resolution of the lab detector, $\Delta E_\gamma$, is indistinguishable from elastic scatterings. 
It should be mentioned that the separation of a photon's phase space into soft and hard regions is somewhat arbitrary.
According to the features of MUSE experiment, we can set $\Delta E_{\gamma} \simeq 1 \% E_{1}$.
It is quite difficult to estimate the soft bremsstrahlung contributions analytically in DR.
The commonly used prescription is SPA\cite{Tsai:1961zz, Maximon:2000hm}.
The following results are given:
\begin{align}
    &\mathcal{M}_{2\gamma, \text{br}}^{(a)}=\frac{e^3\bar{u}(k_2)\gamma^\mu u(k_1)\bar{p}(p_2)\sep^*(k)(\slashed{p_2}+\sk+M)\gamma_\mu p(p_1)}{(q_\ell^2)[(p_2+k)^2-M^2]} \xLongrightarrow[k\to 0]{\mathrm{SPA}}e\left(-\frac{p_2\cdot \epsilon^*}{p_2\cdot k}\right)\mathcal{M}_\gamma^{(1)}\ , \\
    &\mathcal{M}_{2\gamma, \text{br}}^{(b)}=\frac{e^3\bar{u}(k_2)\gamma^\mu u(k_1)\bar{p}(p_2)\gamma_\mu(\slashed{p_1}-\sk+M)\sep^*(k) p(p_1)}{(q_\ell^2)[(p_1-k)^2-M^2]} \xLongrightarrow[k\to 0]{\mathrm{SPA}} e\left(\frac{p_1\cdot \epsilon^*}{p_1\cdot k}\right)\mathcal{M}_\gamma^{(1)} \ , \\
    &\mathcal{M}_{2\gamma, \text{br}}^{(c)}=-\frac{e^3\bar{u}(k_2)\sep^*(\sk_2+\sk+m)\gamma^\mu u(k_1)\bar{p}(p_2)\gamma_\mu p(p_1)}{(q^2)[(k_2+k)^2-m^2]} \xLongrightarrow[k\to 0]{\mathrm{SPA}} e\left(\frac{k_2\cdot \epsilon^*}{k_2\cdot k}\right)\mathcal{M}_\gamma^{(1)}  \ , \\
    &\mathcal{M}_{2\gamma, \text{br}}^{(d)}=-\frac{e^3\bar{u}(k_2)\gamma^\mu(\sk_1-\sk+m)\sep^* u(k_1)\bar{p}(p_2)\gamma_\mu p(p_1)}{(q^2)[(k_1-k)^2-m^2]} \xLongrightarrow[k\to 0]{\mathrm{SPA}} e\left(-\frac{k_1\cdot \epsilon^*}{k_1\cdot k}\right)\mathcal{M}_\gamma^{(1)}  \ , \\
    &\mathcal{M}_{2\gamma, \text{br}}^{(e)}=i e^3\left(c_6+\frac{c_7}{2}\right)\bar{u}(k_2)\gamma^\mu u(k_1)\frac{1}{q_\ell^2}\bar{p}(p_2)\epsilon^*_\alpha \sigma^{\alpha \nu}k_\nu \frac{\left(\sk+\slashed{p_2}+M\right)}{(k+p_2)^2-M^2}\gamma_\mu p(p_1) \xLongrightarrow[k\to 0]{\mathrm{SPA}} 0  \ , \\
    &\mathcal{M}_{2\gamma, \text{br}}^{(f)}=i e^3\left(c_6+\frac{c_7}{2}\right)\bar{u}(k_2)\gamma^\mu u(k_1)\frac{1}{q_\ell^2}\bar{p}(p_2)\gamma^\mu \frac{\left(-\sk+\slashed{p_1}+M\right)}{(p_1-k)^2-M^2}\epsilon^*_\alpha \sigma^{\alpha \nu}k_\nu p(p_1) \xLongrightarrow[k\to 0]{\mathrm{SPA}} 0  \ , \\
    &\mathcal{M}_{2\gamma, \text{br}}^{(g)}=\frac{i e^3\left(c_6+\frac{c_7}{2}\right)\bar{u}(k_2)\gamma^\mu u(k_1) \bar{p}(p_2)\sep^*\left(\sk+\slashed{p_2}+M\right)\sigma_{\mu \nu}q_{\ell}^\nu p(p_1)}{q_\ell^2[(k+p_2)^2-M^2]} \xLongrightarrow[k\to 0]{\mathrm{SPA}} e\left(-\frac{p_2\cdot \epsilon^*}{p_2\cdot k}\right) \mathcal{M}_{\gamma}^{(2)} \ , \\
    &\mathcal{M}_{2\gamma, \text{br}}^{(h)}=\frac{i e^3\left(c_6+\frac{c_7}{2}\right)\bar{u}(k_2)\gamma^\mu u(k_1) \bar{p}(p_2)\sigma_{\mu \nu}q_{\ell}^\nu\left(\slashed{p_1}-\sk+M\right)\sep^* p(p_1)}{q_\ell^2[(p_1-k)^2-M^2]} \xLongrightarrow[k\to 0]{\mathrm{SPA}} e\left(\frac{p_1\cdot \epsilon^*}{p_1\cdot k}\right) \mathcal{M}_{\gamma}^{(2)}  \ , \\
    &\mathcal{M}_{2\gamma, \text{br}}^{(i)}=-\frac{i e^3 \left(c_6+\frac{c_7}{2}\right)\bar{u}(k_2)\sep^*\left(\sk_2+\sk+m\right)\gamma^\mu u(k_1) \bar{p}(p_2)\sigma_{\mu \nu}q^\nu p(p_1)}{q^2[(k_2+k)^2-m^2]} \xLongrightarrow[k\to 0]{\mathrm{SPA}} e\left(\frac{k_2\cdot \epsilon^*}{k_2\cdot k}\right) \mathcal{M}_{\gamma}^{(2)} \ , \\
    &\mathcal{M}_{2\gamma, \text{br}}^{(j)}=-\frac{i e^3 \left(c_6+\frac{c_7}{2}\right)\bar{u}(k_2)\gamma^\mu\left(\sk_1-\sk+m\right)\sep^* u(k_1) \bar{p}(p_2)\sigma_{\mu \nu}q^\nu p(p_1)}{q^2[(k_1-k)^2-m^2]}\xLongrightarrow[k\to 0]{\mathrm{SPA}} e\left(-\frac{k_1\cdot \epsilon^*}{k_1\cdot k}\right) \mathcal{M}_{\gamma}^{(2)} \ ,
\end{align}
where the subscript ``br'' represents bremsstrahlung and $\epsilon^*_\mu(k)$ denotes the polarization of the real emitted photon.

\subsection{Crossed bremsstrahlung contributions}

The square of soft crossed bremsstrahlung amplitudes at LO in lab frame are given by
\begin{align}
    &\md \sigma_{2\gamma, \text{xbr}}^{(1)}=\frac{1}{4M E_1}\frac{E_1}{|\bk_1|}\frac{\md^3 k_2}{(2 \pi)^3 2E_2}\frac{\md^3 p_2}{(2 \pi)^3 2E^\prime_2}\frac{\md^3 k}{(2 \pi)^3 2E_\gamma}  \nonumber\\
    &\times (2\pi)^4 \delta^4(k_1+p_1-k_2-p_2-k) \frac{1}{4}\sum_{\text{spins}}\left[2\mathcal{R}e \left(\mathcal{M}_{2\gamma, \text{br}}^{(a)}+\mathcal{M}_{2\gamma, \text{br}}^{(b)}\right)^\dagger\left(\mathcal{M}_{2\gamma, \text{br}}^{(c)}+\mathcal{M}_{2\gamma, \text{br}}^{(d)}\right)\right]\ ,
\end{align} 
where the subscript ``xbr'' is a crossed bremsstrahlung.
The amplitudes obtained from SPA are
\begin{align}
    \sum_{\text{spins}} &\left[2\mathcal{R}e \left(\mathcal{M}_{2\gamma, \text{br}}^{(a)}+\mathcal{M}_{2\gamma, \text{br}}^{(b)}\right)^\dagger\left(\mathcal{M}_{2\gamma, \text{br}}^{(c)}+\mathcal{M}_{2\gamma, \text{br}}^{(d)}\right)\right]  \nonumber\\
    &\xLongrightarrow[k\to 0]{\mathrm{SPA}} e^2 \sum_{\text{spins}} |\mathcal{M}_{\gamma}^{(0)}|^2 \times 2\left(\frac{p_2\cdot k_2}{p_2\cdot k k_2\cdot k}-\frac{p_2\cdot k_1}{p_2\cdot k k_1\cdot k}-\frac{p_1\cdot k_2}{p_1\cdot k k_2\cdot k}+\frac{p_1\cdot k_1}{p_1\cdot k k_1\cdot k}\right)\ .
\end{align}

The integral of emitted photon phase space is IR divergent, and the standard approach is to consider a special frame, which is sometimes called the S frame\cite{Tsai:1960zz}, to avoid the dependence of the angle of the radiated photon.
The crucial feature of S frame is the setting of $\bp_2+\bk=\bq_\ell +\bp_1=\boldsymbol{0}$.
That is, the CM frame of the final state recoil proton and radiated photon.
The details of the S frame are given in Refs.~\cite{Vanderhaeghen:2000ws,Talukdar:2020aui}, and the results are given directly in the S frame,
\begin{align}\label{int1}
    \left(\frac{\mathrm{d} \sigma^{(1)}}{\mathrm{d} \Omega_{\ell}^{\prime}}\right)_{2\gamma, \text{xbr}}=\left(\frac{\mathrm{d} \sigma^{(1)}}{\mathrm{d} \Omega_{\ell}^{\prime}}\right)_{\gamma} \times e^{2} \int \frac{\mathrm{d}^{3} k}{(2 \pi)^{3} 2 k^{0}} 2\left(\frac{p_{2} \cdot k_{2}}{p_{2} \cdot k k_{2} \cdot k}-\frac{p_{2} \cdot k_{1}}{p_{2} \cdot k k_{1} \cdot k}-\frac{p_{1} \cdot k_{2}}{p_{1} \cdot k k_{2} \cdot k}+\frac{p_{1} \cdot k_{1}}{p_{1} \cdot k k_{1} \cdot k}\right)\ .
\end{align}
It should be mentioned that the integral of radiated photons is calculated in the S frame, and the final results need to be transformed to the lab frame or expressed as Lorentz-invariant form.
When evaluating Eq.~(\ref{int1}), the only problem is how to deal with the integral,
\begin{align}
    I(k_i, p_j) \equiv \int^{k<\frac{\Delta E^{S}}{\nu}} \frac{\mathrm{d}^{D-1} k}{(2 \pi)^{D-1}} \frac{1}{2 k^{0}} \frac{1}{\left(k_{i} \cdot k\right)\left(p_{j} \cdot k\right)}\ ,
\end{align}
where $\Delta E^{S}$ is the upper limit of the integration over the photon energy in S frame.
In this section, $k$ should be understood as the dimensionless variable $k/\nu$.
The calculation of this integral under DR can be found in Appendix.~\ref{appendix:1}.

\begin{figure}[H] 
    \centering
    \subfigure{
        \begin{minipage}[b]{0.98\linewidth}
          
        \begin{minipage}[b]{0.45\linewidth}
        \includegraphics[scale=0.5]{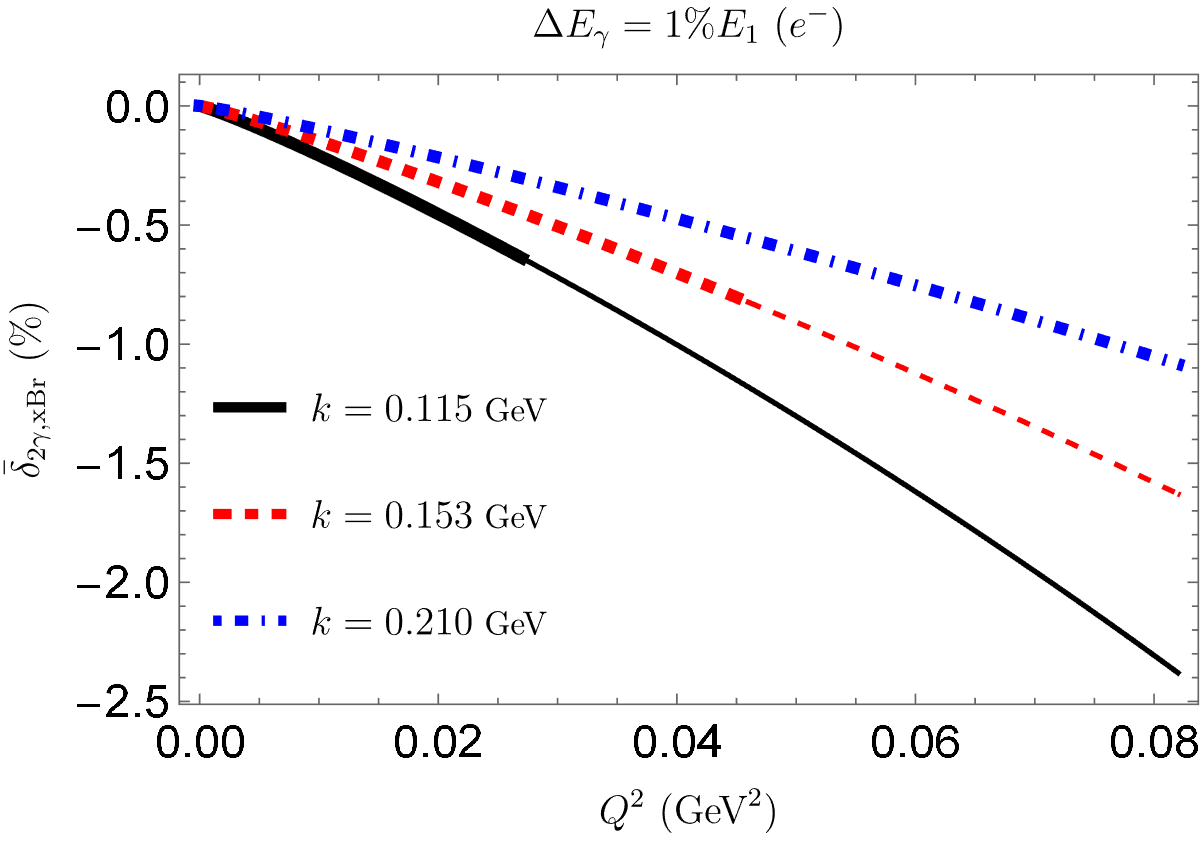}
        \vspace{0.1cm}
        \hspace{0.02cm}
        \end{minipage}
        \qquad
        \begin{minipage}[b]{0.45\linewidth}
        \includegraphics[scale=0.5]{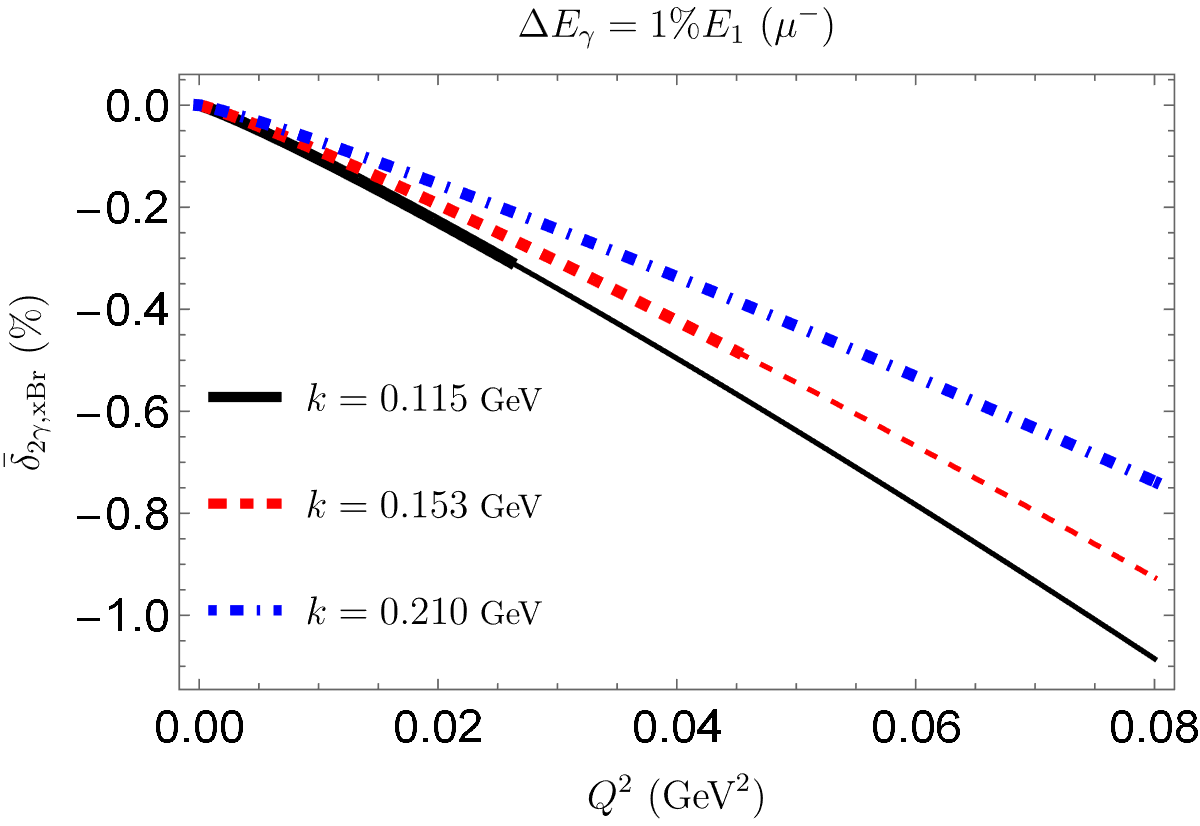}
        \vspace{0.1cm}
        \hspace{0.02cm}
        \end{minipage}

        \end{minipage}
    }
    \caption{The $Q^2$ dependence of the crossed bremsstrahlung corrections to $e^- \text{p}$ and $\mu^- \text{p}$ elastic scatterings.}\label{xbr fig}
    \end{figure}

At this point, the LO crossed bremsstrahlung effects can be written as
\begin{align}\label{br expression}
    \delta_{2\gamma, \text{xbr}}^{(1)}=4 \pi \alpha\left[b_{11} I(k_2, p_2)+b_{11} I(k_1, p_1)-b_{12} I(k_1, p_2)-b_{12} I(k_2, p_1)\right]\ ,
\end{align}
with the corresponding IR divergence,
\begin{align}\label{br IR}
    \delta_{2\gamma, \text{xbr}}^{\mathrm{IR}}=& \frac{2 \alpha}{\pi}\left(\frac{1}{\epsilon_{\mathrm{IR}}}-\gamma_{\mathrm{E}}+\ln \left(\frac{4 \pi \nu^{2}}{Q^{2}}\right)\right)  \nonumber\\
    &\times\left(-\frac{\left(s-m^{2}-M^{2}\right) \ln \left(\frac{\sqrt{\Sigma_{s}}-m^{2}-M^{2}+s}{2 m M}\right)}{\sqrt{\Sigma_{s}}}-\frac{\left(u-m^{2}-M^{2}\right) \ln \left(\frac{\sqrt{\Sigma_{u}}+m^{2}+M^{2}-u}{2 m M}\right)}{\sqrt{\Sigma_{u}}}\right)\ .
\end{align}
Comparing with the IR divergence of TPE amplitudes [cf. Eq.~(\ref{TPE IR})], it can be found that the results  Eq.~(\ref{br IR}) is canceled by $\delta_{2\gamma, \text{TPE}}^{\mathrm{IR}}$ directly.
The calculations of chiral NLO correction of crossed bremsstrahlung are straightforward according to the formula similar to Eq.~(\ref{del chptdef}).
We found that $\delta_{2\gamma, \text{br}}^{(2)}=0$.
The IR divergent part is also $0$, as expected.
The results given in Eq.~(\ref{br expression}) are shown in Fig.~\ref{xbr fig}.

\subsection{Direct bremsstrahlung contributions}

Similar to crossed bremsstrahlung correction, direct bremsstrahlung effects only exist in chiral LO.
Direct bremsstrahlung correction to the differential scattering cross section in lab frame can be written as,
\begin{align}
    \delta_{2\gamma, \text{br}}^{(1)}=& e^{2} \int \frac{\mathrm{d}^{3} k}{(2 \pi)^{3} 2 k^{0}}\left(\frac{Q^2+2m^2}{\left(k_{1} \cdot k\right)\left( k_{2} \cdot k\right)}-\frac{m^2}{(k_1\cdot k)^2}-\frac{m^2}{(k_2\cdot k)^2}\right) \nonumber\\
    +& e^{2} \int \frac{\mathrm{d}^{3} k}{(2 \pi)^{3} 2 k^{0}}\left(\frac{Q^2+2M^2}{\left(p_{1} \cdot k\right)\left( p_{2} \cdot k\right)}-\frac{M^2}{(p_1\cdot k)^2}-\frac{M^2}{(p_2\cdot k)^2}\right) \ ,
\end{align}
where the subscript ``br'' denotes direct bremsstrahlung in this subsection.
There are more types of integral appeared than before, and definitions and explicit expressions are given in Appendix.~\ref{appendix:1}.

\begin{figure}[H] 
    \centering
    \subfigure{
        \begin{minipage}[b]{0.98\linewidth}
          
        \begin{minipage}[b]{0.45\linewidth}
        \includegraphics[scale=0.5]{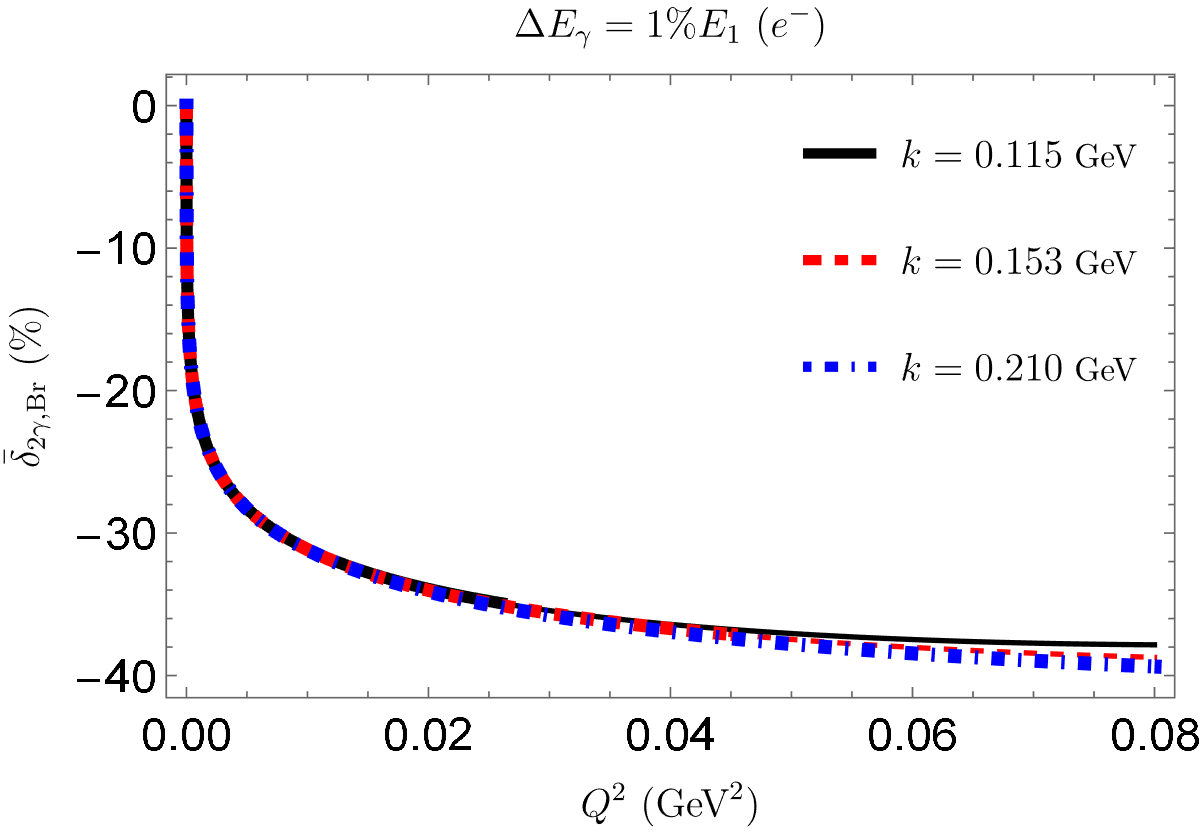}
        \vspace{0.1cm}
        \hspace{0.02cm}
        \end{minipage}
        \qquad
        \begin{minipage}[b]{0.45\linewidth}
        \includegraphics[scale=0.5]{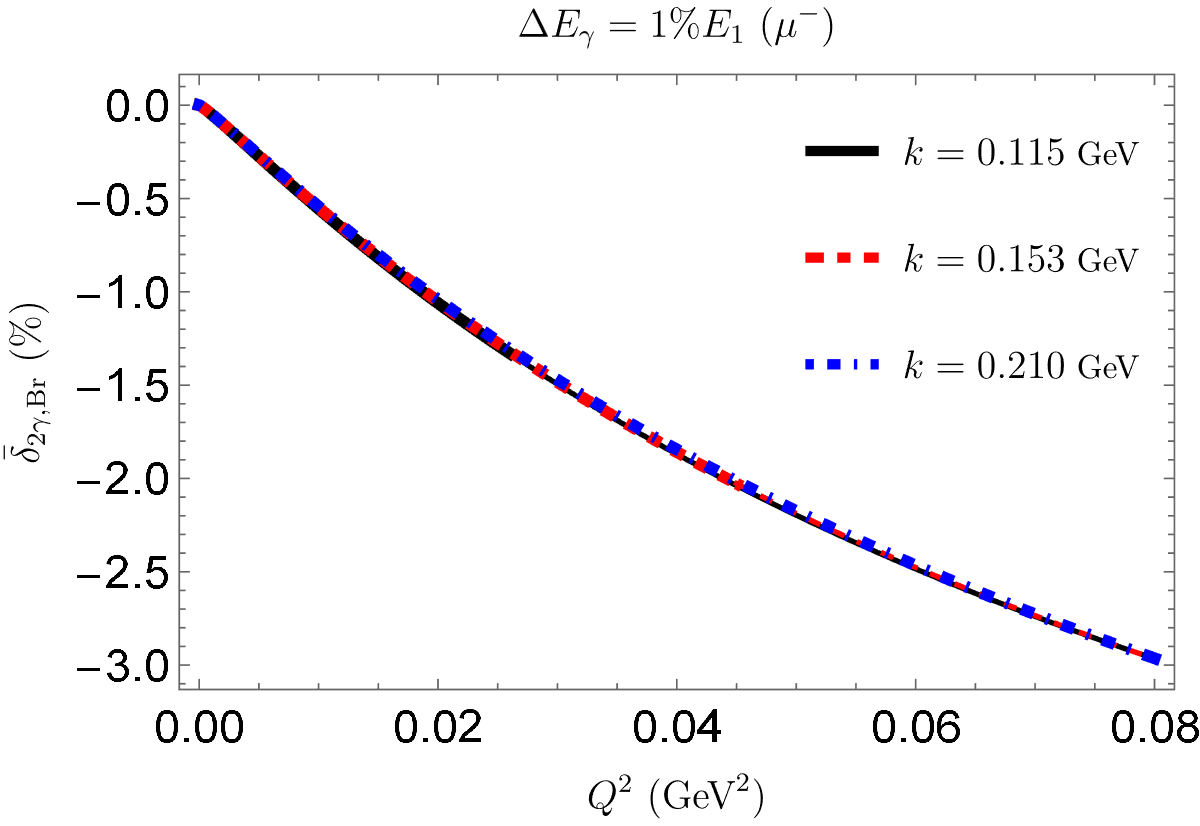}
        \vspace{0.1cm}
        \hspace{0.02cm}
        \end{minipage}

        \end{minipage}
    }
    \caption{The $Q^2$ dependence of the direct bremsstrahlung corrections to $e^- \text{p}$ and $\mu^- \text{p}$ elastic scatterings.}\label{br fig}
\end{figure}

Thus, the chiral LO direct bremsstrahlung correction is obtained (NLO is 0),
\begin{align}
    \delta_{2\gamma, \text{br}}^{(1)}=&4\pi \alpha m^{2}\left(\frac{2\left(v_{\ell}^{2}+1\right)}{v_{\ell}-1} I(k_1, k_2)-I(k_1)-I(k_2)\right) \nonumber\\
    +&4\pi \alpha M^{2}\left(\frac{2\left(v_{N}^{2}+1\right)}{v_{N}-1} I(p_1, p_2)-I(p_1)-I(p_2)\right)\ .
\end{align}
The IR divergent part can be written as
\begin{align}\label{IR br}
    \delta_{2\gamma, \text{br}}^{\mathrm{IR}}=& \frac{\alpha}{2 \pi}\left(\frac{1}{\epsilon_{\mathrm{IR}}}-\gamma_{\mathrm{E}}+\ln \left(\frac{4 \pi \nu^{2}}{Q^{2}}\right)\right)\left(2-\frac{v_{\ell}^{2}+1}{v_{\ell}} \ln \frac{v_{\ell}+1}{v_{\ell}-1}\right) \nonumber\\
    +&\frac{\alpha}{2 \pi}\left(\frac{1}{\epsilon_{\mathrm{IR}}}-\gamma_{\mathrm{E}}+\ln \left(\frac{4 \pi \nu^{2}}{Q^{2}}\right)\right)\left(2-\frac{v_{N}^{2}+1}{v_{N}} \ln \frac{v_{N}+1}{v_{N}-1}\right)\ ,
\end{align}
which is similar to the calculation of Ref.~\cite{Vanderhaeghen:2000ws}.
The numerical results are shown in Fig.~\ref{br fig}.
A remarkable observation is that the results are almost invariant within the variation of incoming lepton momentum in MUSE kinematical region.
The $e \text{p}$ and $\mu \text{p}$ direct bremsstrahlung corrections are both negative, but the latter is over one order magnitude smaller.
Because the Sudakov double-log terms $\ln^2 (Q^2/m^2)$ dominate in the case of $e \text{p}$ scattering~\cite{Talukdar:2020aui}.

\section{The calculation of the vertex correction diagrams}\label{sec:5}

\begin{figure} [H]
    \centering
    \subfigure[]{
    \begin{tikzpicture}[scale = 0.7]
        \draw[nucleon](-2,1)to(-1,1);
        \draw[nucleon](-1,1)to(0,1);
        \draw[nucleon](0,1)to(1,1);
        \draw[nucleon](1,1)to(2,1);
        \draw[photon, very thick](1,1) arc (0: 180: 1);
        \draw[nucleon, very thick](-2,0)to(0,0);
        \draw[nucleon, very thick](0,0)to(2,0);
        \draw[photon, very thick](0,1)to(0,0);
        
        \draw[very thick](0,0)circle(0.25);
        \fill[white](0,0) circle(0.25);
        \node at (0,0) {$1$};
    \end{tikzpicture}
    } 
    \qquad
    \subfigure[]{
    \begin{tikzpicture}[scale = 0.7]
        \draw[nucleon](-2,1)to(-1,1);
        \draw[nucleon](-1,1)to(0,1);
        \draw[nucleon](0,1)to(1,1);
        \draw[nucleon](1,1)to(2,1);
        \draw[photon, very thick](1,1) arc (0: 180: 1);
        \draw[nucleon, very thick](-2,0)to(0,0);
        \draw[nucleon, very thick](0,0)to(2,0);
        \draw[photon, very thick](0,1)to(0,0);
        
        \draw[very thick](0,0)circle(0.25);
        \fill[white](0,0) circle(0.25);
        \node at (0,0) {$2$};
    \end{tikzpicture}
    }
    \qquad
    \subfigure[]{
        \begin{tikzpicture}[scale = 0.7]
        \draw[nucleon](-2,1)to(0,1);
        \draw[nucleon](0,1)to(2,1);
        \draw[nucleon, very thick](-2,0)to(-1,0);
        \draw[nucleon, very thick](-1,0)to(0,0);
        \draw[nucleon, very thick](0,0)to(1,0);
        \draw[nucleon, very thick](1,0)to(2,0);
        \draw[photon, very thick](0,1)to(0,0);            
        \draw[photon, very thick](-1,0) arc (180: 360: 1);
            
        \draw[very thick](0,0)circle(0.25);
        \fill[white](0,0) circle(0.25);
        \node at (0,0) {$1$};
        \draw[very thick](-1,0)circle(0.25);
        \fill[white](-1,0) circle(0.25);
        \node at (-1,0) {$1$};
        \draw[very thick](1,0)circle(0.25);
        \fill[white](1,0) circle(0.25);
        \node at (1,0) {$1$};
        \end{tikzpicture}
    } 

    \subfigure[]{
        \begin{tikzpicture}[scale = 0.7]
        \draw[nucleon](-2,1)to(0,1);
        \draw[nucleon](0,1)to(2,1);
        \draw[nucleon, very thick](-2,0)to(-1,0);
        \draw[nucleon, very thick](-1,0)to(0,0);
        \draw[nucleon, very thick](0,0)to(1,0);
        \draw[nucleon, very thick](1,0)to(2,0);
        \draw[photon, very thick](0,1)to(0,0);            
        \draw[photon, very thick](-1,0) arc (180: 360: 1);
            
        \draw[very thick](0,0)circle(0.25);
        \fill[white](0,0) circle(0.25);
        \node at (0,0) {$1$};
        \draw[very thick](-1,0)circle(0.25);
        \fill[white](-1,0) circle(0.25);
        \node at (-1,0) {$2$};
        \draw[very thick](1,0)circle(0.25);
        \fill[white](1,0) circle(0.25);
        \node at (1,0) {$1$};
        \end{tikzpicture}
    }
    \qquad
    \subfigure[]{
        \begin{tikzpicture}[scale = 0.7]
        \draw[nucleon](-2,1)to(0,1);
        \draw[nucleon](0,1)to(2,1);
        \draw[nucleon, very thick](-2,0)to(-1,0);
        \draw[nucleon, very thick](-1,0)to(0,0);
        \draw[nucleon, very thick](0,0)to(1,0);
        \draw[nucleon, very thick](1,0)to(2,0);
        \draw[photon, very thick](0,1)to(0,0);            
        \draw[photon, very thick](-1,0) arc (180: 360: 1);
            
        \draw[very thick](0,0)circle(0.25);
        \fill[white](0,0) circle(0.25);
        \node at (0,0) {$2$};
        \draw[very thick](-1,0)circle(0.25);
        \fill[white](-1,0) circle(0.25);
        \node at (-1,0) {$1$};
        \draw[very thick](1,0)circle(0.25);
        \fill[white](1,0) circle(0.25);
        \node at (1,0) {$1$};
        \end{tikzpicture}
    }
    \qquad
    \subfigure[]{
        \begin{tikzpicture}[scale = 0.7]
        \draw[nucleon](-2,1)to(0,1);
        \draw[nucleon](0,1)to(2,1);
        \draw[nucleon, very thick](-2,0)to(-1,0);
        \draw[nucleon, very thick](-1,0)to(0,0);
        \draw[nucleon, very thick](0,0)to(1,0);
        \draw[nucleon, very thick](1,0)to(2,0);
        \draw[photon, very thick](0,1)to(0,0);            
        \draw[photon, very thick](-1,0) arc (180: 360: 1);
            
        \draw[very thick](0,0)circle(0.25);
        \fill[white](0,0) circle(0.25);
        \node at (0,0) {$1$};
        \draw[very thick](-1,0)circle(0.25);
        \fill[white](-1,0) circle(0.25);
        \node at (-1,0) {$1$};
        \draw[very thick](1,0)circle(0.25);
        \fill[white](1,0) circle(0.25);
        \node at (1,0) {$2$};
        \end{tikzpicture}
    }
    \caption{Vertex correction diagrams of chiral LO and NLO.}\label{vertex fig}
\end{figure}
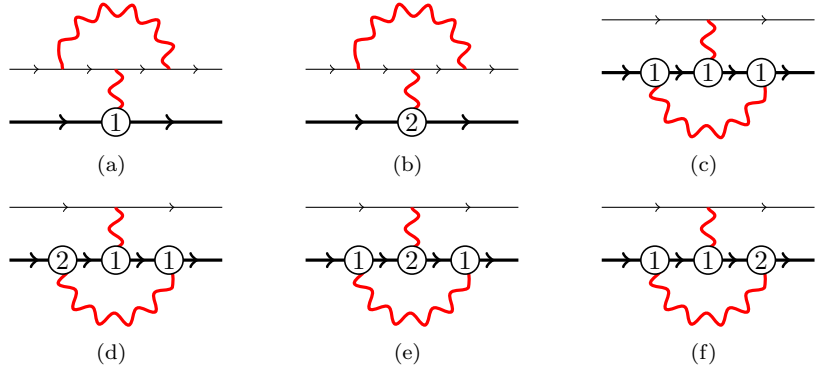

The vertex correction diagrams for chiral LO and NLO are shown in Fig.~\ref{vertex fig}.
In most cases, previous investigations ignored the Pauli FF $F^\ell_2$, but it is included in our calculation.
It will become important for the case of $\mu \text{p}$ scattering.
In general, an amplitude of any order of vertex diagram can be written as,
\begin{align}
    i \mathcal{M}_{\text {vertex }}=e^2 \bar{u}\left(k_{2}\right)\left(F_{1}^{\ell} \gamma^{\mu}+F_{2}^{\ell} \frac{i}{2 m} \sigma^{\mu \nu}\left(-q_{\nu}\right)\right) u\left(k_{1}\right)\left(\frac{-i}{q^{2}}\right) \bar{p}\left(p_{2}\right)\left(F_{1}^{p} \gamma_{\mu}+F_{2}^{p} \frac{i}{2 M} \sigma_{\mu \rho} q^{\rho}\right) p\left(p_{1}\right)\ ,
\end{align}
where the $F_1^{\ell, p}, F_2^{\ell, p}$ corresponds to Dirac and Pauli FFs for lepton or proton, respectively.
For simplicity, the definition $F_{1}^{\ell}=1+\delta F_{1}^{\ell}$ will be used.

\begin{figure} [H]
    \centering
    \subfigure{
        \begin{minipage}[b]{0.99\linewidth}
          
        \begin{minipage}[b]{0.45\linewidth}
        \includegraphics[scale=0.5]{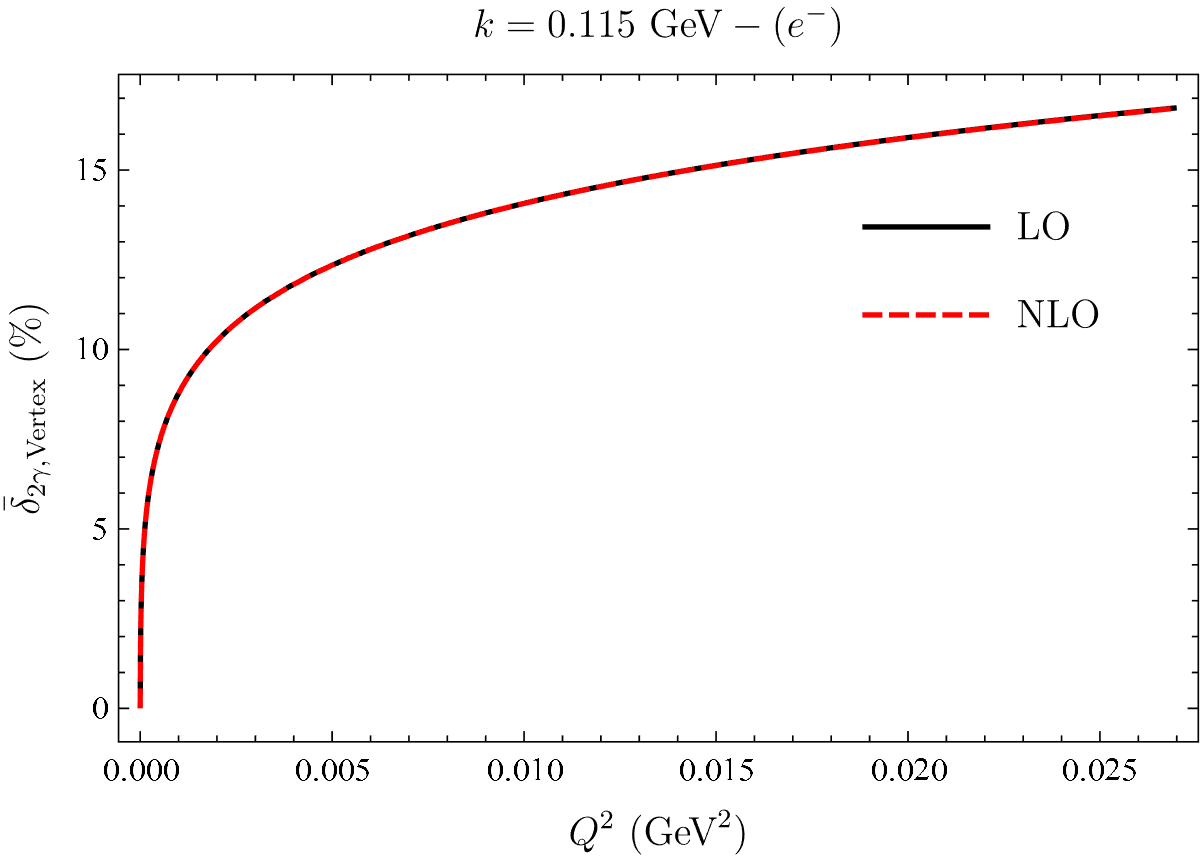}
        \vspace{0.1cm}
        \hspace{0.02cm}
        \end{minipage}
        \qquad
        \begin{minipage}[b]{0.45\linewidth}
        \includegraphics[scale=0.5]{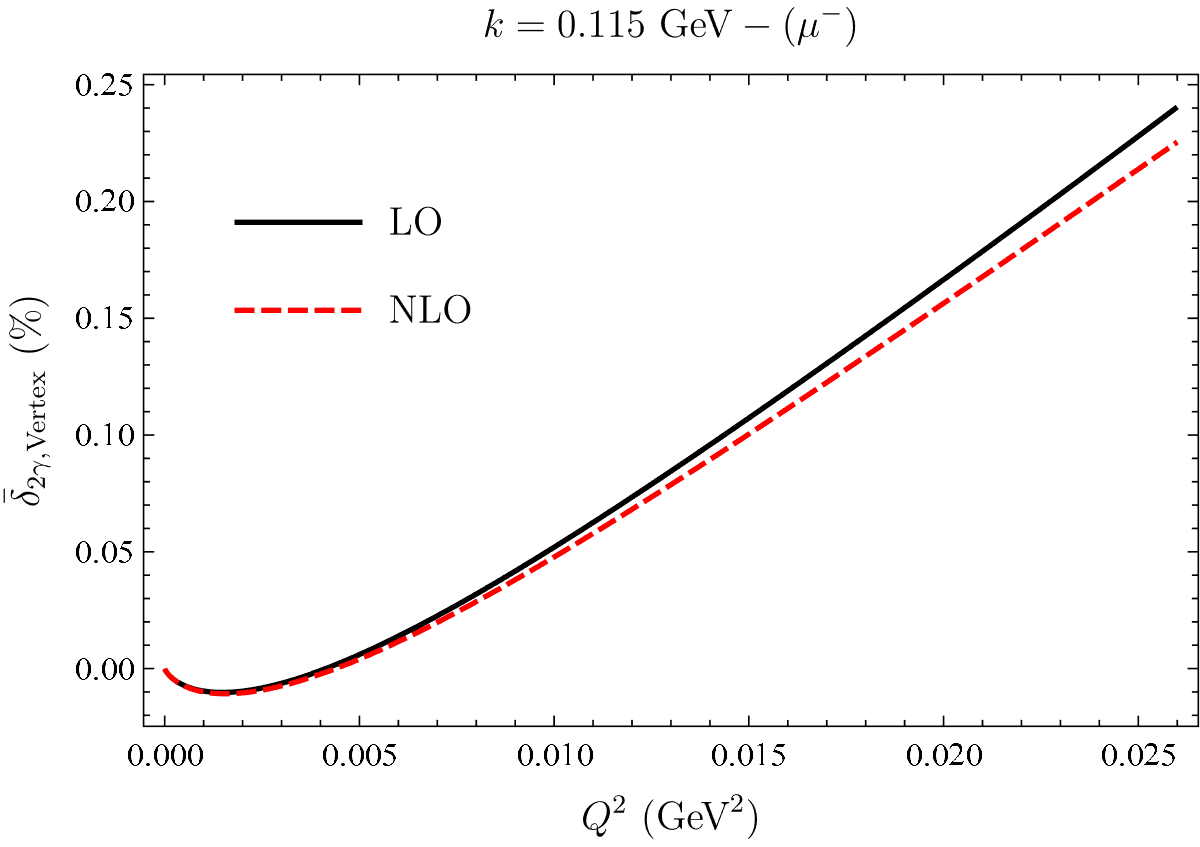}
        \vspace{0.1cm}
        \hspace{0.02cm}
        \end{minipage}
    
        \begin{minipage}[b]{0.45\linewidth}
        \includegraphics[scale=0.5]{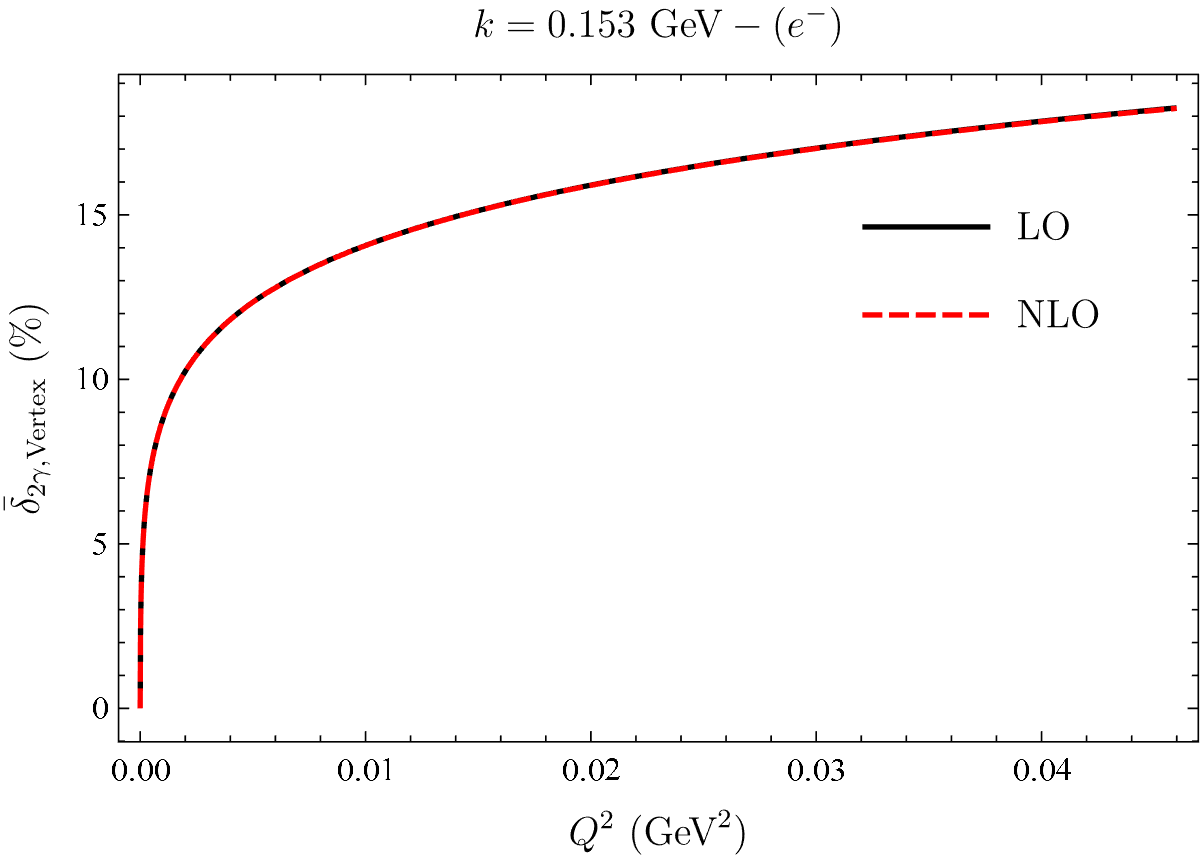}
        \vspace{0.1cm}
        \hspace{0.05cm}
        \end{minipage} 
        \qquad 
        \begin{minipage}[b]{0.45\linewidth}
        \includegraphics[scale=0.5]{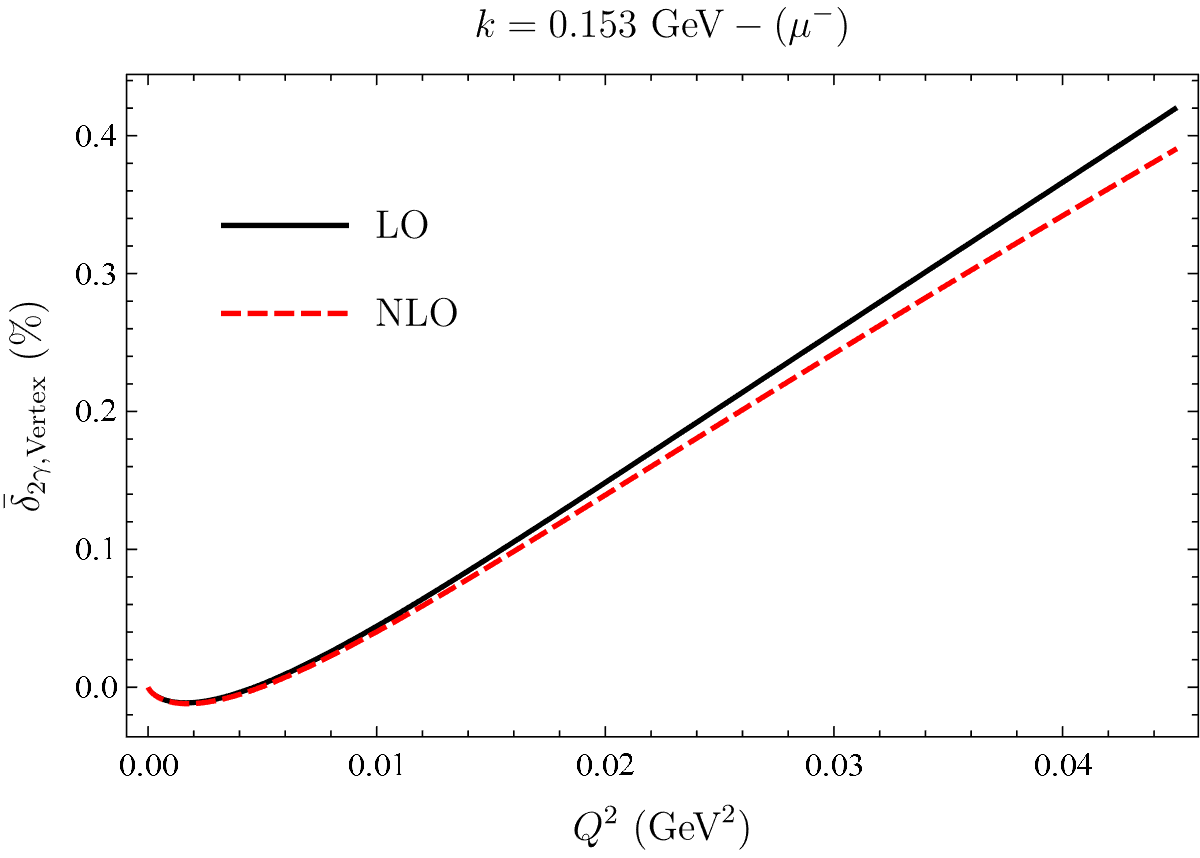}
        \vspace{0.1cm}
        \hspace{0.05cm}
        \end{minipage}

        \begin{minipage}[b]{0.45\linewidth}
            \includegraphics[scale=0.5]{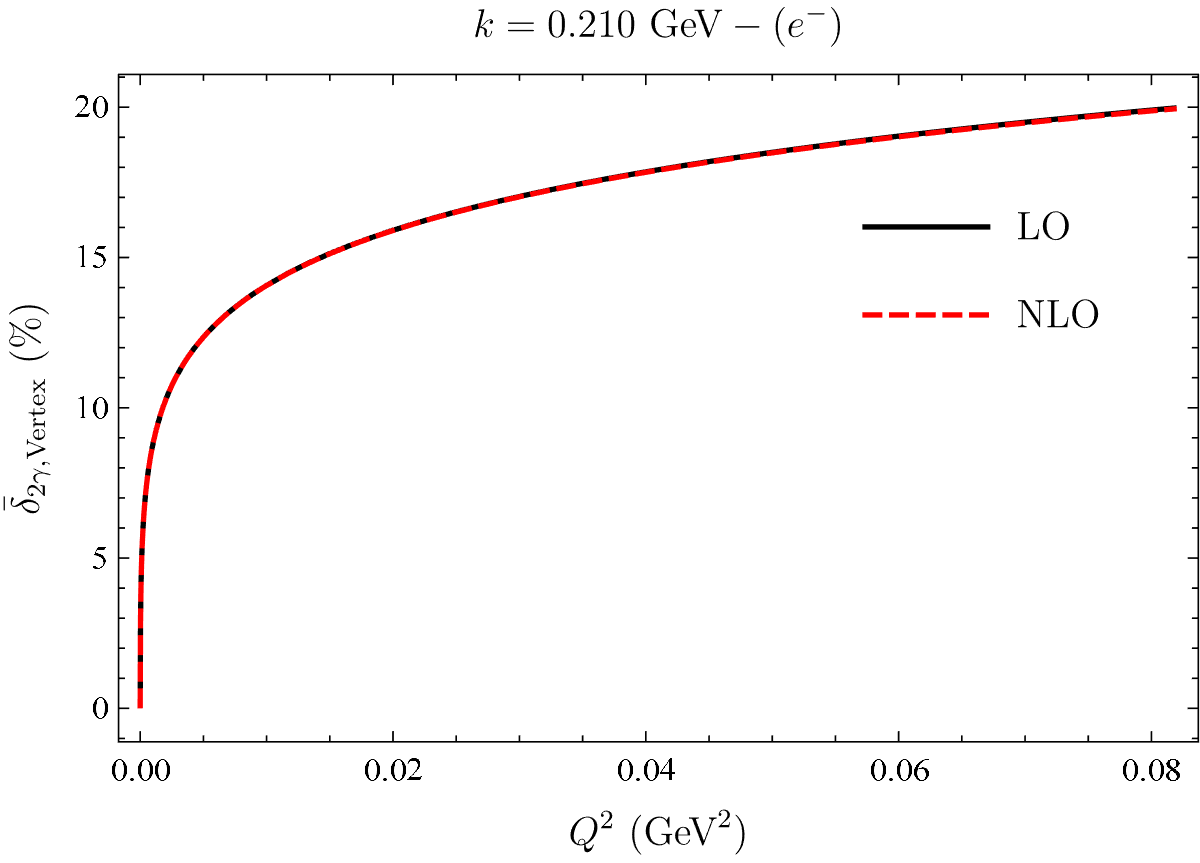}
            \vspace{0.1cm}
            \hspace{0.05cm}
            \end{minipage} 
            \qquad 
            \begin{minipage}[b]{0.45\linewidth}
            \includegraphics[scale=0.5]{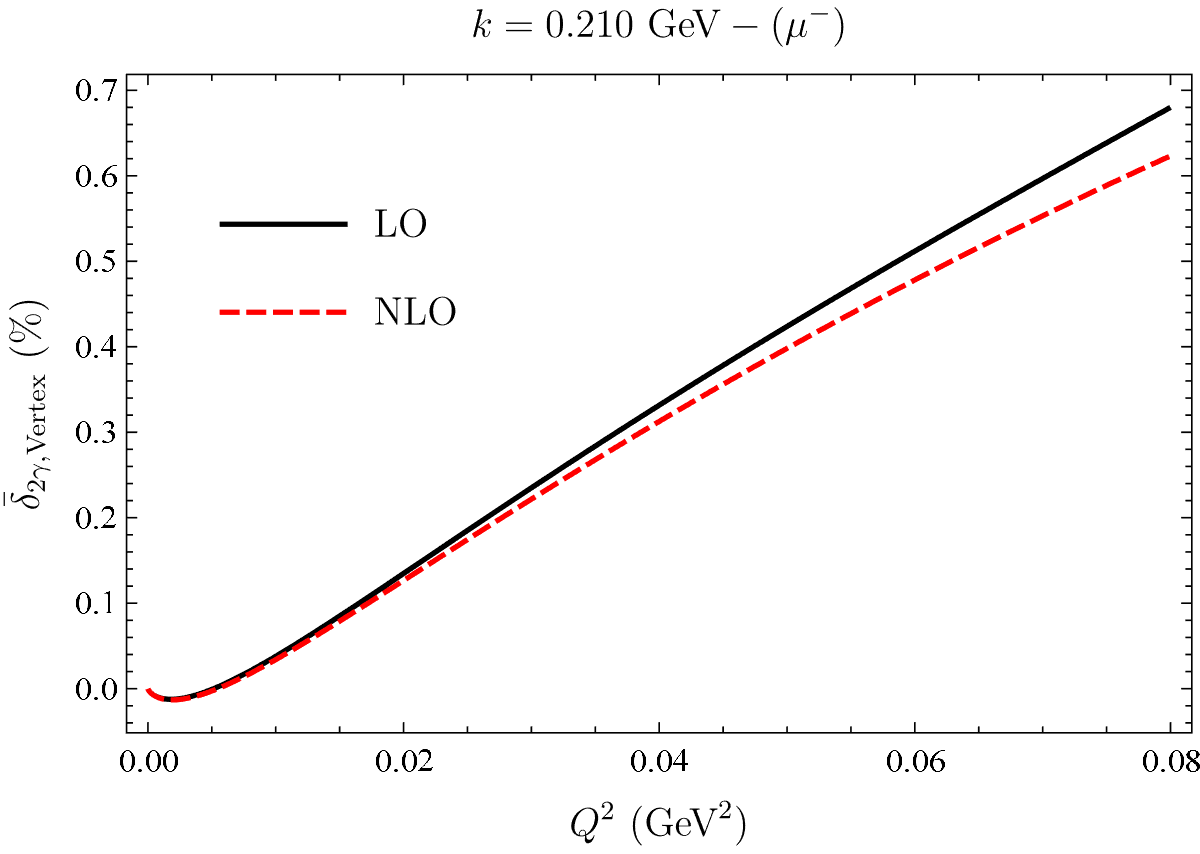}
            \vspace{0.1cm}
            \hspace{0.05cm}
            \end{minipage}
   
        \end{minipage}
        }
    \caption{The one-loop chiral LO and NLO vertex corrections.}\label{vertex}
\end{figure}

The calculation of any vertex correction diagram is to obtain $F_1^{\ell, p}, F_2^{\ell, p}$, so that the estimation can be simplified by using the projection operator method. 
Once the FFs are known, the contribution of the vertex correction to the differential cross section will be obtained straightforwardly through the interference terms.
All the contributions are evaluated in Appendix.~\ref{appendix:2} using DR.
Adding all the nonvanishing contributions of the vertex correction, the IR divergence is canceled by the direct bremsstrahlung correction [Eq.~(\ref{IR br})].
Fig.~\ref{vertex} displays the chiral LO and NLO contributions stemming from B$\chi$PT.

The important feature of Fig.~\ref{vertex} is that the chiral correction can be ignored as for $e \text{p}$ scatterings, but it has significant effects on $\mu \text{p}$.  
The huge numerical difference (about two orders of magnitude) between the corrections in $e \text{p}$ case and $\mu \text{p}$ case comes from Sudakov double logarithm, similar to the bremsstrahlung contribution.

\section{The calculation of vacuum polarizations}\label{sec:6}

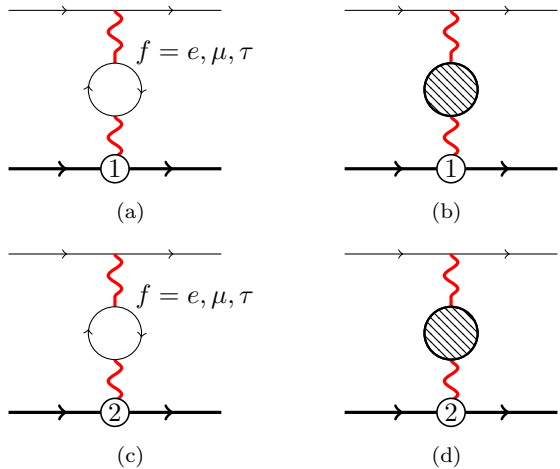
\begin{figure} [H]
    \centering
    \subfigure[]{
    \begin{tikzpicture}[scale = 0.7]
        \draw[nucleon](-2,3)to(0,3);
        \draw[nucleon](0,3)to(2,3);
        \draw[nucleon, very thick](-2,0)to(0,0);
        \draw[nucleon, very thick](0,0)to(2,0);
        \draw[photon, very thick](0,1)to(0,0);
        \draw[photon, very thick](0,3)to(0,2);
        
        \draw[nucleon] (0,1) arc (270: 90: 0.5);
        \draw[nucleon] (0,2) arc (90: -90: 0.5);
        
        \draw[very thick](0,0)circle(0.25);
        \fill[white](0,0) circle(0.25);
        \node at (0,0) {$1$};
        
        \node at (1.5, 2.2) {$f=e,\mu,\tau$};
    \end{tikzpicture}
    } 
    \qquad
    \subfigure[]{
    \begin{tikzpicture}[scale = 0.7]
        \draw[nucleon](-2,3)to(0,3);
        \draw[nucleon](0,3)to(2,3);
        \draw[nucleon, very thick](-2,0)to(0,0);
        \draw[nucleon, very thick](0,0)to(2,0);
        \draw[photon, very thick](0,1)to(0,0);
        \draw[photon, very thick](0,3)to(0,2);
        
        \draw[black, thick] (0,1) arc (270: 90: 0.5);
        \draw[black, thick] (0,2) arc (90: -90: 0.5);
        \draw[pattern=north west lines, thick](0,1.5) circle(0.5);
        
        \draw[very thick](0,0)circle(0.25);
        \fill[white](0,0) circle(0.25);
        \node at (0,0) {$1$};
        
    \end{tikzpicture}
    }

    \subfigure[]{
    \begin{tikzpicture}[scale = 0.7]
        \draw[nucleon](-2,3)to(0,3);
        \draw[nucleon](0,3)to(2,3);
        \draw[nucleon, very thick](-2,0)to(0,0);
        \draw[nucleon, very thick](0,0)to(2,0);
        \draw[photon, very thick](0,1)to(0,0);
        \draw[photon, very thick](0,3)to(0,2);
        
        \draw[nucleon] (0,1) arc (270: 90: 0.5);
        \draw[nucleon] (0,2) arc (90: -90: 0.5);
        
        \draw[very thick](0,0)circle(0.25);
        \fill[white](0,0) circle(0.25);
        \node at (0,0) {$2$};
        
        \node at (1.5, 2.2) {$f=e,\mu,\tau$};
    \end{tikzpicture}
    } 
    \qquad
    \subfigure[]{
    \begin{tikzpicture}[scale = 0.7]
        \draw[nucleon](-2,3)to(0,3);
        \draw[nucleon](0,3)to(2,3);
        \draw[nucleon, very thick](-2,0)to(0,0);
        \draw[nucleon, very thick](0,0)to(2,0);
        \draw[photon, very thick](0,1)to(0,0);
        \draw[photon, very thick](0,3)to(0,2);
        
        \draw[black, thick] (0,1) arc (270: 90: 0.5);
        \draw[black, thick] (0,2) arc (90: -90: 0.5);
        \draw[pattern=north west lines, thick](0,1.5) circle(0.5);
        
        \draw[very thick](0,0)circle(0.25);
        \fill[white](0,0) circle(0.25);
        \node at (0,0) {$2$};
        
    \end{tikzpicture}
    }
    \caption{Vacuum polarization diagrams of chiral LO and NLO.
    The shaded parts represent the contribution of hadronic HVP.}\label{vp fig}
\end{figure}

\begin{figure} 
    \centering
    \subfigure{
        \includegraphics[scale=0.5]{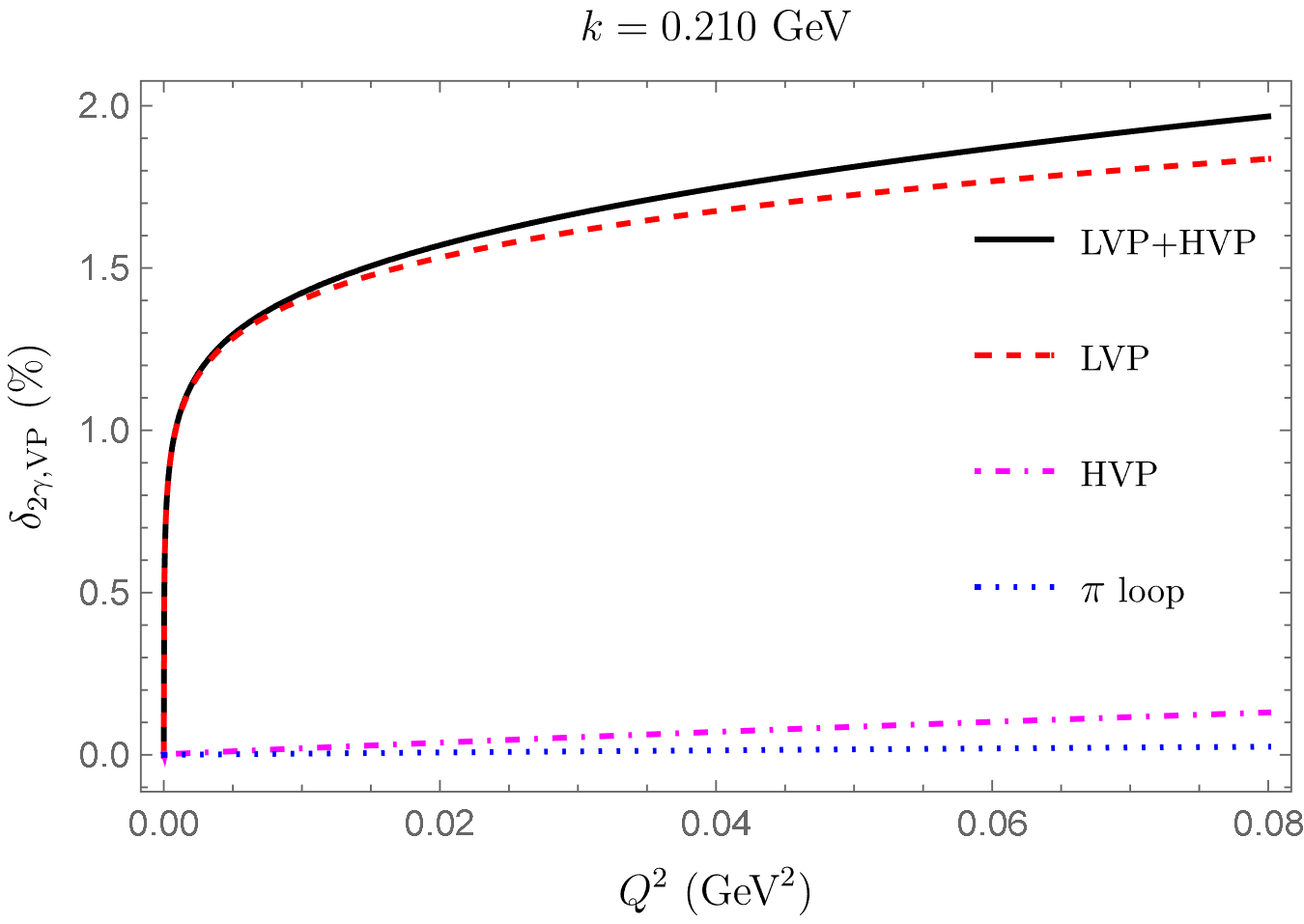}
    }
    \caption{LVP and HVP contributions to $\ell \text{p}$ scatterings.}\label{vp}
\end{figure}

The one-loop photon vacuum polarization contribution is IR finite, which has been extensively studied in the literature.
According to Refs.~\cite{Maximon:2000hm,Talukdar:2020aui}, we consider two kinds of important contributions.
At low energies, it is dominated by QED lepton vacuum polarization (LVP) and hadron vacuum polarization (HVP).
LVP contributions have been calculated to sufficiently high precision.
The QED LO and NLO contributions are known as analytic expressions including the full mass dependence~\cite{PhysRev.76.769, Kallen:1955fb}.
For our applications the LO LVP contribution (one-loop with $e,\mu$ and $\tau$) can be easily implemented with a sufficient accuracy.
But HVP cannot be reliably calculated from perturbation QCD. 
HVP must use experimental data from $e^+e^-$ annihilation to hadrons as input for calculation.
We use a package provided by Jegerlehner~\cite{Jegerlehner_web} and a table provided by Ignatov~\cite{Ignatov_web} (their results are identical in MUSE kinematical region) to obtain the complete hadronic HVP (for a review see~\cite{WorkingGrouponRadiativeCorrections:2010bjp}).
In Fig.~\ref{vp fig} we display the diagrams of LVP and HVP,
The results of lepton loops are given in terms of photon self-energy function, in the compact form~\cite{PhysRev.76.769, Kallen:1955fb}
\begin{align}
    \Pi_{\text {LVP }}\left(Q^{2}\right)= \frac{\alpha}{4 \pi} \sum_{f=e, \mu, \tau}\left[\frac{4}{3}\left(v_f^{2}-\frac{8}{3}\right)+2 v_f\left(\frac{3-v_f^{2}}{3}\right) \ln \left(\frac{v_f+1}{v_f-1}\right)\right]\ .
\end{align}
It could be also useful to perform the numerical comparison between the HVP by $\pi$ loop calculated by Tsai~\cite{Tsai:1960zz}
\begin{align}
    \Pi_{\pi~\text{loop}}\left(Q^{2}\right)=\frac{\alpha}{4 \pi}\left[-\frac{4}{3}\left(v_\pi^{2}+\frac{1}{3}\right)+\frac{2 v_\pi^{3}}{3} \ln \left(\frac{v_\pi+1}{v_\pi-1}\right)\right]\ ,
\end{align}
where $v_{f,\pi}=1+4m_{f,\pi}^2/Q^2$, and modern approach to HVP in Fig.~\ref{vp}.
The total renormalized chiral LO VP contribution (chiral NLO result is exactly $0$) is given by
\begin{align}
    \delta_{\mathrm{VP}}^{(1)}=2 \Pi_{\mathrm{LVP}}\left(Q^{2}\right)+2 \Pi_{\mathrm{HVP}}\left(Q^{2}\right) = \sum_{f=e, \mu, \tau} \delta_{\mathrm{VP} ; f}^{(1)}+\delta_{\mathrm{VP} ; \pi}^{(1)}\equiv\delta_{2\gamma, \text{VP}}\ ,
\end{align} 
and Fig.~\ref{vp} shows the numerical results of the largest kinematical incoming momentum in MUSE kinematical region.
It should be noted that the vacuum polarization correction is independent of the flavours of lepton, i.e., it is similar for $e \text{p}$ and $\mu \text{p}$ scatterings.
In Fig.~\ref{vp}, we conclude that vacuum polarization is dominated by LVP in MUSE kinematical region, furthermore, one has to include the the effects of hadronic HVP instead of $\pi$ loop for $Q^2$ above a few times $0.01~\text{GeV}^2$.

\section{Numerical results and discussions}\label{sec:7}

\begin{figure} [H]
    \centering
    \subfigure{
        \begin{minipage}[b]{0.98\linewidth}
          
        \begin{minipage}[b]{0.45\linewidth}
        \includegraphics[scale=0.45]{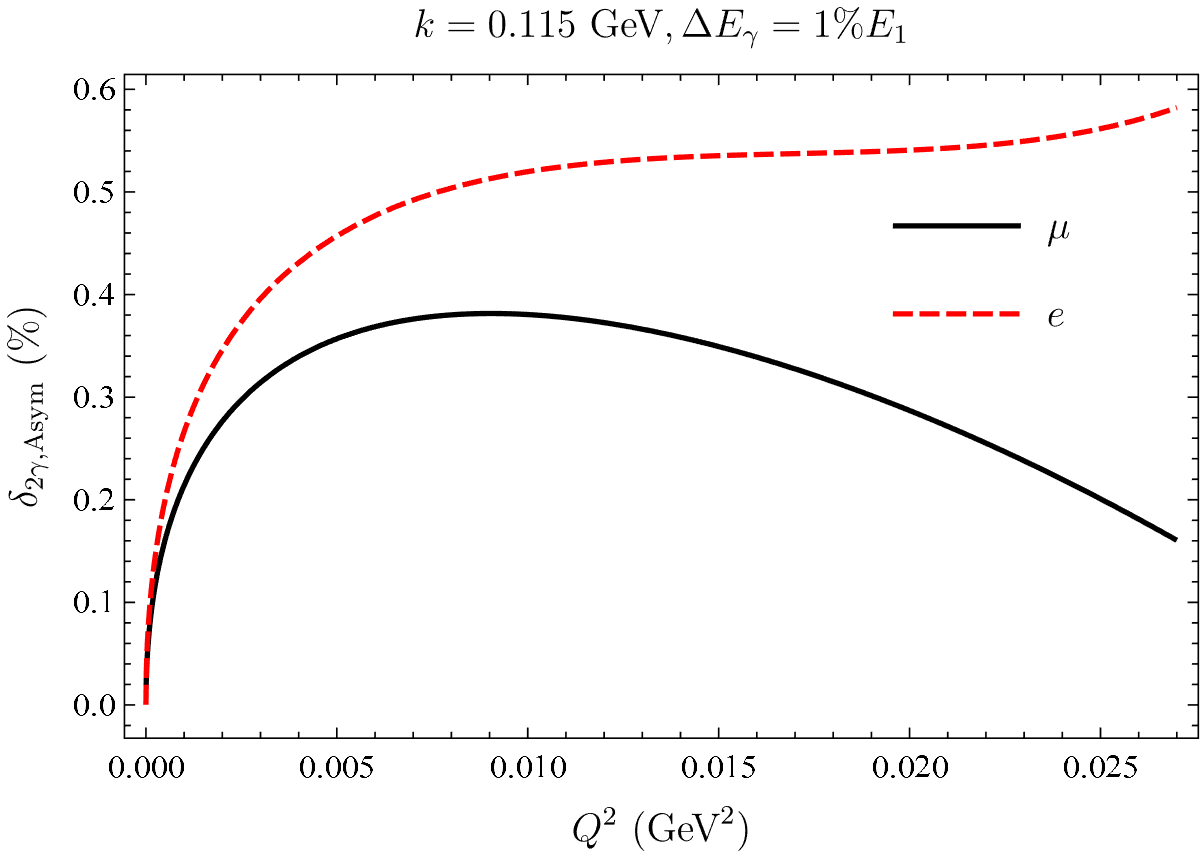}
        \vspace{0.1cm}
        \hspace{0.02cm}
        \end{minipage}
        \qquad
        \begin{minipage}[b]{0.45\linewidth}
        \includegraphics[scale=0.45]{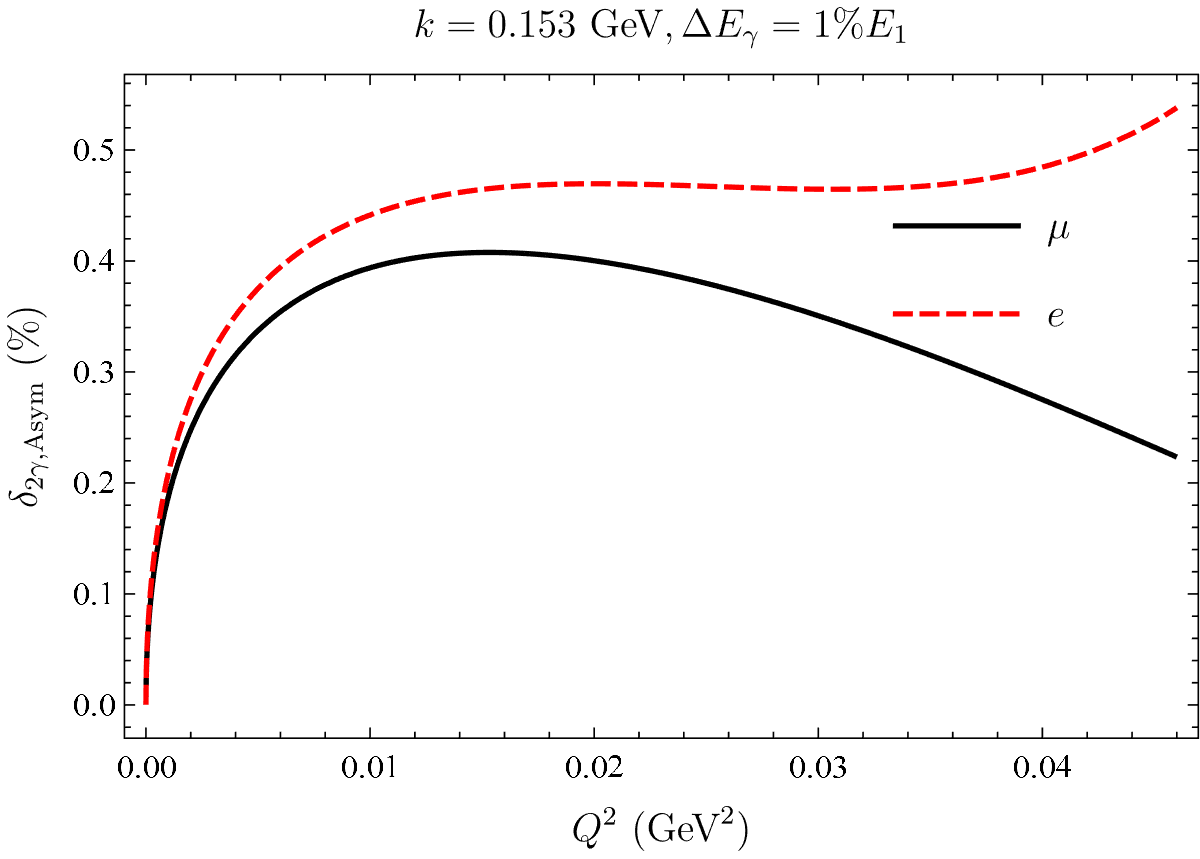}
        \vspace{0.1cm}
        \hspace{0.02cm}
        \end{minipage}
        
        \begin{minipage}[b]{0.98\linewidth}
        \centering
        \includegraphics[scale=0.45]{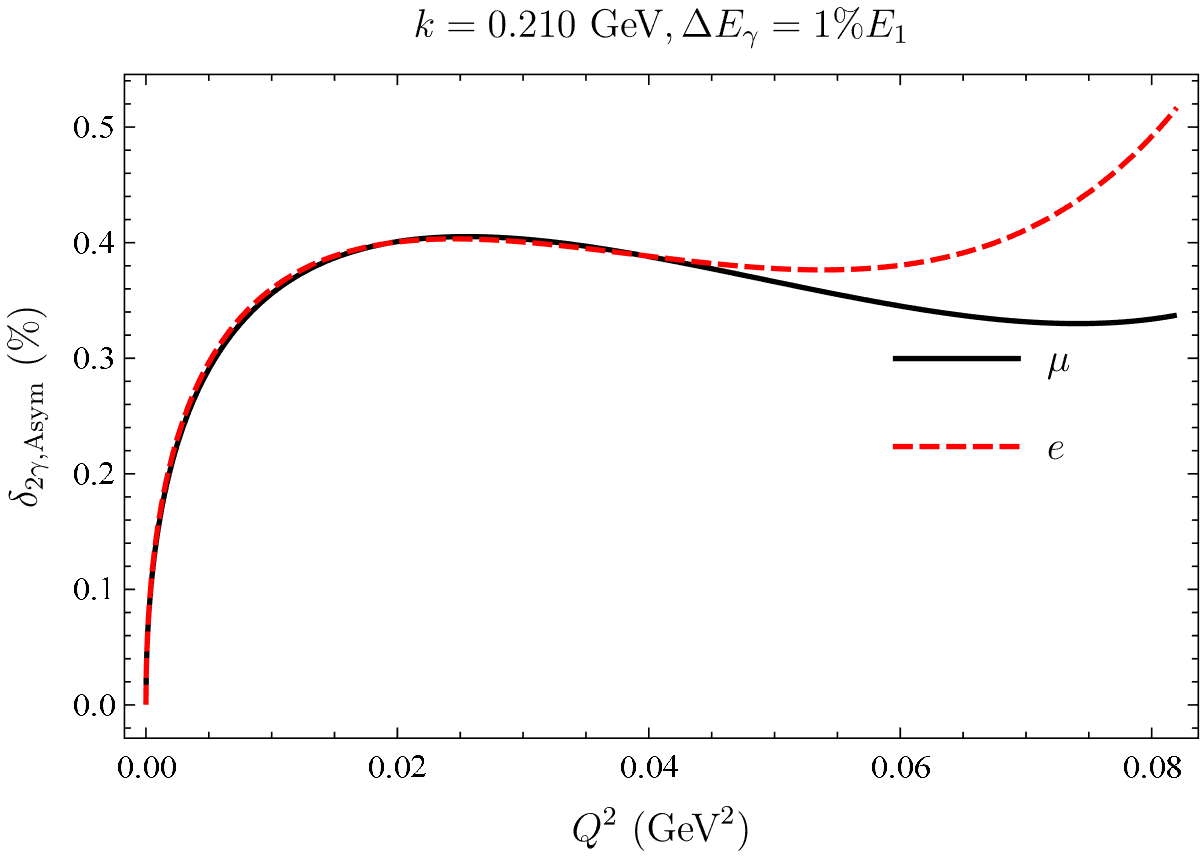}
        \vspace{0.1cm}
        \hspace{0.02cm}
        \end{minipage}
        
        \end{minipage}
    }
    \caption{The $Q^2$ dependence of the charge asymmetry of $e \text{p}$ and $\mu \text{p}$ elastic scatterings.}\label{charge sym Q}
\end{figure}

\begin{figure} [H]
    \centering
    \subfigure{
        \begin{minipage}[b]{0.99\linewidth}
          
        \begin{minipage}[b]{0.45\linewidth}
        \includegraphics[scale=0.45]{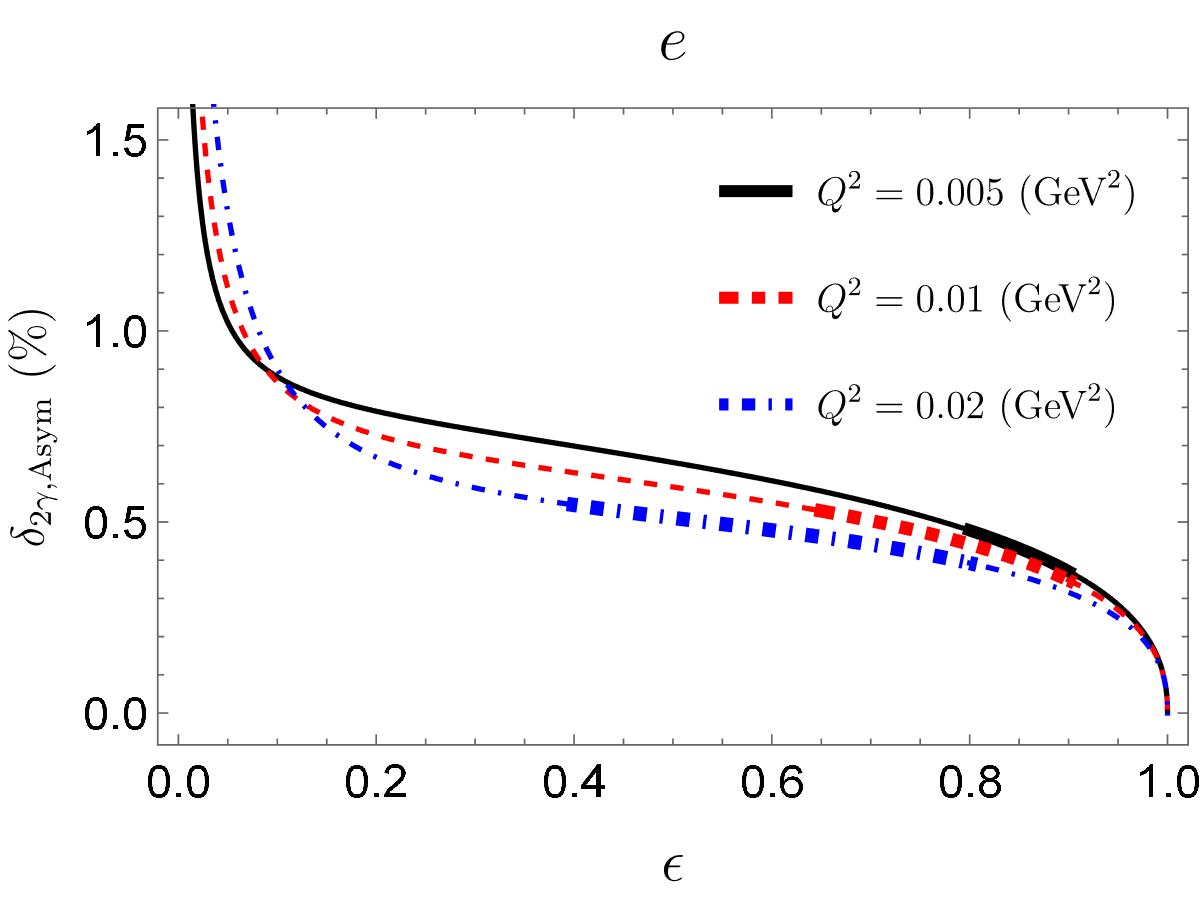}
        \vspace{0.1cm}
        \hspace{0.02cm}
        \end{minipage}
        \qquad
        \begin{minipage}[b]{0.45\linewidth}
        \includegraphics[scale=0.45]{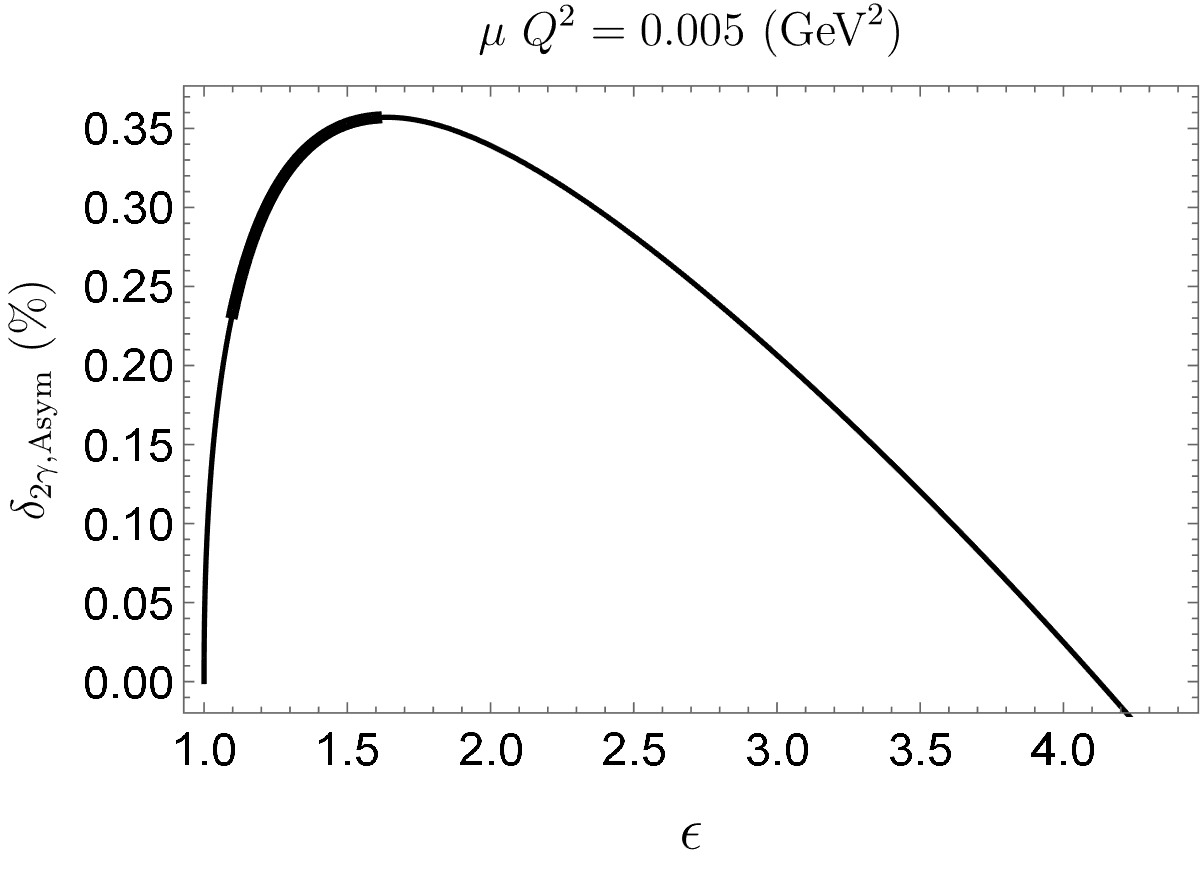}
        \vspace{0.1cm}
        \hspace{0.02cm}
        \end{minipage}
    
        \begin{minipage}[b]{0.45\linewidth}
        \includegraphics[scale=0.45]{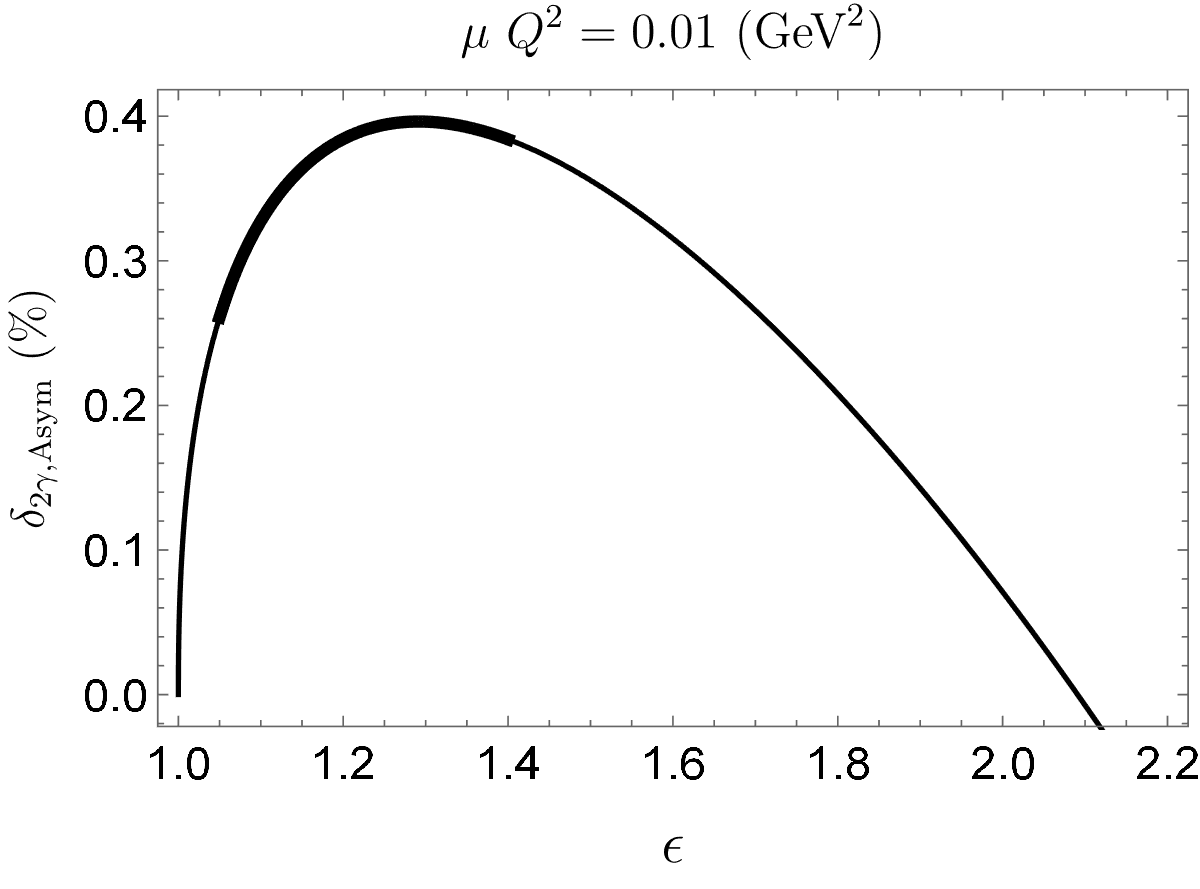}
        \vspace{0.1cm}
        \hspace{0.05cm}
        \end{minipage} 
        \qquad 
        \begin{minipage}[b]{0.45\linewidth}
        \includegraphics[scale=0.45]{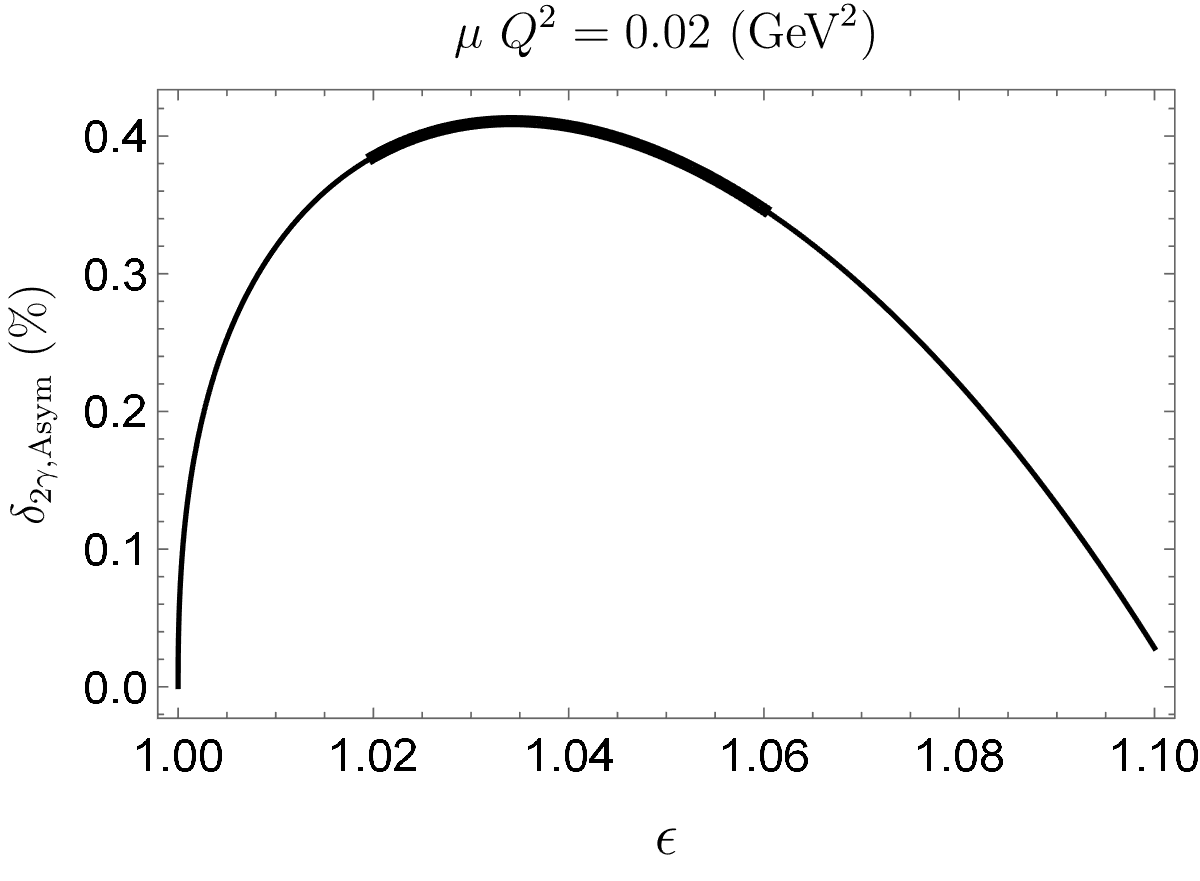}
        \vspace{0.1cm}
        \hspace{0.05cm}
        \end{minipage}
   
        \end{minipage}
    }
    \caption{The $\epsilon$ dependence of the charge asymmetry for $e \text{p}$ and $\mu \text{p}$ elastic scatterings.
    The results correspond to $\Delta E_\gamma=1\% E_1$.}\label{charge sym ep}
\end{figure}

\begin{figure} [H]
    \centering
    \subfigure{
        \begin{minipage}[b]{0.98\linewidth}
          
        \begin{minipage}[b]{0.45\linewidth}
        \includegraphics[scale=0.45]{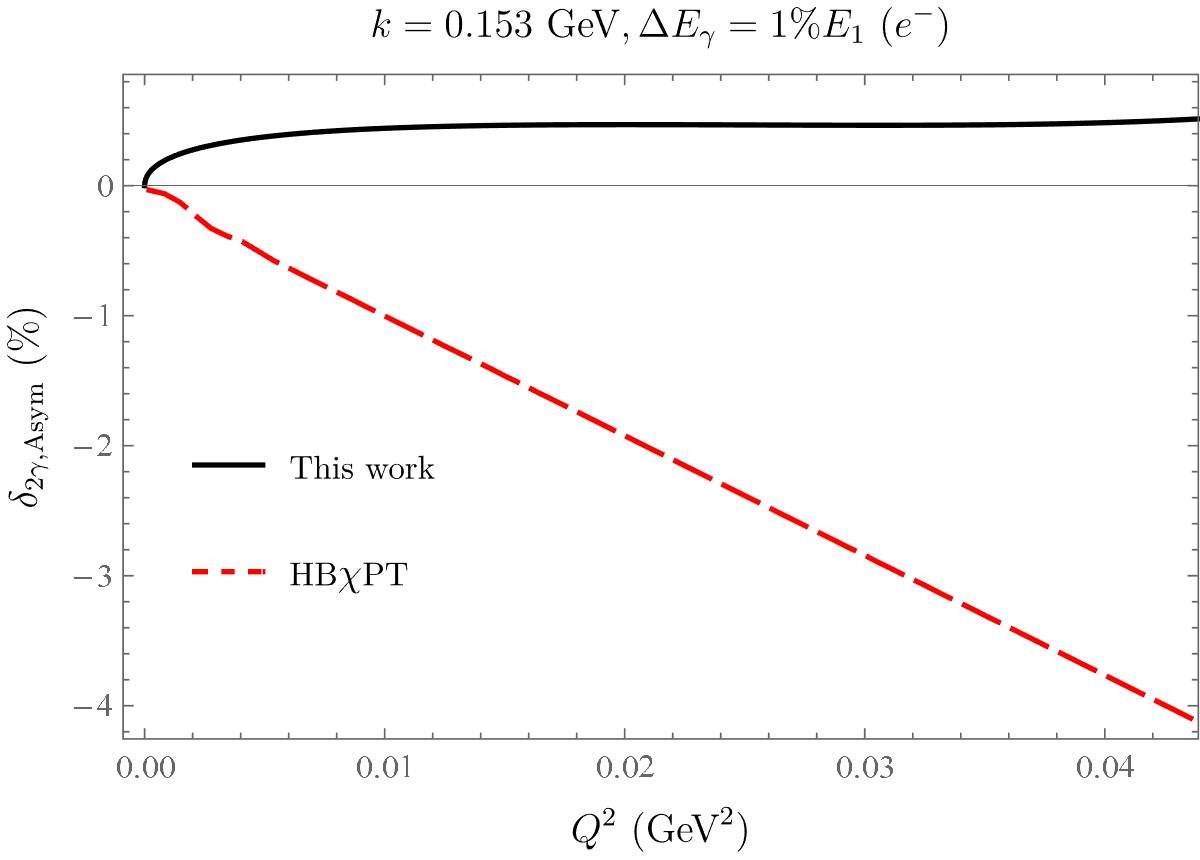}
        \vspace{0.1cm}
        \hspace{0.02cm}
        \end{minipage}
        \qquad
        \begin{minipage}[b]{0.45\linewidth}
        \includegraphics[scale=0.45]{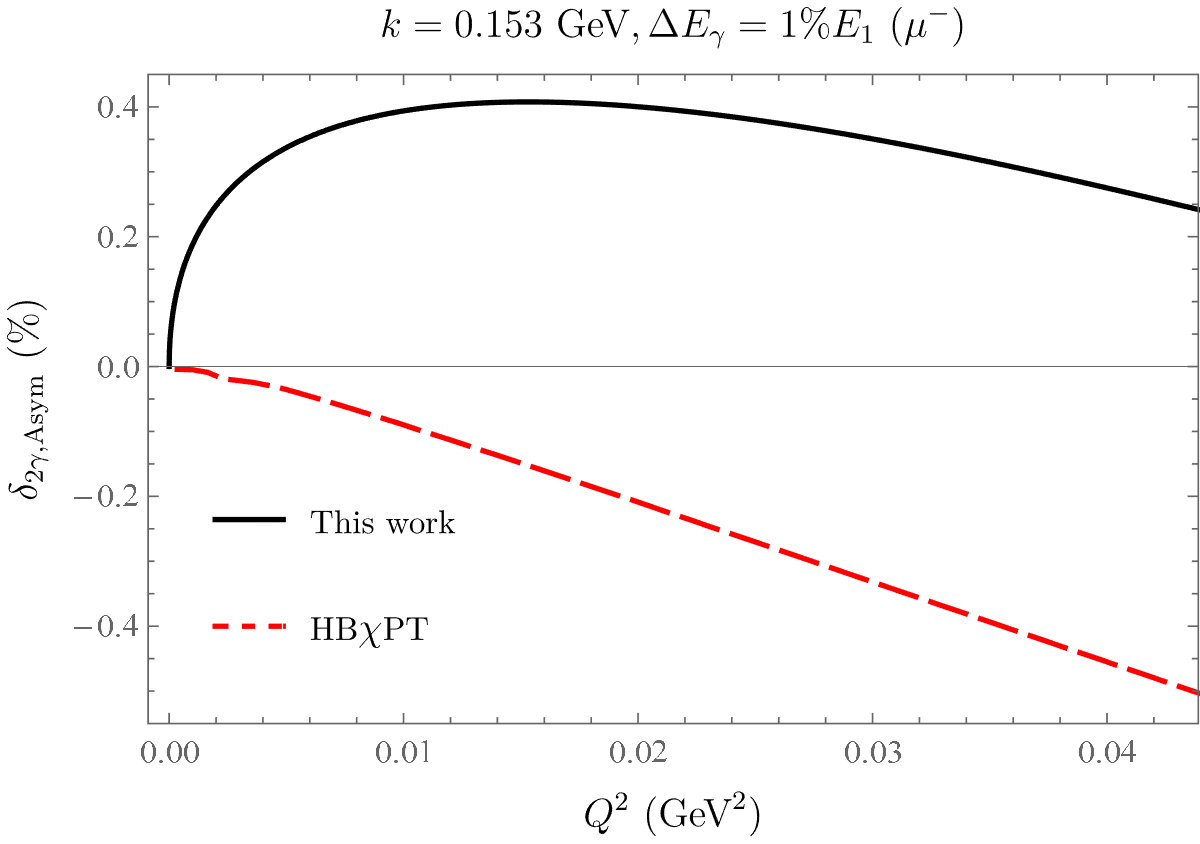}
        \vspace{0.1cm}
        \hspace{0.02cm}
        \end{minipage}

        \end{minipage}
    }
    \caption{Comparison of charge asymmetry to $e \text{p}$ and $\mu \text{p}$ elastic scatterings.
    }\label{Asym comparison fig}
\end{figure}

As previewed in Sec.~\ref{sec:3}, we only consider $\ell^- p$ scatterings.
On the other hand, by comparing $\ell^- p$ and $\ell^+ p$ scattering cross sections, we can obtain the charge asymmetry of radiative correction, which could be measured in modern experiments directly.
As mentioned above, only the TPE and the crossed bremsstrahlung corrections $\delta_{2\gamma, \text{Asym}}=\delta_{2\gamma, \text{TPE}}+\delta_{2\gamma, \text{xbr}}$ have charge dependence, i.e., the charge dependent differential cross section is given by $\md \sigma^{\ell^\mp} \simeq \md \sigma_{\gamma}\left(1\pm \delta_{2\gamma, \text{Asym}}+\cdots\right)$.
The charge asymmetry is defined as\cite{Koshchii:2017dzr}
\begin{align}
    \delta_{2\gamma, \text{Asym}}=\frac{\md \sigma^{\ell^-}-\md \sigma^{\ell^+}}{\md \sigma^{\ell^-}+\md \sigma^{\ell^+}}\ .
\end{align}
It can be connected to $\ell^{+} \text{p}/ \ell^{-} \text{p}$ ratio~\cite{Arrington:2011dn} by the definition,
$R^{\ell^{+} \ell^{-}}=\frac{\md \sigma^{\ell^{+}}}{\md \sigma^{\ell^{-}}} \simeq 1-2\delta_{2\gamma, \text{Asym}}$.
The predictions on charge asymmetry are shown in Figs.~\ref{charge sym Q} and \ref{charge sym ep}.
It is surprised to see in Fig.~\ref{Asym comparison fig} that the HB$\chi$PT calculations of charge asymmetry has opposite sign compared with our results.
The reason may lie in two facets: one is that HB$\chi$PT underestimates TPE effects by means of SPA, the other is that there are more diagrams contributing to bremsstrahlung process in HB$\chi$PT than B$\chi$PT and their power counting of $\delta_{2\gamma}$ is different from Eq.~(\ref{del chptdef}). 
This difference needs to be investigated in the future.

\begin{figure} 
    \centering
    \subfigure{
        \begin{minipage}[b]{0.99\linewidth}
          
        \begin{minipage}[b]{0.45\linewidth}
        \includegraphics[scale=0.5]{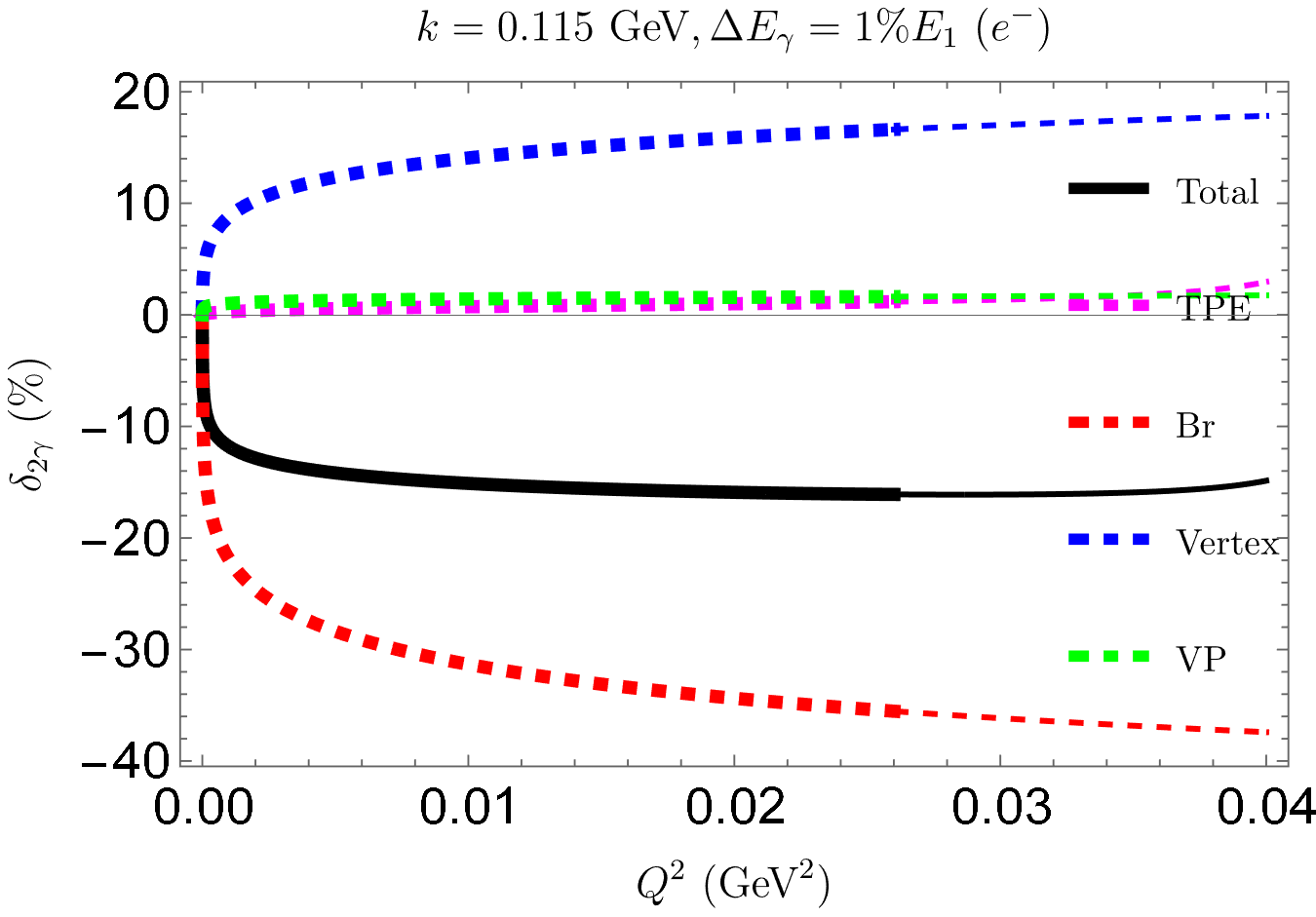}
        \vspace{0.1cm}
        \hspace{0.02cm}
        \end{minipage}
        \qquad
        \begin{minipage}[b]{0.45\linewidth}
        \includegraphics[scale=0.5]{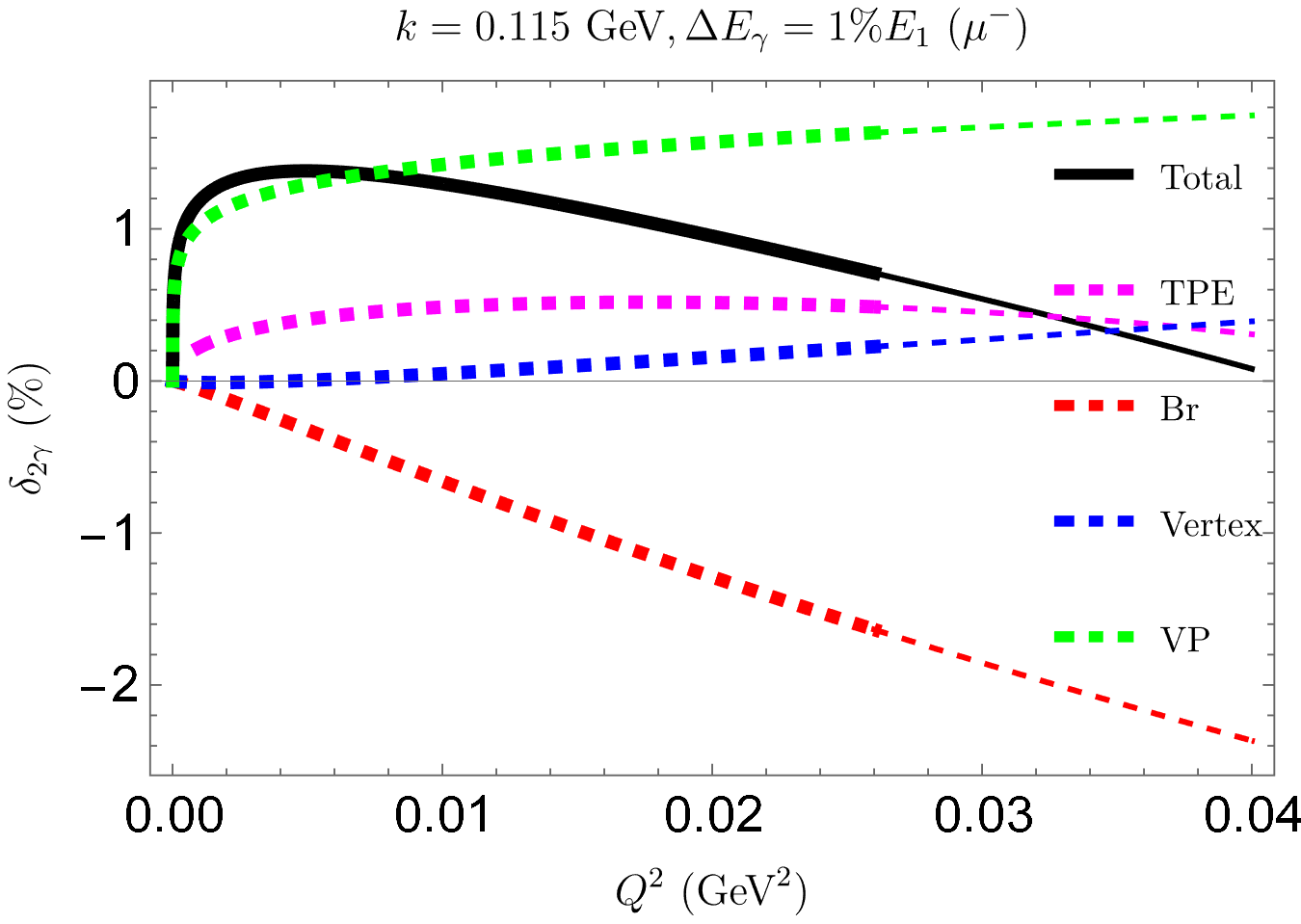}
        \vspace{0.1cm}
        \hspace{0.02cm}
        \end{minipage}
    
        \begin{minipage}[b]{0.45\linewidth}
        \includegraphics[scale=0.5]{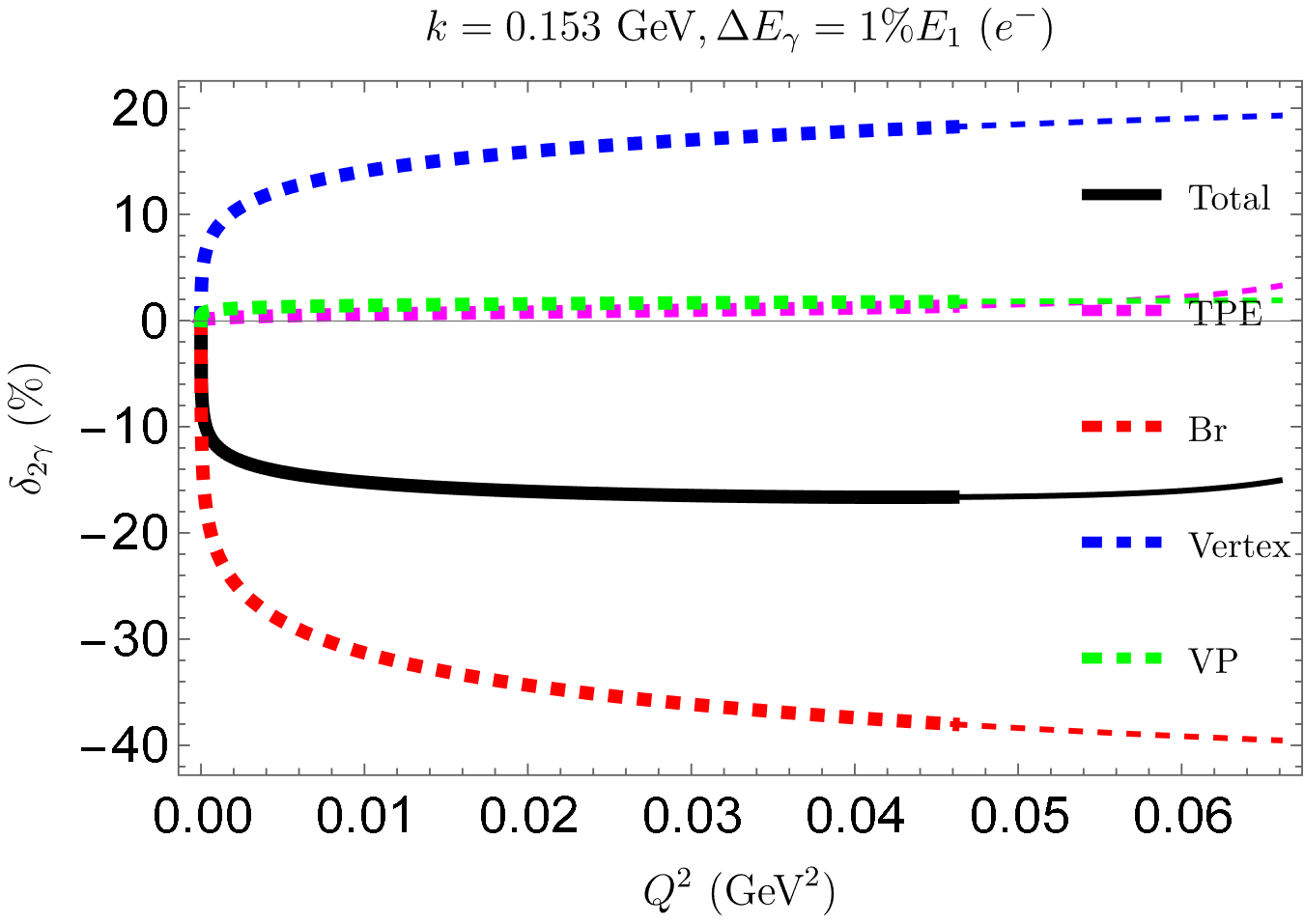}
        \vspace{0.1cm}
        \hspace{0.02cm}
        \end{minipage}
        \qquad
        \begin{minipage}[b]{0.45\linewidth}
        \includegraphics[scale=0.5]{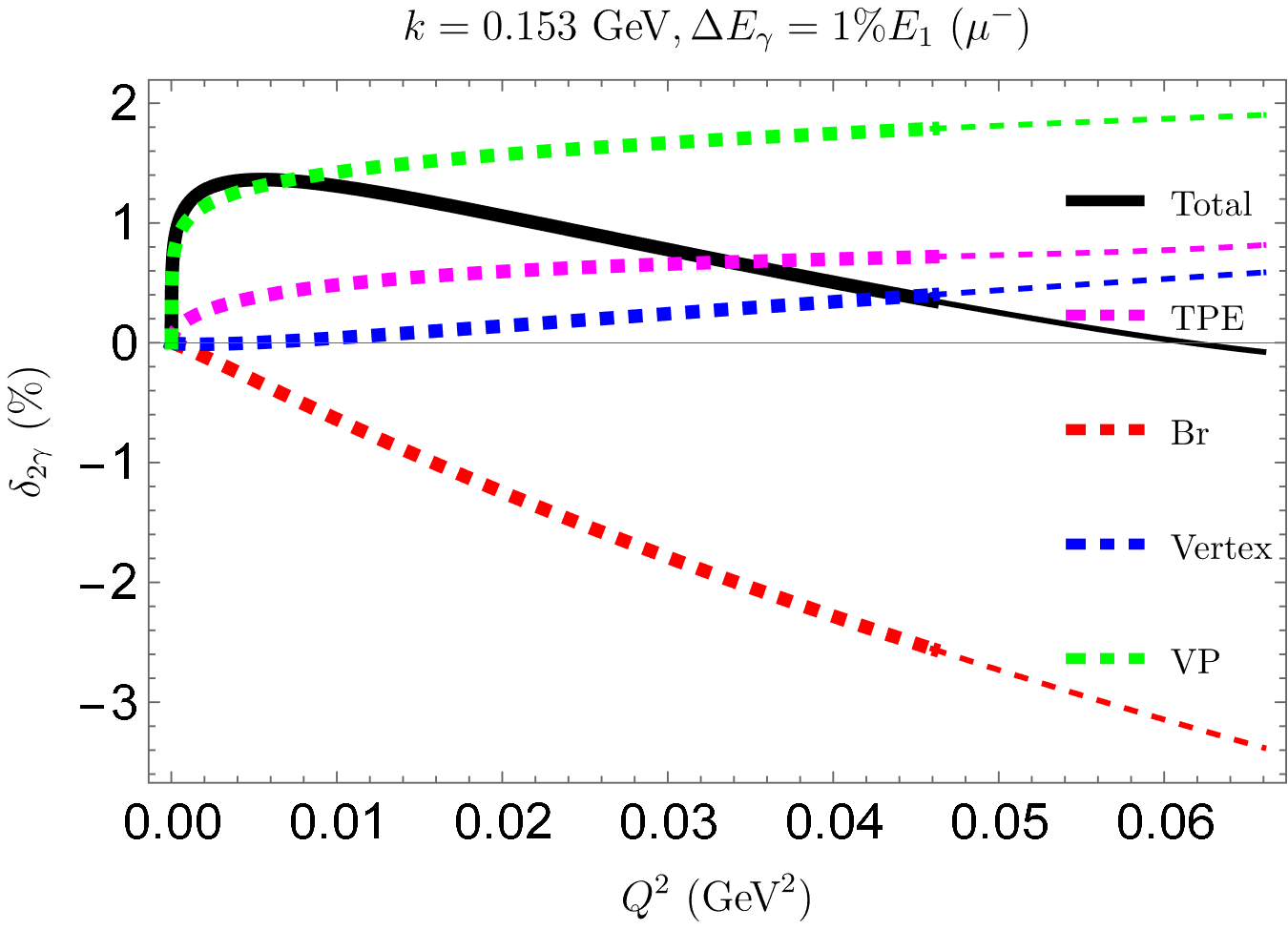}
        \vspace{0.1cm}
        \hspace{0.02cm}
        \end{minipage}

        \begin{minipage}[b]{0.45\linewidth}
        \includegraphics[scale=0.5]{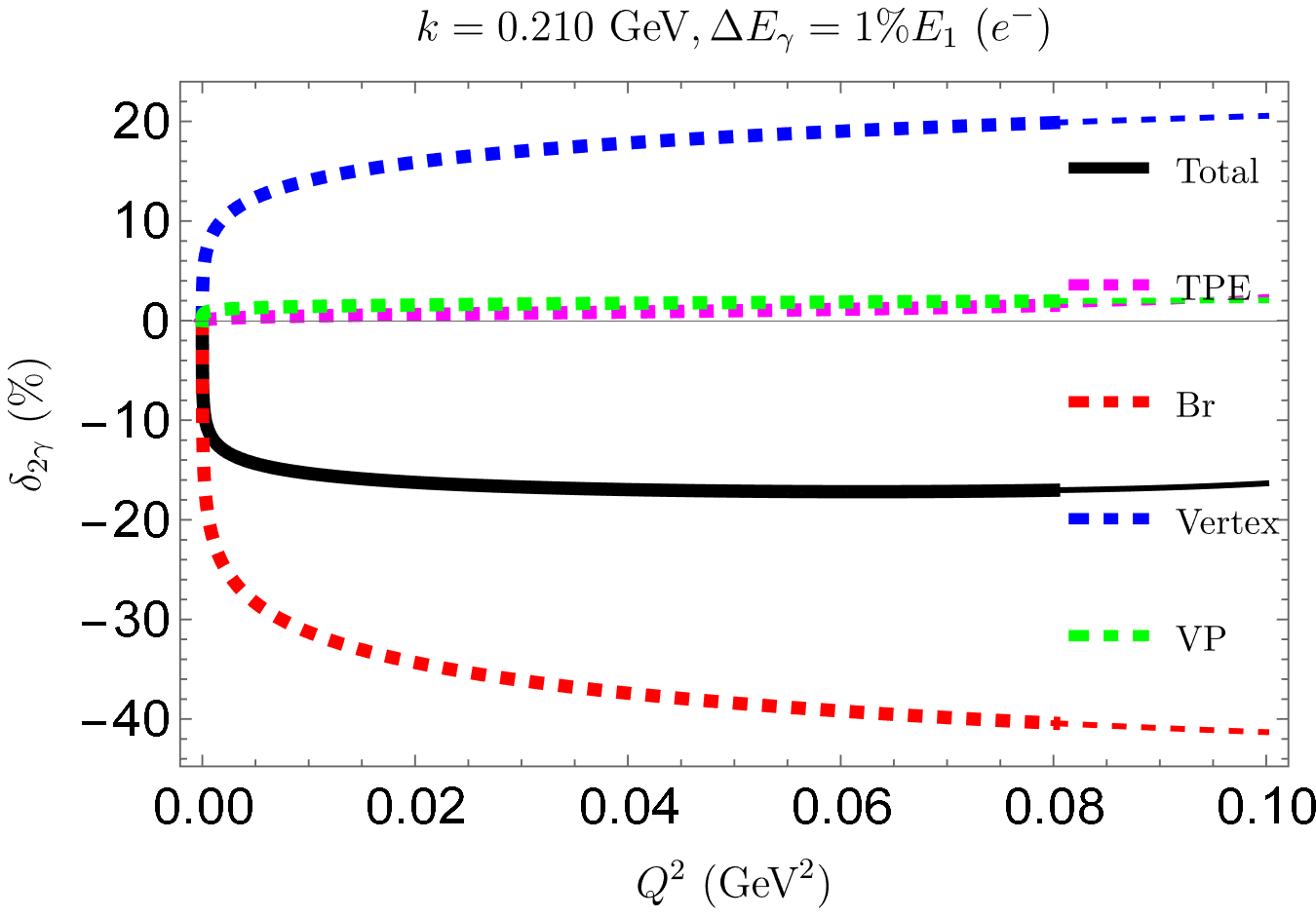}
        \vspace{0.1cm}
        \hspace{0.02cm}
        \end{minipage}
        \qquad
        \begin{minipage}[b]{0.45\linewidth}
        \includegraphics[scale=0.5]{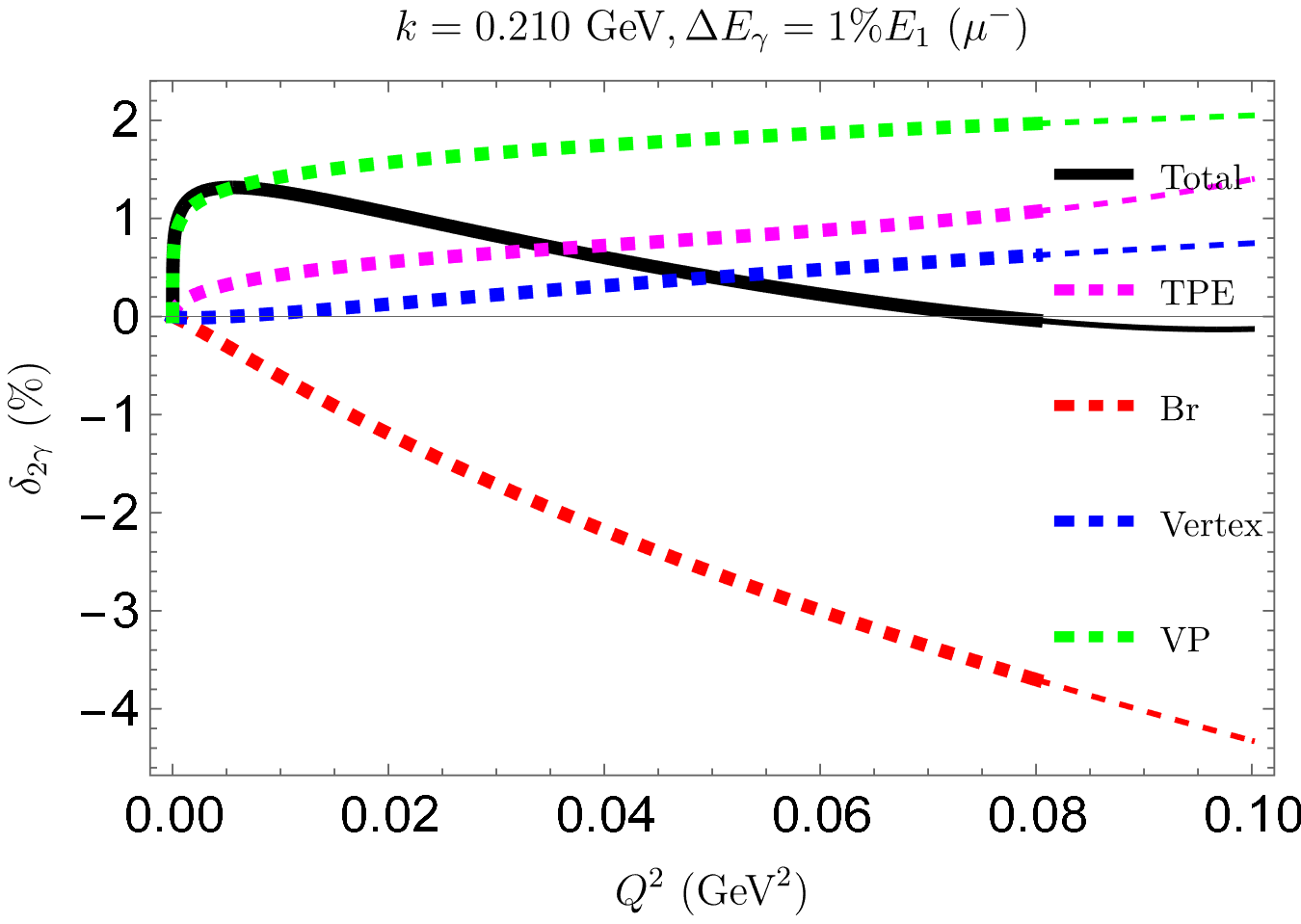}
        \vspace{0.1cm}
        \hspace{0.02cm}
        \end{minipage}    
   
        \end{minipage}
        }
    \caption{The correction from different sources up to NLO, the thickened part of each curve corresponds to MUSE kinematical region.
    }\label{total individual}
\end{figure}

In Fig.~\ref{total individual}, we summarize all the contributions of $\ell^- \text{p}$ elastic scatterings up to NLO.
We just note that in Fig.~\ref{total individual} large cancellations occur between the vertex correction and bremsstrahlung contribution at LO in $e \text{p}$ scatterings, which was discovered in Ref.~\cite{Talukdar:2020aui}.

\begin{figure} 
    \centering
    \subfigure{
        \begin{minipage}[b]{0.98\linewidth}
          
        \begin{minipage}[b]{0.45\linewidth}
        \includegraphics[scale=0.5]{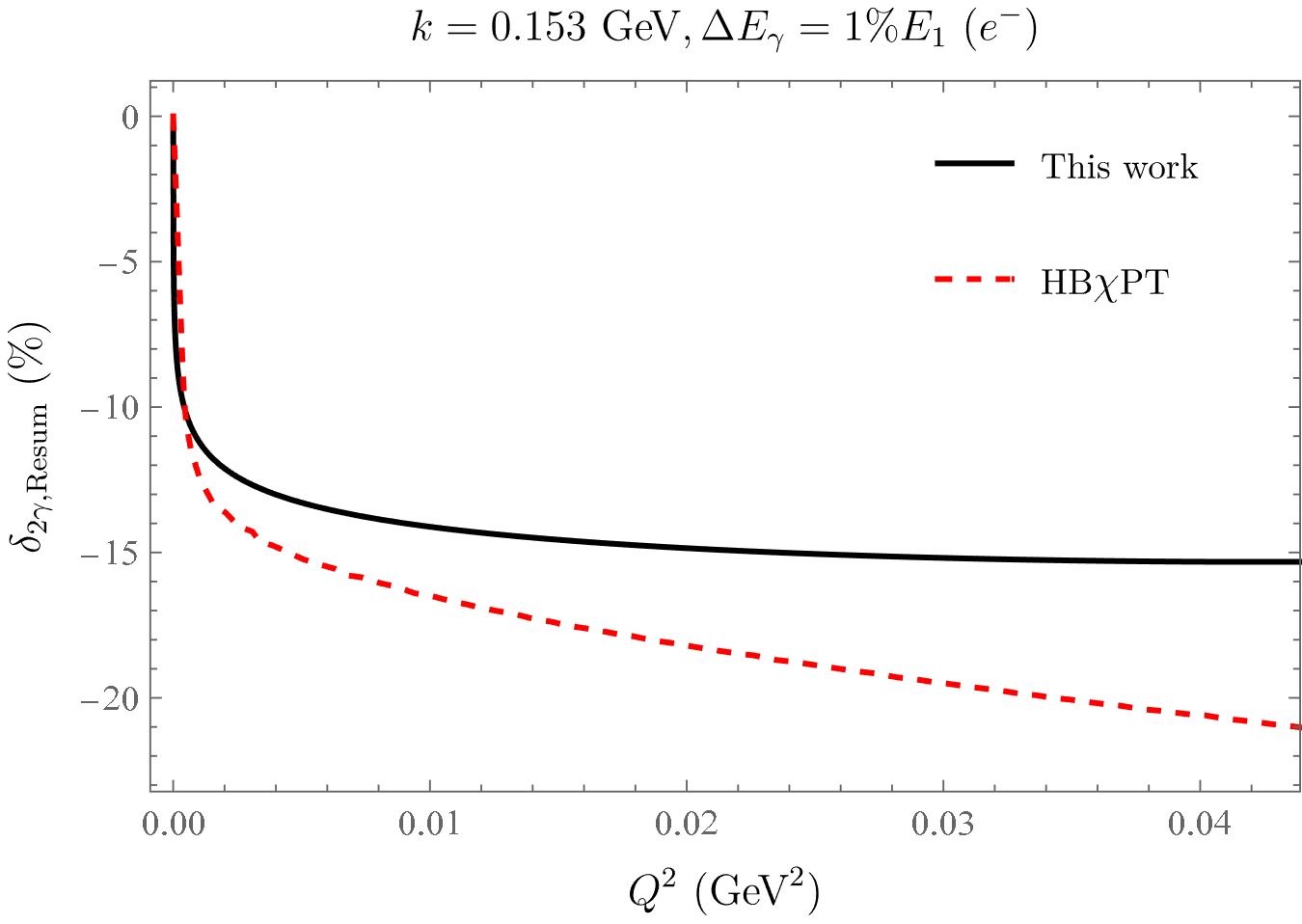}
        \vspace{0.1cm}
        \hspace{0.02cm}
        \end{minipage}
        \qquad
        \begin{minipage}[b]{0.45\linewidth}
        \includegraphics[scale=0.5]{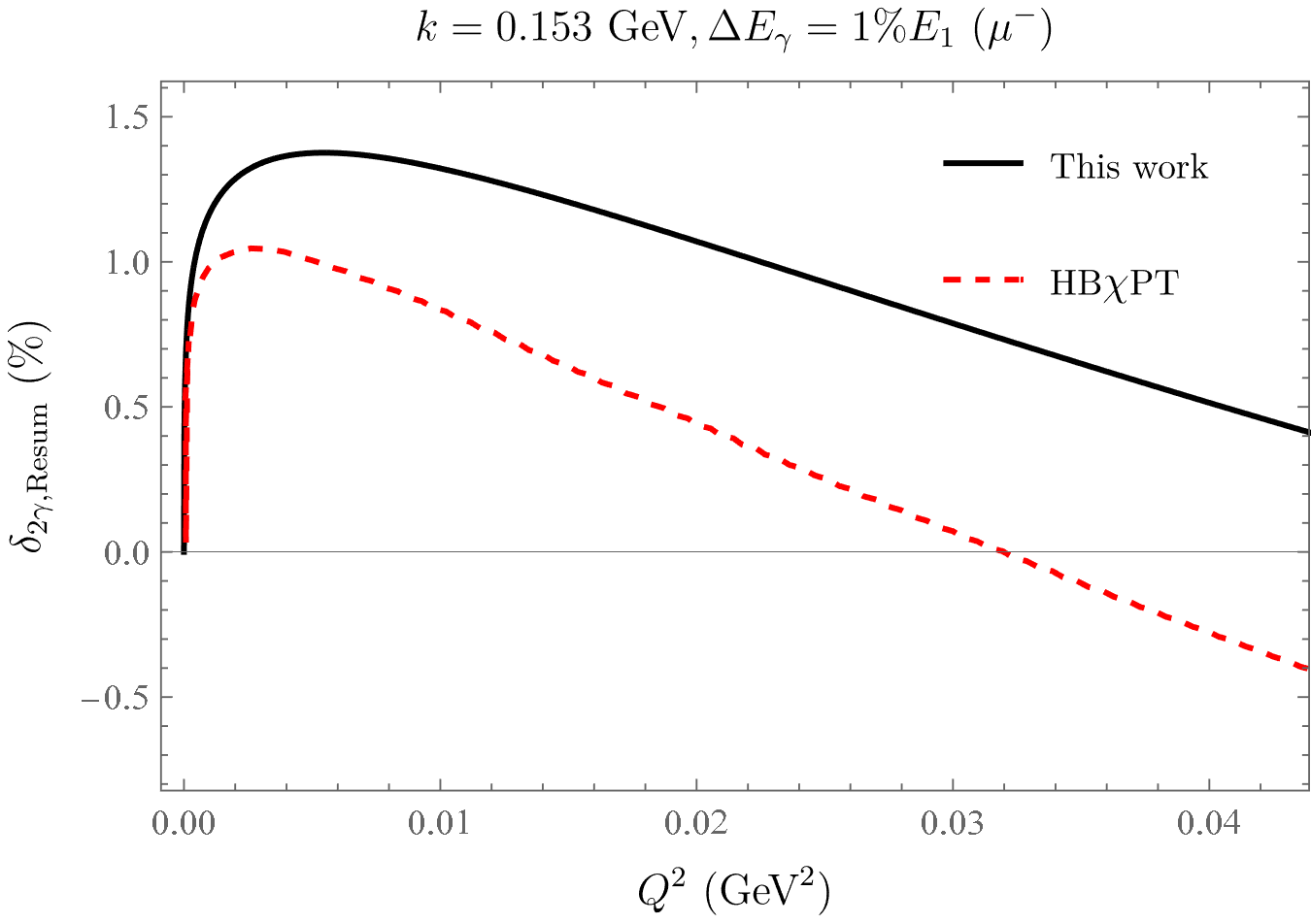}
        \vspace{0.1cm}
        \hspace{0.02cm}
        \end{minipage}

        \end{minipage}
    }
    \caption{Comparison of radiative corrections to $e \text{p}$ and to $\mu \text{p}$ elastic scatterings.
    }\label{Total comparison fig}
\end{figure}

\begin{figure} 
    \centering
    \subfigure{
        \begin{minipage}[b]{0.99\linewidth}
          
        \begin{minipage}[b]{0.45\linewidth}
        \includegraphics[scale=0.5]{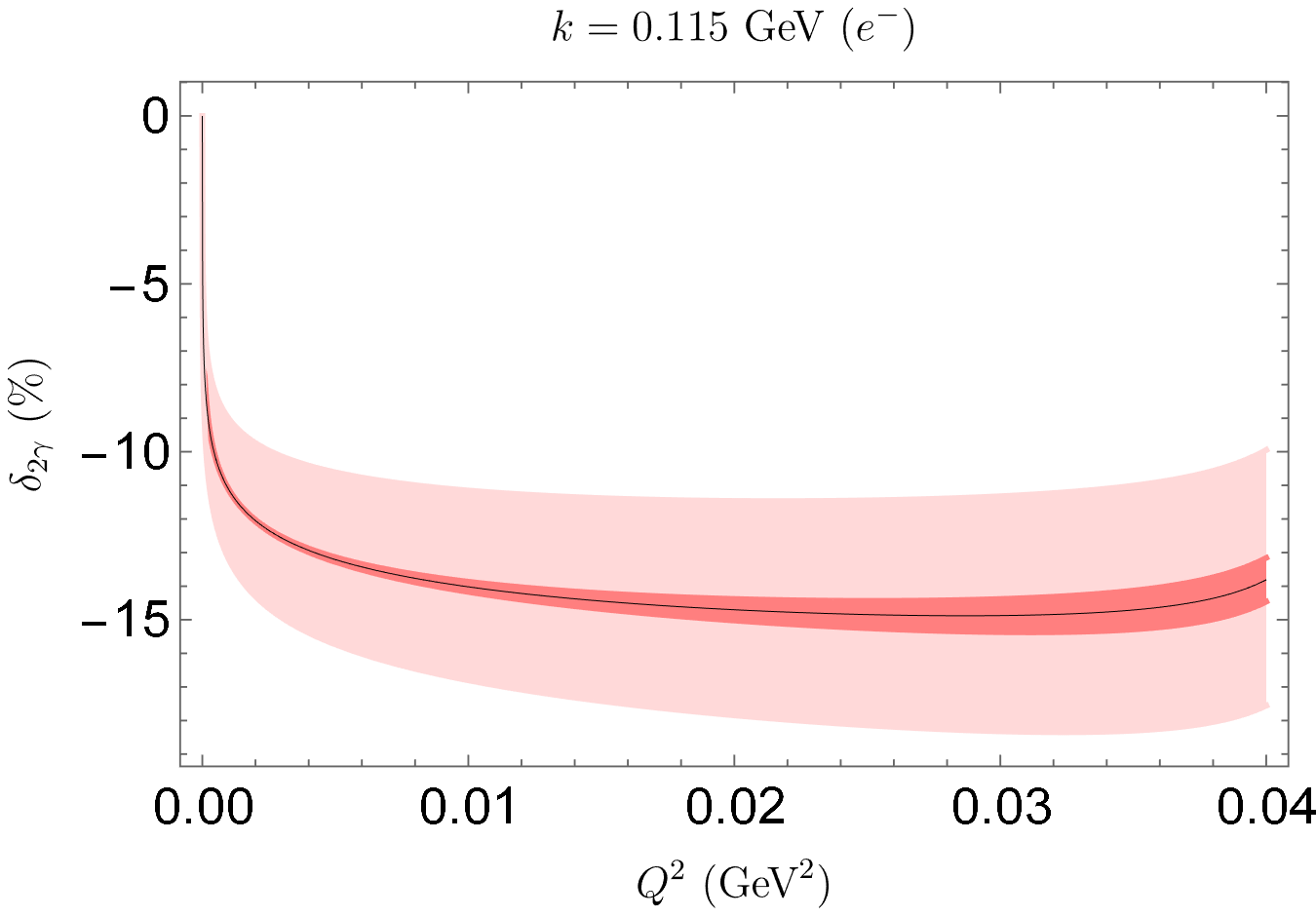}
        \vspace{0.1cm}
        \hspace{0.02cm}
        \end{minipage}
        \qquad
        \begin{minipage}[b]{0.45\linewidth}
        \includegraphics[scale=0.5]{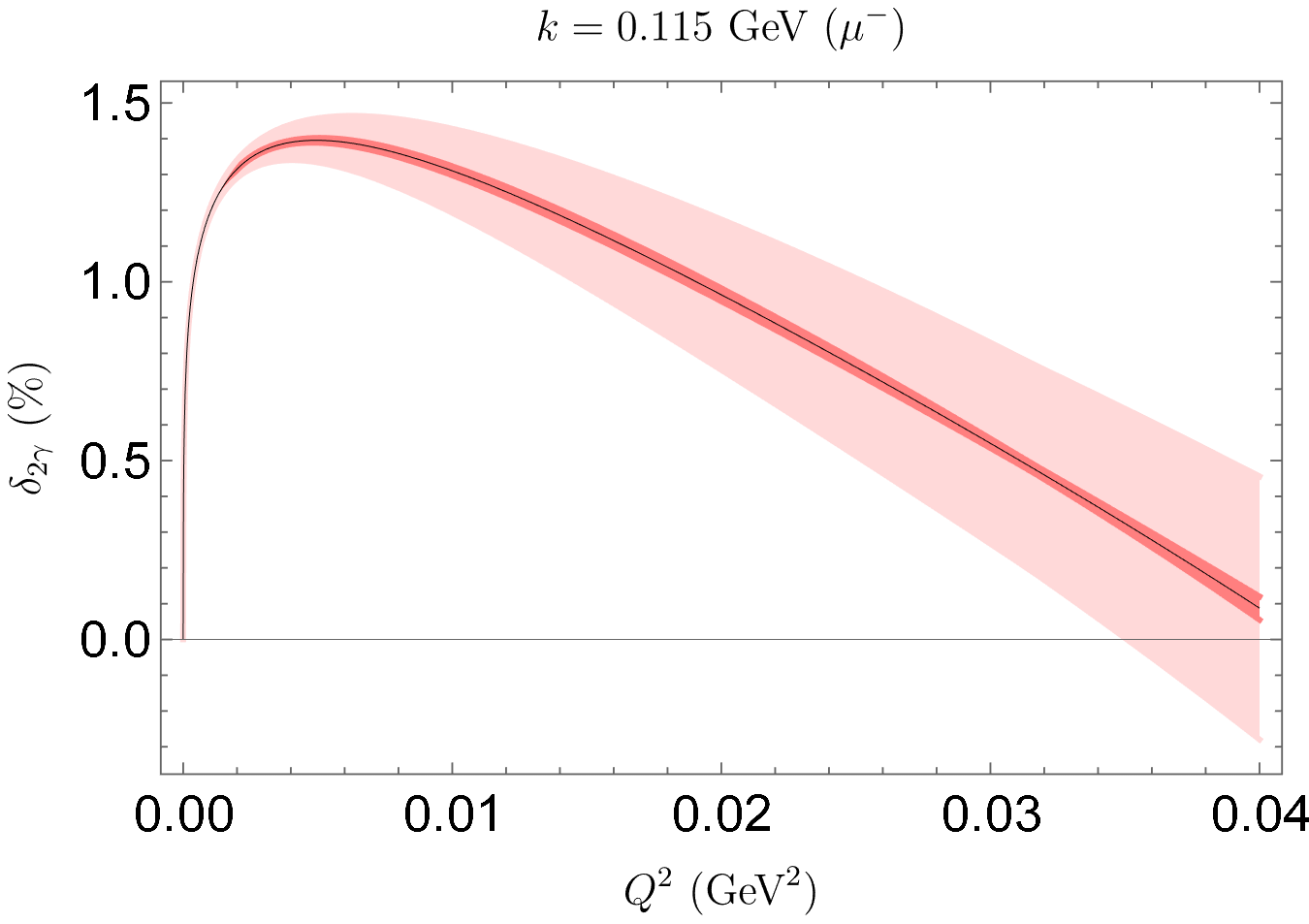}
        \vspace{0.1cm}
        \hspace{0.02cm}
        \end{minipage}
    
        \begin{minipage}[b]{0.45\linewidth}
        \includegraphics[scale=0.5]{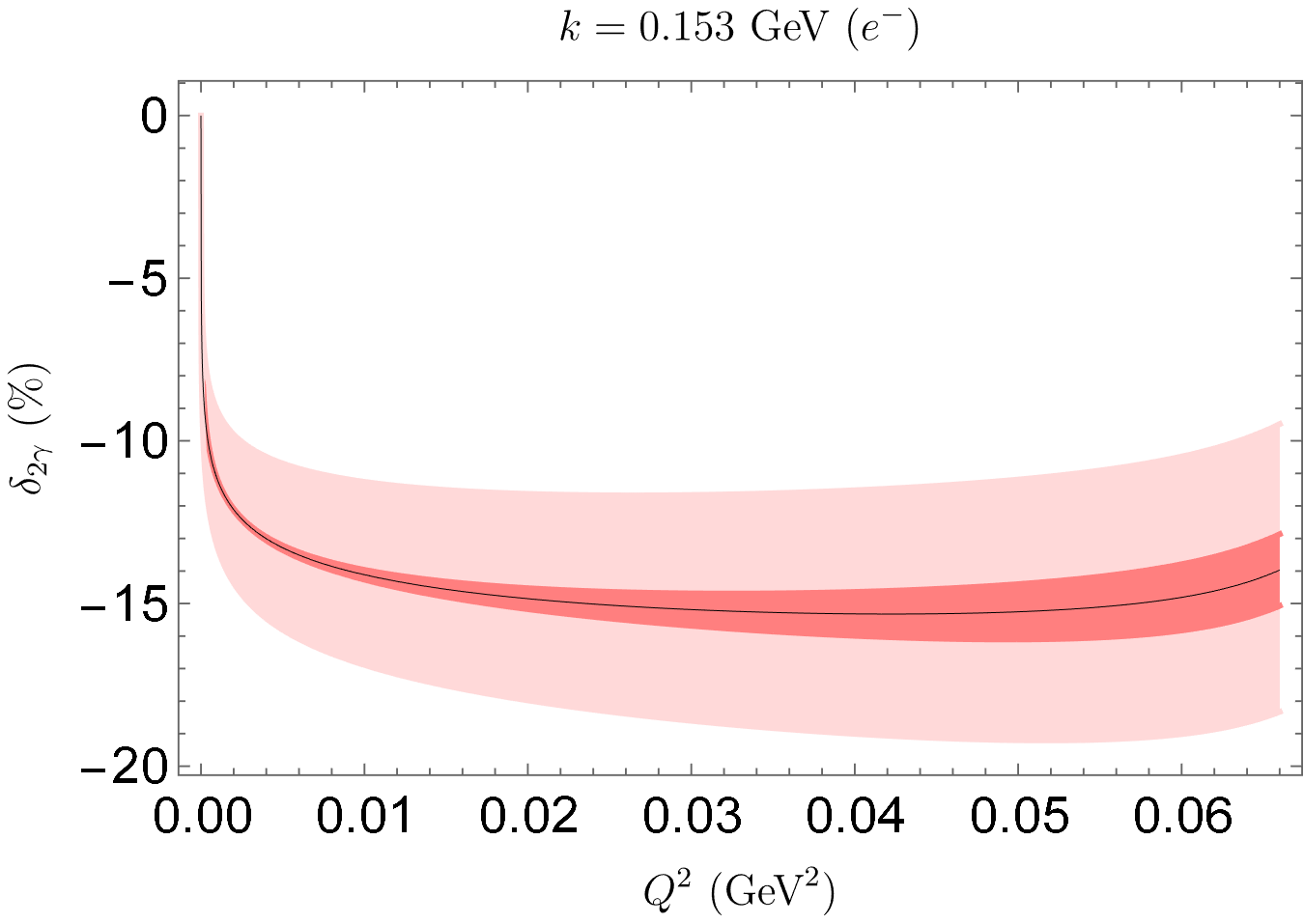}
        \vspace{0.1cm}
        \hspace{0.02cm}
        \end{minipage}
        \qquad
        \begin{minipage}[b]{0.45\linewidth}
        \includegraphics[scale=0.5]{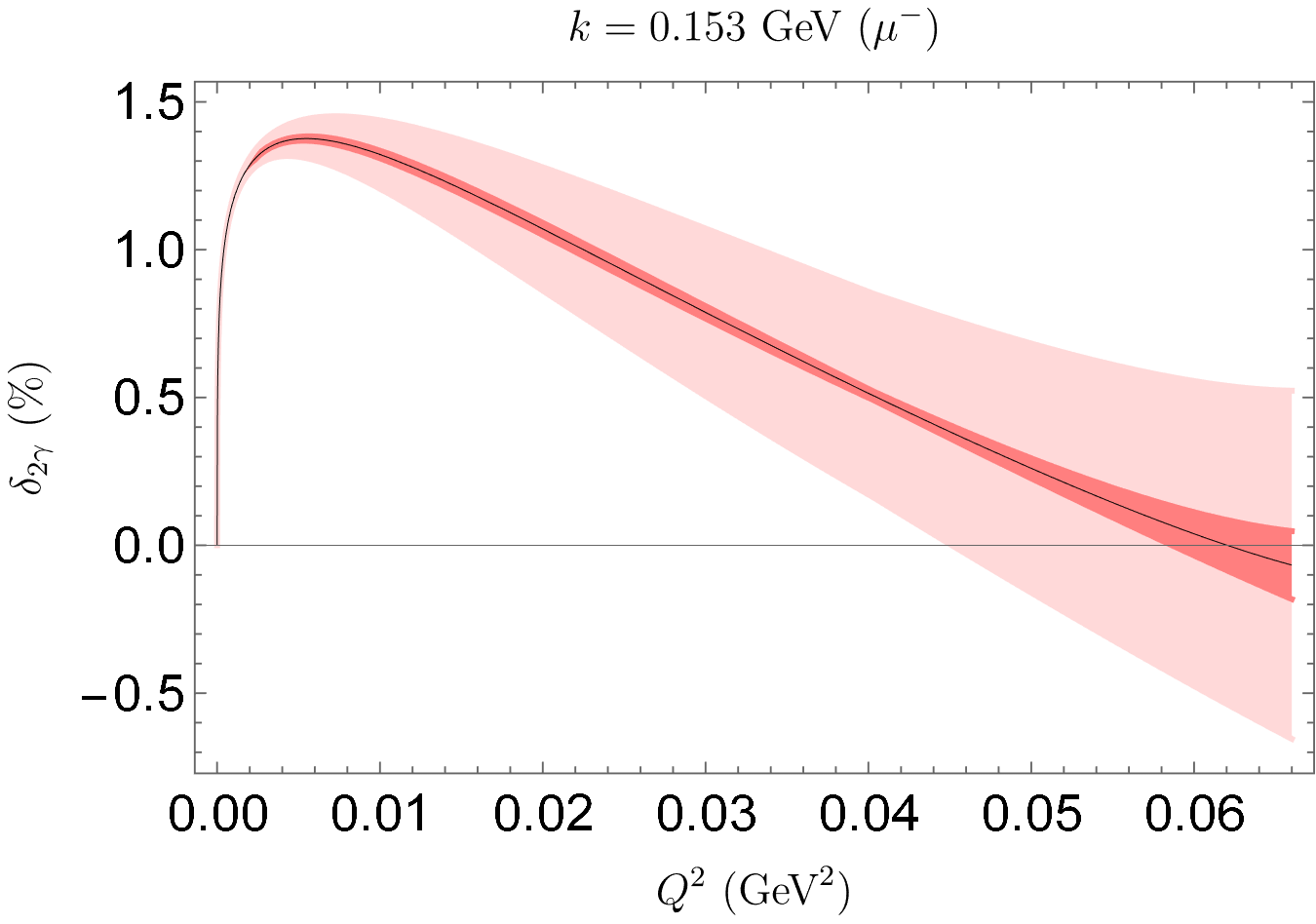}
        \vspace{0.1cm}
        \hspace{0.02cm}
        \end{minipage}

        \begin{minipage}[b]{0.45\linewidth}
        \includegraphics[scale=0.5]{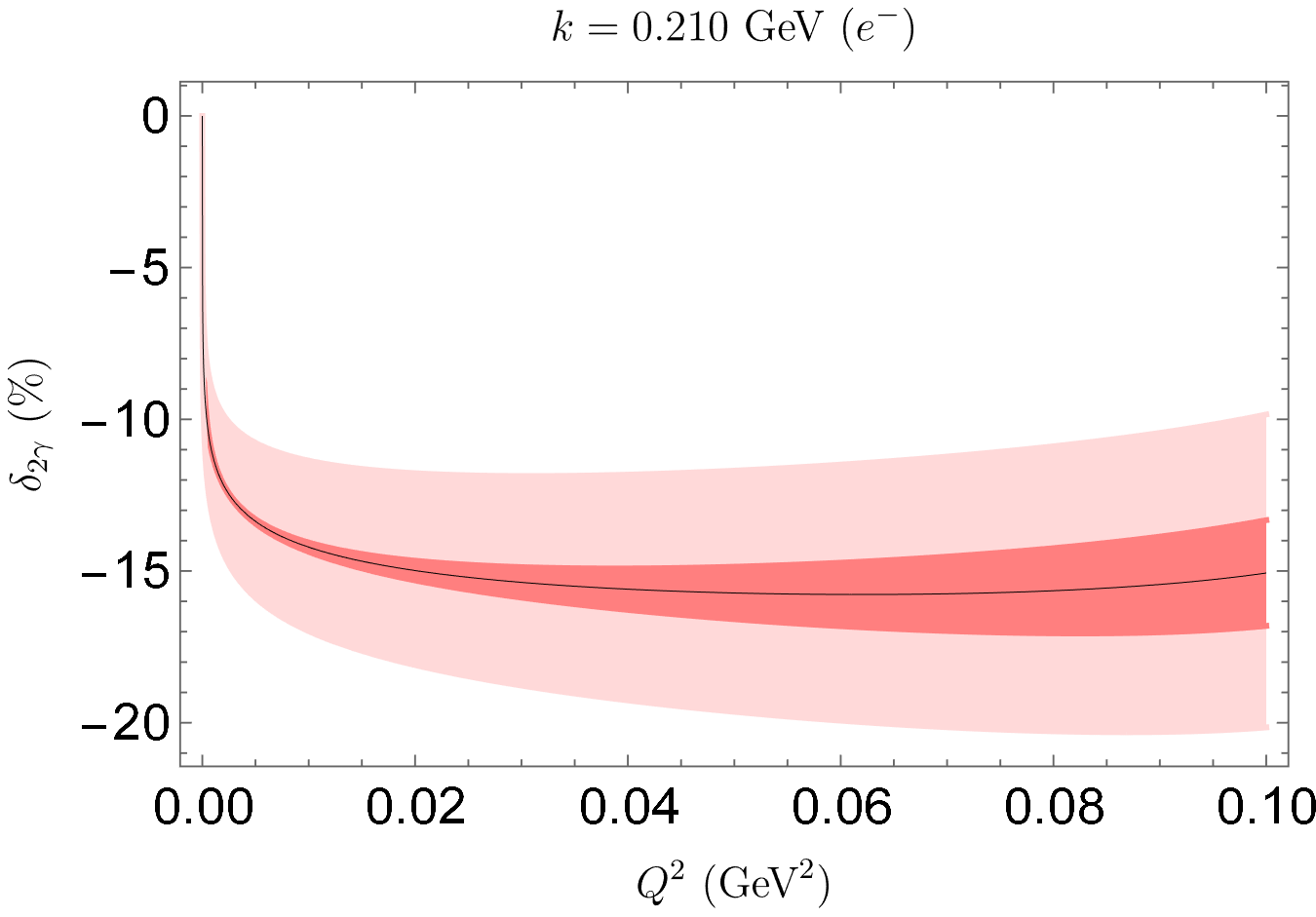}
        \vspace{0.1cm}
        \hspace{0.02cm}
        \end{minipage}
        \qquad
        \begin{minipage}[b]{0.45\linewidth}
        \includegraphics[scale=0.5]{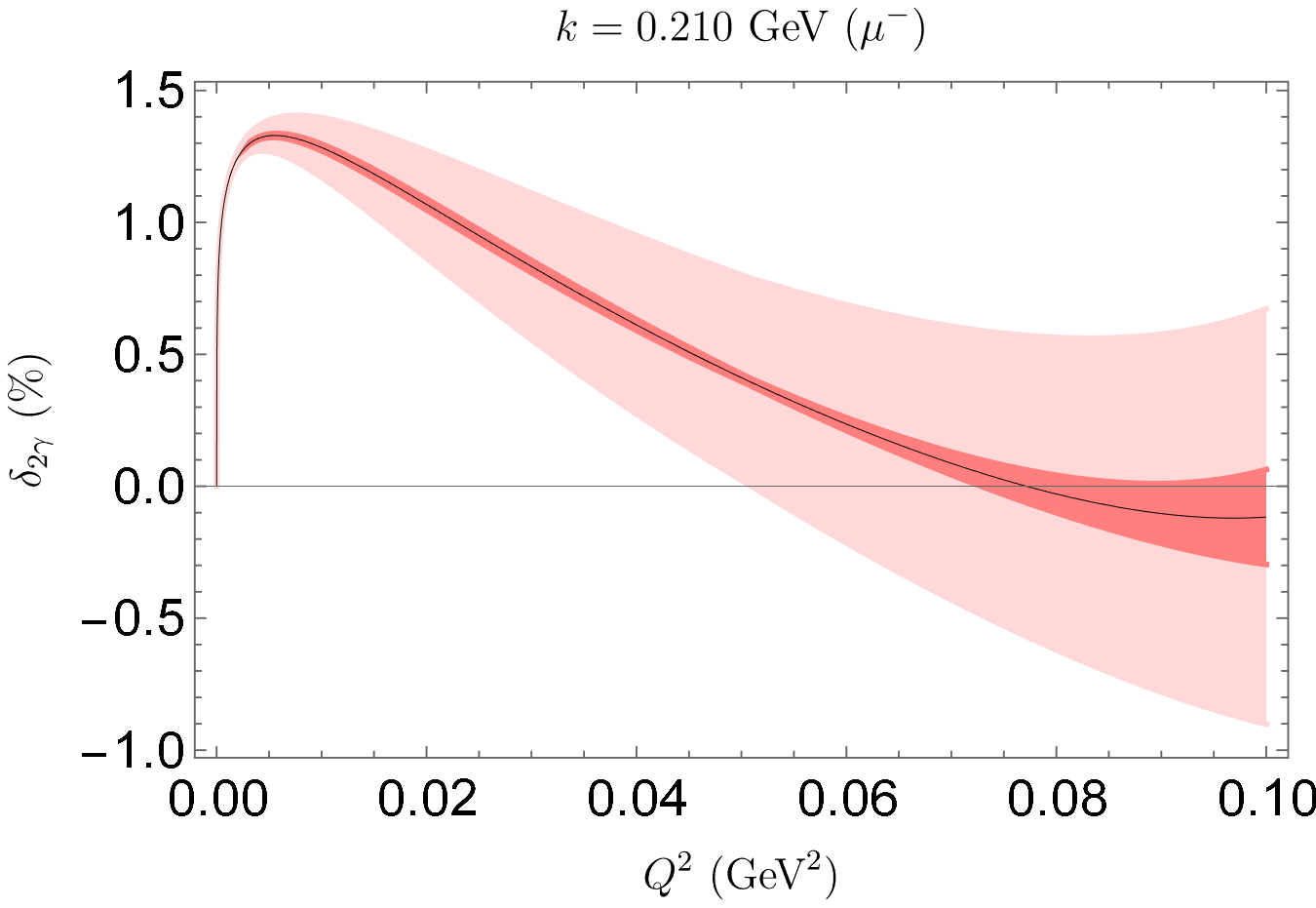}
        \vspace{0.1cm}
        \hspace{0.02cm}
        \end{minipage}    
   
        \end{minipage}
        }
    \caption{The inner band represents the NNLO chiral trunction uncertainties obtained by Eq.(~\ref{uncertainty NNLO}).
    The outer band stands for the variation $0.5\% E_1<\Delta E_\gamma<2\% E_1$; upper limit corresponds to $\Delta E_\gamma=2\%$; lower limit corresponds to $\Delta E_\gamma=0.5\%$.}\label{uncertainty}
\end{figure}

The so-called Sudakov double-log term that appeared in an IR divergent part has a significant enhancement at $Q^2 \gg  m^2$, which will obviously make the perturbation expansion invalid under large transfer momentum. 
It means that more than one soft photon radiation needs to be considered.
So we can approximately take into account the high order by exponentiating the LO QED corrections.
It was firstly proposed in Refs.~\cite{Bloch:1937pw, Schwinger:1949ra, Yennie:1961ad}.
Therefore, the differential cross section can be written as:\footnote{This approximation can be checked by comparing the result with the first order of exponential expansion,
$\left[\frac{\mathrm{d} \sigma_{\text{el}}\left(Q^{2}\right)}{\mathrm{d} \Omega_{\ell}}\right]_{\text{lab}} \simeq \left[\frac{\mathrm{d} \sigma_{\text{el}}\left(Q^{2}\right)}{\mathrm{d} \Omega_{\ell}}\right]_{\gamma}^{(1)} \times\left(1+\delta_{2 \gamma}(Q^2)\right)$.}
\begin{align}
    \left[\frac{\mathrm{d} \sigma_{\text{el}}\left(Q^{2}\right)}{\mathrm{d} \Omega_{\ell}}\right]_{\text{lab}} \simeq \left[\frac{\mathrm{d} \sigma_{\text{el}}\left(Q^{2}\right)}{\mathrm{d} \Omega_{\ell}}\right]_{\gamma}^{(1)} \times\left(1+\delta^{\text{el}}_{\text{resum}}(Q^2)\right) \ ,
\end{align}
where the resumed contribution is~\cite{Vanderhaeghen:2000ws}
\begin{align}
    \delta_{\text {resum }}^{\text{el}}\left(Q^{2}\right)=\frac{\exp \left[\delta_{2 \gamma}\left(Q^{2}\right)-\delta_{2\gamma, \mathrm{vp}}(Q^2)\right]}{\left[1-\delta_{2\gamma, \mathrm{vp}}\left(Q^{2}\right) / 2\right]^{2}}-1\ .
\end{align}
\begin{figure} 
    \centering
    \subfigure{
        \begin{minipage}[b]{0.99\linewidth}
          
        \begin{minipage}[b]{0.45\linewidth}
        \includegraphics[scale=0.5]{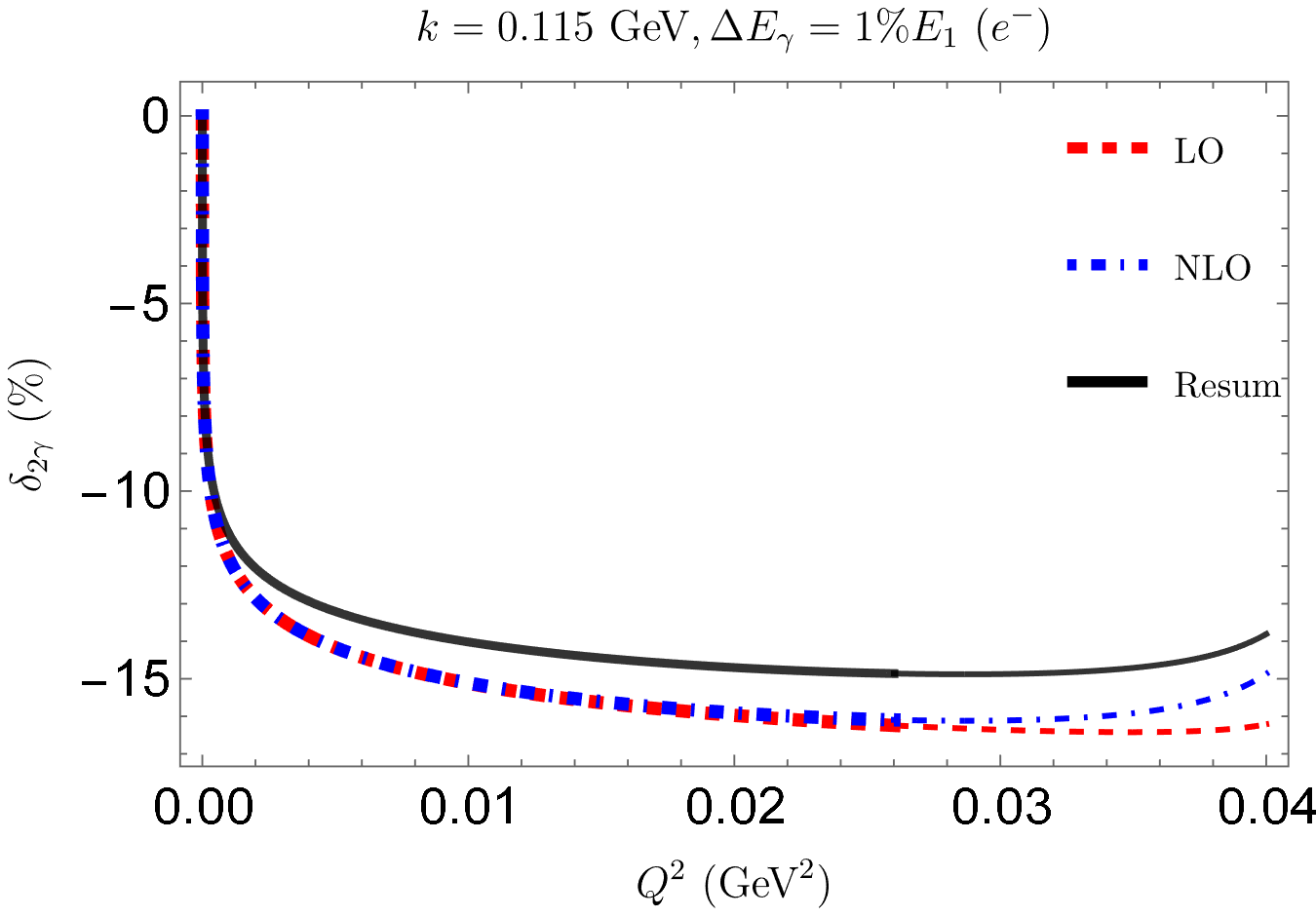}
        \vspace{0.1cm}
        \hspace{0.02cm}
        \end{minipage}
        \qquad
        \begin{minipage}[b]{0.45\linewidth}
        \includegraphics[scale=0.5]{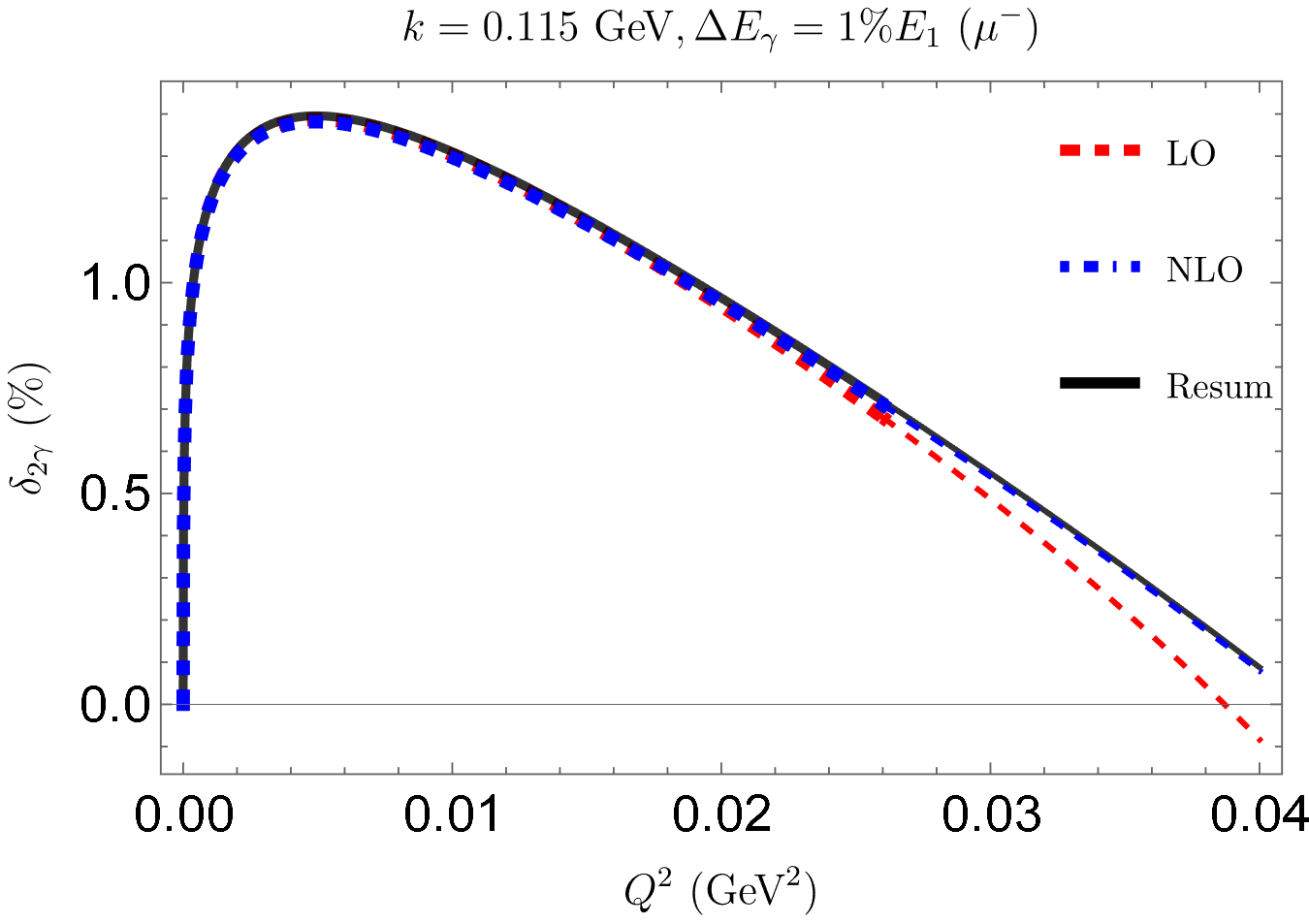}
        \vspace{0.1cm}
        \hspace{0.02cm}
        \end{minipage}
    
        \begin{minipage}[b]{0.45\linewidth}
        \includegraphics[scale=0.5]{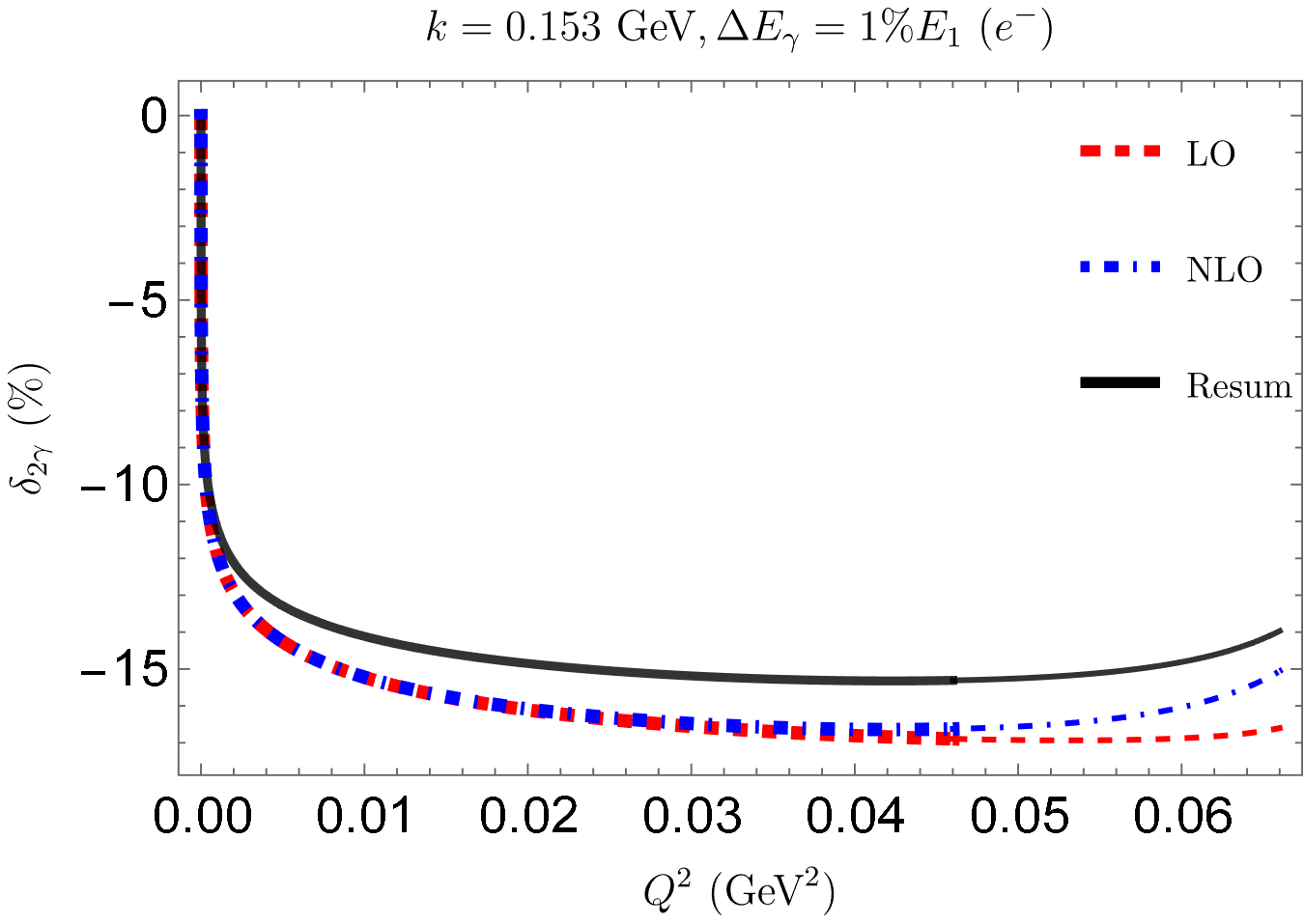}
        \vspace{0.1cm}
        \hspace{0.02cm}
        \end{minipage}
        \qquad
        \begin{minipage}[b]{0.45\linewidth}
        \includegraphics[scale=0.5]{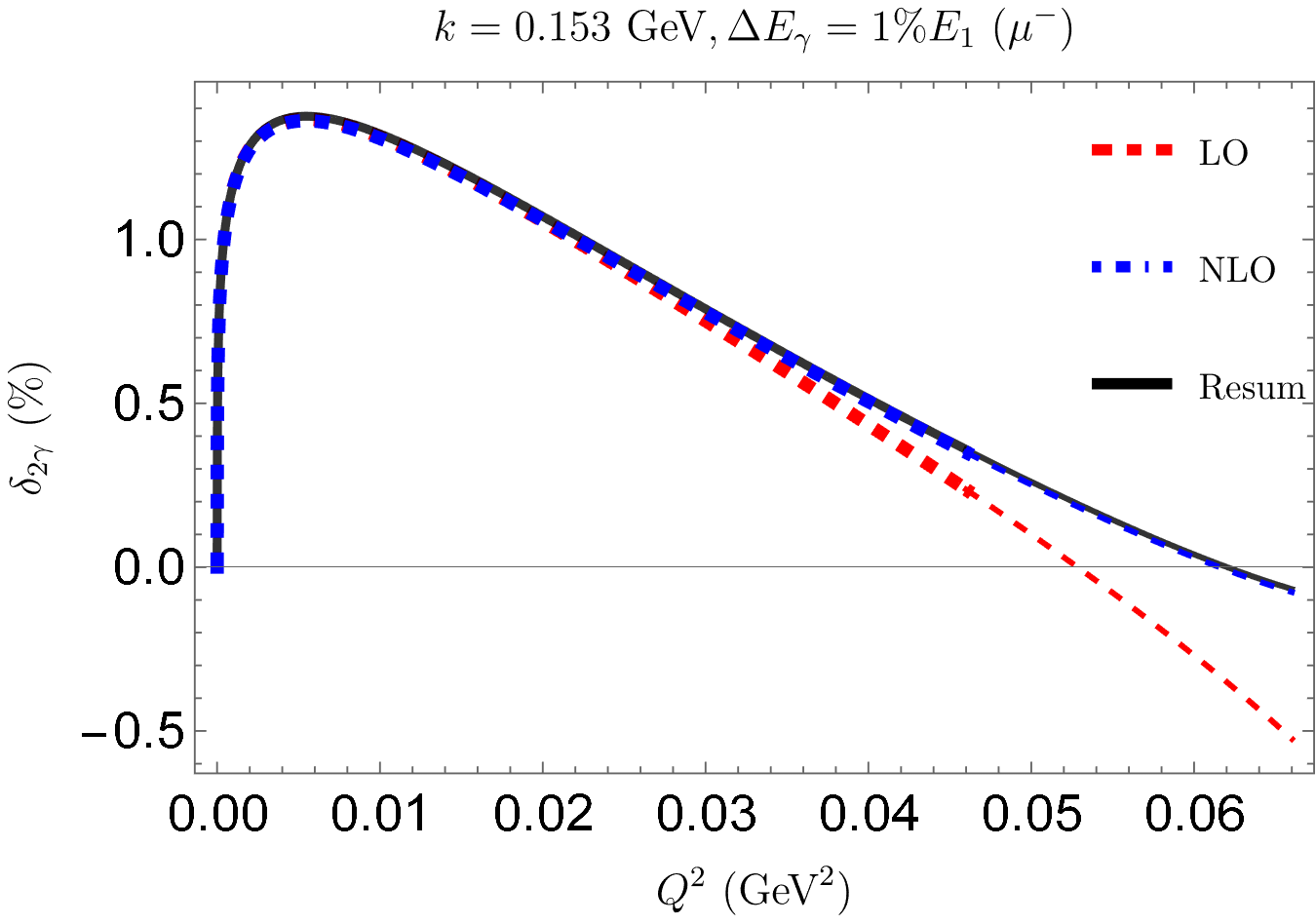}
        \vspace{0.1cm}
        \hspace{0.02cm}
        \end{minipage}

        \begin{minipage}[b]{0.45\linewidth}
        \includegraphics[scale=0.5]{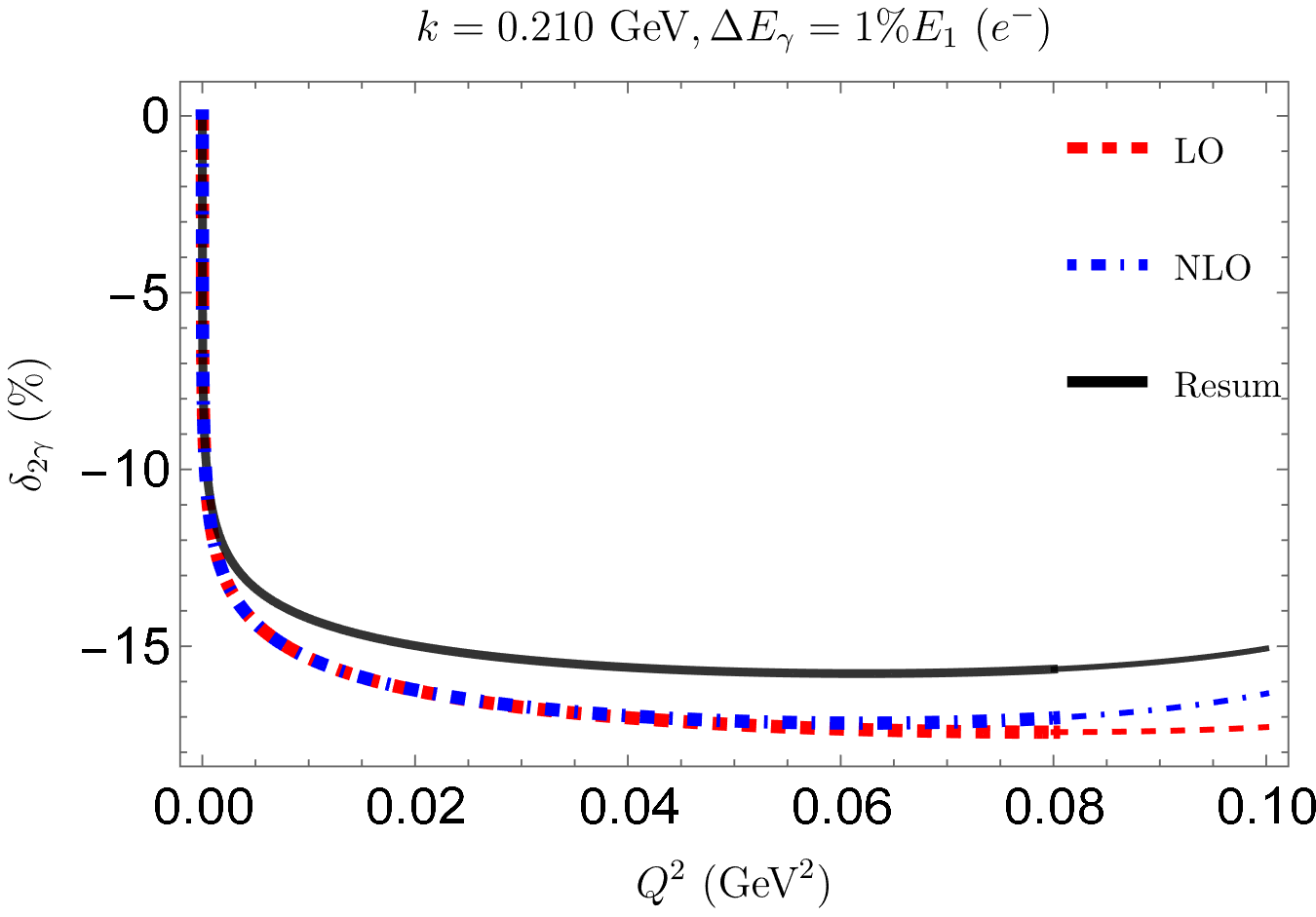}
        \vspace{0.1cm}
        \hspace{0.02cm}
        \end{minipage}
        \qquad
        \begin{minipage}[b]{0.45\linewidth}
        \includegraphics[scale=0.5]{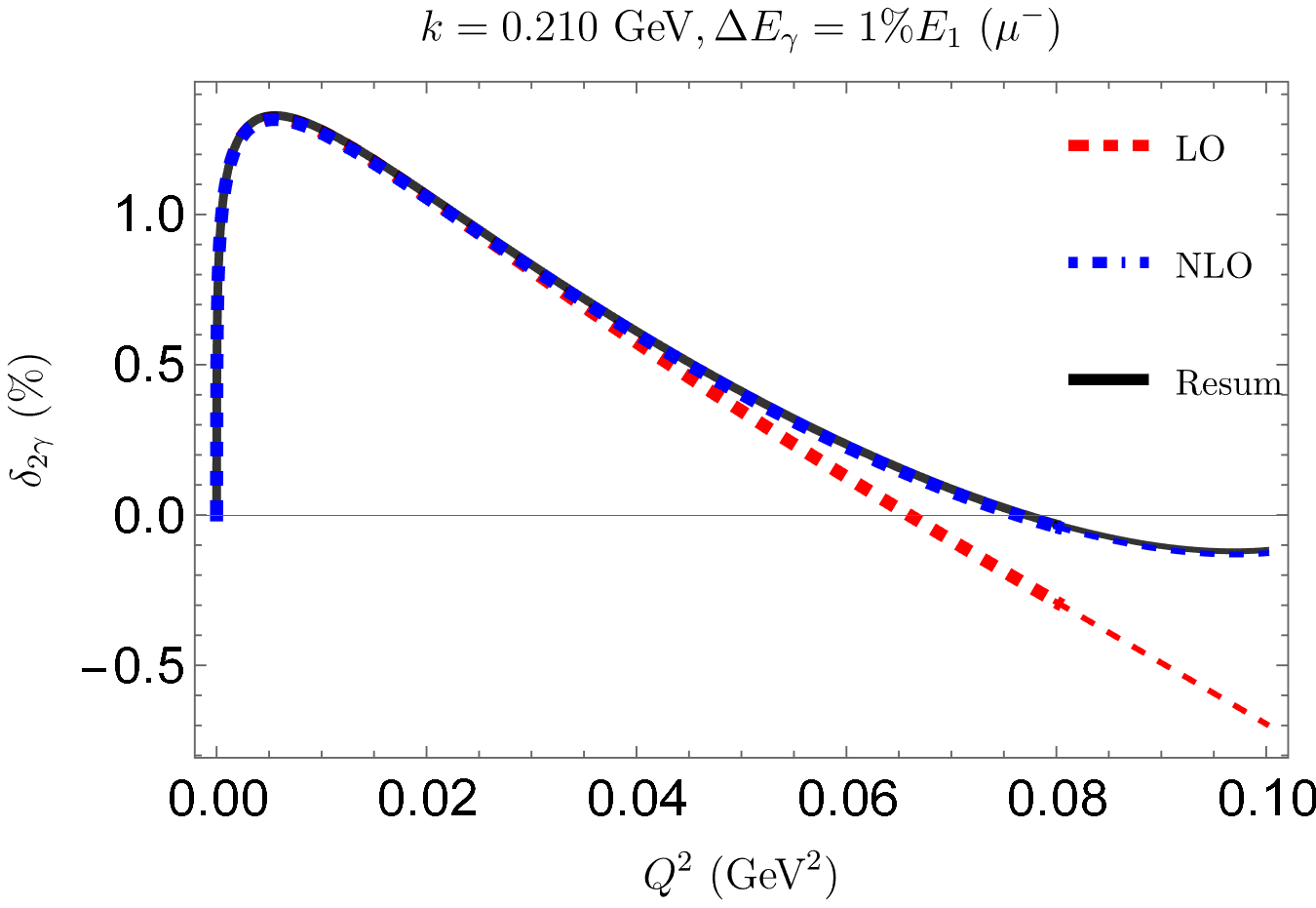}
        \vspace{0.1cm}
        \hspace{0.02cm}
        \end{minipage}    
   
        \end{minipage}
        }
    \caption{The total radiative corrections for $e \text{p}$ and $\mu \text{p}$ scatterings.}\label{resum}
\end{figure}
In Fig.~\ref{resum}, we compare the LO, NLO and resumed NLO results for $e \text{p}$ and $\mu \text{p}$ scatterings.
The total contributions vary between $15\%$ and $20\%$ in MUSE kinematical region for $e \text{p}$ scatterings.
As for $\mu \text{p}$ scatterings, the total radiative correction does not exceed $1.5\%$ in the limit region of MUSE.
One can immediately discover that the lepton incoming momentum dependence of total radiative correction is not obvious, especially in the MUSE kinematical region.
We also compare the resumed results with HB$\chi$PT in Fig.~\ref{Total comparison fig}.
The magnitude of complete radiative corrections from B$\chi$PT are the same as HB$\chi$PT basically, 
and the numerical difference shows that both schemes may need a complete calculation of chiral next-to-next-leading-order correction to clarify this point.
Finally, the theoretical uncertainties of the total radiative correction mainly comes from two aspects.
First, the detector acceptance $\Delta E_\gamma$ is not known exactly, which relies on the structure of detector.
According to Ref.~\cite{Talukdar:2020aui}, we assume that $\Delta E_\gamma$ varies between $0.5\% E_1$ and $2\% E_1$.
Second, the chiral truncation up to NLO is another unknown uncertainty.
Using the method of Ref.~\cite{Epelbaum:2014efa}, for NLO calculation, an estimation of uncertainty is expressed as,
\begin{align}\label{uncertainty NNLO}
    \delta \mathcal{O}^{(2)}=\max \left\{\left|\mathcal{O}^{\left(1\right)}\right| B^{2},\left|\mathcal{O}^{(2)}-\mathcal{O}^{(1)}\right| B\right\}\ ,
\end{align}
where $B=Q/M$. 
The uncertainty originated from contributions including pion loops, excited states of nucleon and etc.
Fig.~\ref{uncertainty} shows the error bands of above uncertainties.

\section{Discussion and conclusion}\label{sec:8}

In this work, based on Lorentz-invariant B$\chi$PT, we have investigated $\ell \text{p}$ elastic and inelastic scatterings including a nonzero lepton mass in low momentum transfer $Q^2$.
Our approach involves virtual QED loops and soft photon bremsstrahlung corrections.
In particular, the TPE contribution is calculated beyond SPA.
Consequently, we have found that the SPA scheme misses the hard region of loop integrals~\cite{Talukdar:2020aui}, and it has significant effects on numerical results.

The charge asymmetry is also calculated to chiral NLO analytically, and the present result improves previous theoretical predictions such as Ref.~\cite{Koshchii:2016muj}, where the authors use SPA to estimate the TPE results.
The model-independent charge asymmetry $\delta_{2\gamma, \text{Asym}}$ (or $\ell^{+} \text{p} / \ell^{-} \text{p}$ ratio) can be tested in future precision experiments such as MUSE.
It is worth noting that regardless of $e \text{p}$ or $\mu \text{p}$ scatterings, the order of magnitude is about $1\%$.

The total radiative corrections are resumed by exponentiation method.
The estimation of total radiative corrections for $e \text{p}$ scattering cross section vary between $10\%$ and $20\%$.
But for $\mu \text{p}$ scatterings, it does not exceed $1.5\%$ in MUSE kinematical region.
These radiative corrections, especially the TPE correction, are valuable in providing an improved fit results of electric and magnetic FFs from elastic $\ell \text{p}$ scatterings.
However, for experiments, using different approximations of TPE correction has little influence on final differential cross section in analysis\cite{Lorenz:2014yda}.
In Refs\cite{Horbatsch:2015qda, Horbatsch:2016ilr}, the authors argued that different parametrizations of FFs are the most critical point in fitting low-$Q^2$ cross section data of $e \text{p}$ elastic scattering experiment. 
Radiative corrections will only give a small correction to the value of charge radius, but can not change the value from ``large'' (about 0.88 fm) to ``small'' (about 0.84 fm), or vice versa\cite{Horbatsch:2015qda}.
Nevertheless, for $\mu \text{p}$ scatterings, radiative corrections, especially TPE corrections, may play an important role in the extraction of FFs and charge radius.
It is thus instructive to investigate the validity of our results in elastic $\mu \text{p}$ scatterings such as MUSE\cite{MUSE:2013uhu, MUSE:2017dod}.
For future planned $e \text{p}$ and $\mu \text{p}$ scattering experiments, we recommend the recent report\cite{GaoHY}.
Finally, the extension to the description of other radiative corrections such as $e^+ e^- \to \pi^+ \pi^-, N \bar{N}$\cite{Zhao:2020naq} will advance these studies even further, while offering the possibility of making reliable and accurate predictions for future precision experiments.

\section*{Acknowledgments}
The author X. H. C would like to thank Hao Chen for helpful discussions.
We are also grateful to Guang-Peng Zhang and Zhi-Guang Xiao for a careful reading of the manuscript and valuable comments.
This work is supported in part by National Nature Science Foundations of China (NSFC) under Contract N0. 11975028 and N0. 10925522.

\newpage

\begin{appendices}
 
\section{Bremsstrahlung integrals}\label{appendix:1}

The complete calculation of bremsstrahlung integrals utilizing DR can be found in Appendix A.5 of Ref.~\cite{Vanderhaeghen:2000ws}, since the expressions are rather lengthy and are inconvenient to use. 
Here we adopt another method,\cite{tHooft:1978jhc} \footnote{There is a slight difference that they introduced a photon mass $\lambda$ to regulate the IR divergent instead of DR.} which gives a more compact result:
\begin{align}\label{integral-I-kp}
    I(k_i, p_j)=\frac{R^{(1)}\left(k_{i}, p_{j}\right)}{8 \pi^{2}}\left(-\frac{1}{\epsilon_{\mathrm{IR}}}+\gamma_{E}-\ln \frac{4 \pi \nu^{2}}{Q^{2}}-\ln \frac{Q^{2}}{4\left(\Delta E^{S}\right)^{2}}\right)+\frac{R^{(2)}\left(k_{i}, p_{j}\right)}{8 \pi^{2}}\ ,
\end{align}
where ($m_{i}^{2}=k_{i}^{2}, m_{j}^{2}=p_{j}^{2}$)
\begin{align}
    R^{(1)}(k_i,p_j)=&\frac{1}{\gamma_{i j}} \ln \left(\frac{k_{i} \cdot p_{j}+\gamma_{i j}}{m_i m_j}\right)\ ,  \\
    R^{(2)}(k_i,p_j)=&\frac{1}{\gamma_{i j}}\left[\ln ^{2}\left(\frac{\beta\left(k_{i}\right)}{m_i M}\right)-\ln ^{2}\left(\frac{\beta\left(p_{j}\right)}{m_j M}\right)\right.  \nonumber\\
    &+\mathrm{Li}_2\left(1-\frac{\beta\left(k_{i}\right) l_{ij} \cdot p_2}{M^{2} \gamma_{i j}}\right)+\mathrm{Li}_2\left(1-\frac{m_i^{2} l_{ij} \cdot p_2}{\beta\left(k_{i}\right) \gamma_{i j}}\right)  \nonumber\\
    &\left.-\mathrm{Li}_2\left(1-\frac{\beta\left(p_{j}\right) l_{ij} \cdot p_2}{M^2 \alpha_{ij}\gamma_{i j}}\right)-\mathrm{Li}_2\left(1-\frac{m_j^{2} l_{ij} \cdot p_2}{\beta\left(p_{j}\right) \alpha_{ij}\gamma_{i j}}\right)\right]\ ,  \\
    \gamma_{i j}=&\sqrt{\left(k_{i} \cdot p_{j}\right)^{2}-m_i^{2} m_j^{2}}\ ,  \\
    \beta(p)=&p\cdot p_2+\sqrt{(p\cdot p_2)^2-kp^2M^2}\ ,  \\
    \alpha_{i j}=& \frac{k_{i} \cdot p_{j} + \sqrt{\left(k_{i} \cdot p_{j}\right)^{2}-m_i^{2} m_j^{2}}}{m_i^{2}}\ ,  \\
    l_{ij}=& \alpha_{i j}k_i-p_j\ .
\end{align} 
For simplicity, the abbreviations $\gamma_{ij}, \alpha_{ij}, l_{ij}$ denote $\gamma(k_i, p_j), \alpha(k_i, p_j), l(k_i, p_j)$, respectively.
It is well known that $\mathrm{Li}_2(z)$ is the simplest polylogarithm function, also known as Spence function $\mathrm{Sp}(z)$ in some literatures, defined as
\begin{align}
    \mathrm{Li}_{2}(z)=\operatorname{Sp}(z)=-\int_{0}^{z} \mathrm{~d} t \frac{\ln (1-t)}{t}\ , \quad z \in \mathbb{R}\ .
\end{align}

Further we just note that $\Delta E^S$ in Eq.(\ref{integral-I-kp}) is defined in the S frame. 
It is convenient to connect $\Delta E^S$ with some energy scales in lab frame.
According to Refs.~\cite{Maximon:2000hm, Vanderhaeghen:2000ws, Talukdar:2020aui},
\begin{align}
    \Delta E^{S}=\eta\left(\tilde{E}_{2}-E_{2}\right) \simeq \eta \Delta E_{\gamma}\ ,
\end{align}
where $\Delta E_{\gamma}$ is defined in the lab frame.
$\eta=E_1/\tilde{E_2}$ is called the ``inelastic" lab system recoil factor, and the lab recoil lepton energy is $\tilde{E_2}$ in the elastic process.
When the radiated photon is soft, then $\tilde{E}_{2} \simeq E_{2}$ and $\eta$ can be understood as the ``elastic" lab system recoil factor, $\eta=E_1/E_2$.
The only parameter that can be adjusted is $\Delta E_{\gamma}$.
In principle, this depends on the acceptance of the detector. 
According to the characteristics of the MUSE experiment, we can set $\Delta E_{\gamma} \simeq 1 \% E_{1}$.

At the same time, similar integrals used in this paper are given:
\begin{align}
    I(k_1, k_2)& \equiv \int^{k<\frac{\Delta E^{S}}{\nu}} \frac{\mathrm{d}^{D-1} k}{(2 \pi)^{D-1}} \frac{1}{2 k^0} \frac{1}{\left(k_{1} \cdot k\right)\left(k_{2} \cdot k\right)} \nonumber\\
    &=\frac{1}{8 \pi^{2}} \frac{v_{\ell}^{2}-1}{2 m^{2} v_{\ell}} \ln \frac{v_{\ell}+1}{v_{\ell}-1}\left(-\frac{1}{\epsilon_{\mathrm{I R}}}+\gamma_{E}-\ln \frac{4 \pi \nu^{2}}{Q^{2}}-\ln \frac{Q^{2}}{4\left(\Delta E^{S}\right)^{2}}\right)+\frac{R^{(2)}\left(k_{1}, k_{2}\right)}{8 \pi^{2}} \ , \\
    I(p_1, p_2)& \equiv \int^{k<\frac{\Delta E^{S}}{\nu}} \frac{\mathrm{d}^{D-1} k}{(2 \pi)^{D-1}} \frac{1}{2 k^0} \frac{1}{\left(p_{1} \cdot k\right)\left(p_{2} \cdot k\right)} \nonumber\\
    &=\frac{1}{8 \pi^{2}} \frac{v_{N}^{2}-1}{2 M^{2} v_{N}} \ln \frac{v_{N}+1}{v_{N}-1}\left(-\frac{1}{\epsilon_{\mathrm{I R}}}+\gamma_{E}-\ln \frac{4 \pi \nu^{2}}{Q^{2}}-\ln \frac{Q^{2}}{4\left(\Delta E^{S}\right)^{2}}\right)+\frac{R^{(2)}\left(p_{1}, p_{2}\right)}{8 \pi^{2}}\ , \\
    I(k_i)& \equiv \int^{k<\frac{\Delta E^{S}}{\nu}} \frac{\mathrm{d}^{D-1} k}{(2 \pi)^{D-1}} \frac{1}{2 k^0} \frac{1}{\left(k_{i} \cdot k\right)^2} \nonumber\\
    &=\frac{1}{8 \pi^{2} m^{2}}\left(-\frac{1}{\epsilon_{\mathrm{I R}}}+\gamma_{E}-\ln \frac{4 \pi \nu^{2}}{Q^{2}}-\ln \frac{Q^{2}}{4\left(\Delta E^{S}\right)^{2}}\right)+\frac{1}{4 \pi^{2} m^{2}} \frac{k_{i} \cdot p_2}{\sqrt{\left(k_{i} \cdot p_2\right)^{2}-m^{2} M^{2}}} \ln \frac{m M}{\beta\left(k_{i}\right)}\ , \\
    I(p_i)& \equiv \int^{k<\frac{\Delta E^{S}}{\nu}} \frac{\mathrm{d}^{D-1} k}{(2 \pi)^{D-1}} \frac{1}{2 k^0} \frac{1}{\left(p_{i} \cdot k\right)^2} \nonumber\\
    &=\frac{1}{8 \pi^{2} M^{2}}\left(-\frac{1}{\epsilon_{\mathrm{I R}}}+\gamma_{E}-\ln \frac{4 \pi \nu^{2}}{Q^{2}}-\ln \frac{Q^{2}}{4\left(\Delta E^{S}\right)^{2}}\right)+\frac{1}{4 \pi^{2} M^{2}} \frac{p_{i} \cdot p_2}{\sqrt{\left(p_{i} \cdot p_2\right)^{2}-M^{4}}} \ln \frac{M^{2}}{\beta\left(p_{i}\right)}\ ,
\end{align}
where $v_\ell^2=1+4m^2/Q^2\ \text{and}\ v_N=1+4M^2/Q^2$ are invariant kinematical variables.

\section{Vertex corrections}\label{appendix:2}

\subsection{Lepton photon vertex corrections}

The one-loop chiral LO calculation for FFs evaluated using DR are given~\cite{Vanderhaeghen:2000ws}:
\begin{align}
    \delta F^{\ell,(a), 0}_1\left(Q^{2}\right)=& \frac{\alpha}{4 \pi}\left\{\left[\frac{1}{\epsilon_{\mathrm{U V}}}-\gamma_{E}+\ln \frac{4 \pi \nu^{2}}{m^{2}}\right]\right. \nonumber\\
    &+\left[\frac{1}{\epsilon_{\mathrm{I R}}}-\gamma_{E}+\ln \frac{4 \pi \nu^{2}}{Q^{2}}+\ln \frac{Q^{2}}{m^{2}}\right] \frac{v_\ell^{2}+1}{v_\ell} \ln \frac{v_\ell+1}{v_\ell-1} \nonumber\\
    &+\frac{v_\ell^{2}+1}{2 v_\ell} \ln \frac{v_\ell+1}{v_\ell-1} \ln \frac{v_\ell^{2}-1}{4 v_\ell^{2}}+\frac{2 v_\ell^{2}+1}{v_\ell} \ln \frac{v_\ell+1}{v_\ell-1} \nonumber\\
    &\left.+\frac{v_\ell^{2}+1}{v_\ell}\left[\mathrm{Li}_2\left(\frac{v_\ell+1}{2 v_\ell}\right)-\mathrm{Li}_2\left(\frac{v_\ell-1}{2 v_\ell}\right)\right]\right\}\ ,\\
    F^{\ell,(a), 0}_2\left(Q^{2}\right)=&\frac{\alpha}{4 \pi} \frac{v_\ell^{2}-1}{v_\ell} \ln \frac{v_\ell+1}{v_\ell-1}\ ,\\
    F^{p,(a), 0}_1\left(Q^{2}\right)=&1\ ,\\
    F^{p,(a), 0}_2\left(Q^{2}\right)=&0\ ,
\end{align}
where superscript $0$ represents the bare FFs and the superscripts (a)-(f) represent the the number of the subfigure in Fig.~\ref{vertex fig}.
The NLO result of subfigure (b) is similar to that of subfigure (a),
\begin{align}
    \delta F^{\ell,(b), 0}_1\left(Q^{2}\right)=&\delta F^{\ell,(a)}_1\left(Q^{2}\right)\ ,\\
    F^{\ell,(b), 0}_2\left(Q^{2}\right)=&F^{\ell,(a)}_2\left(Q^{2}\right)\ ,\\
    F^{p,(b), 0}_1\left(Q^{2}\right)=&0\ ,\\
    F^{p,(b), 0}_2\left(Q^{2}\right)=&(2c_6+c_7)M\ .
\end{align}
The UV divergence can be renormalized by the standard renormalization method.
In the case of one-loop diagrams, it is convenient to obtain the renormalized results by adding the counterterm Lagrangian.
The renormalized lepton photon vertex correction is well known~\cite{Vanderhaeghen:2000ws},
\begin{align}
    &F_{1}^{\ell}\left(Q^{2}\right)=1+\delta F_{1}^{\ell, 0}\left(Q^{2}\right)-\delta F_{1}^{\ell, 0}(0)  \nonumber\\
    &=1+\frac{\alpha}{4 \pi}\left\{\left[\frac{1}{\epsilon_{\mathrm{I R}}}-\gamma_{E}+\ln \frac{4 \pi \nu^{2}}{Q^{2}}+\ln \frac{Q^{2}}{m^{2}}\right] \left(\frac{v_{\ell}^{2}+1}{v_{\ell}} \ln \frac{v_{\ell}+1}{v_{\ell}-1}-2\right)\right. \nonumber\\
    &+\frac{v_{\ell}^{2}+1}{2 v_{\ell}} \ln \frac{v_{\ell}+1}{v_{\ell}-1} \ln \frac{v_{\ell}^{2}-1}{4 v_{\ell}^{2}}+\frac{2 v_{\ell}^{2}+1}{v_{\ell}} \ln \frac{v_{\ell}+1}{v_{\ell}-1}-4 \nonumber\\
    &\left.+\frac{v_{\ell}^{2}+1}{v_{\ell}}\left[\operatorname{Li}_{2}\left(\frac{v_{\ell}+1}{2 v_{\ell}}\right)-\operatorname{Li}_{2}\left(\frac{v_{\ell}-1}{2 v_{\ell}}\right)\right]\right\}\ ,
\end{align}  
and only the $F_1^\ell$ needs to be renormalized.

\subsection{Proton photon vertex corrections}

In the scheme of B$\chi$PT, the interaction between proton and photon is constructed in a gauge invariant way order by order in contrast to traditional on shell FFs approximation~\cite{Mo:1968cg, Maximon:2000hm}.
The one-loop calculation of chiral LO diagram (c) is given,
\begin{align}
    \delta F_{1}^{\ell,(c), 0}\left(Q^{2}\right) &=0\ , \\
    F_{2}^{\ell,(c), 0}\left(Q^{2}\right) &=0\ , \\
    F_{1}^{p,(c), 0}\left(Q^{2}\right) &=1+\left.\delta F_{1}^{\ell,(a), 0}\left(Q^{2}\right)\right|_{v_{\ell} \rightarrow v_{N}, m \rightarrow M}\ , \\
    F_{2}^{p,(c), 0}\left(Q^{2}\right) &=\left.F_{2}^{\ell,(a), 0}\left(Q^{2}\right)\right|_{v_{\ell} \rightarrow v_{N}, m \rightarrow M}\ ,
\end{align}
where the result is just the same as lepton photon vertex correction when the lepton mass is replaced by proton mass.

The only nontrivial contribution is derived from diagrams (d),(e), and (f),\footnote{For the sake of simplicity, the notation $(\mathrm{d,e,f})$ imply the summations of diagrams (d),(e), and (f).} which are the chiral NLO contribution,
\begin{align}
    \delta F^{\ell,(\mathrm{d,e,f}), 0}_1\left(Q^{2}\right)=&0\ ,\\
    F^{\ell,(\mathrm{d,e,f}), 0}_2\left(Q^{2}\right)=&0\ ,\\
    F^{p,(\mathrm{d,e,f}), 0}_1\left(Q^{2}\right)=& \left(2 c_6+c_7\right)M\frac{\alpha}{4 \pi}\left\{ \left[\frac{1}{\epsilon_\mathrm{UV}}-\gamma_E+\ln \left(\frac{4 \pi \nu^{2}}{M^{2}}\right)\right]\left(\frac{3}{2}\right)+\frac{3}{v_N}\ln \left(\frac{v_N-1}{v_N+1}\right)+\frac{1}{2}\right\} \ ,\\
    F^{p,(\mathrm{d,e,f}), 0}_2\left(Q^{2}\right)=&(2c_6+c_7)M\frac{\alpha}{4\pi}\left\{\left[\frac{1}{\epsilon_\mathrm{UV}}-\gamma_{E}+\ln \frac{4 \pi \nu^{2}}{Q^{2}}+\ln \frac{Q^{2}}{M^{2}}\right](-4)\right. \nonumber\\
    &+\left[\frac{1}{\epsilon_\mathrm{IR}}-\gamma_{E}+\ln \frac{4 \pi \nu^{2}}{Q^{2}}+\ln \frac{Q^{2}}{M^{2}}\right] \frac{v_N^{2}+1}{v_N} \ln \frac{v_N+1}{v_N-1} \nonumber\\
    &+\frac{v_N^{2}+1}{2 v_N} \ln \frac{v_N+1}{v_N-1} \ln \frac{v_N^{2}-1}{4 v_N^{2}}+\frac{5 v_N^{2}+3}{v_N} \ln \frac{v_N+1}{v_N-1}-8 \nonumber\\
    &\left.+\frac{v_N^{2}+1}{v_N}\left[\mathrm{Li}_2\left(\frac{v_N+1}{2 v_N}\right)-\mathrm{Li}_2\left(\frac{v_N-1}{2 v_N}\right)\right]\right\}\ .
\end{align}
Especially note that the UV divergence in chiral NLO not only appears in $F_1^{p}$, but also in $F_2^p$.
They need to be renormalized.
In the following, the notation is similar to QED renormalization, and the related convention can be found, for example, in Ref.~\cite{Peskin:1995ev}.
The bare chiral Lagrangian is written explicity as
\begin{align}
    \mathcal{L}_{\gamma pp}^0=&\bar{p}\left(i \slashed{\partial}-M\right)p+e \bar{p}\gmu pA_\mu+\frac{e}{4}(2c_6+c_7)\bar{p}\sigma^{\mu\nu}F_{\mu\nu}p  \nonumber\\
    &+(Z_2-1)\bar{p}\left(i \slashed{\partial}\right)p-(Z_2Z_M-1)\bar{p}\left(M\right)p+(Z_1-1)e \bar{p}\gmu pA_\mu  \nonumber\\
    &+(Z_1Z_c-1)\frac{e}{4}(2c_6+c_7)\bar{p}\sigma^{\mu\nu}F_{\mu\nu}p\ ,
\end{align}
where $Z_c$ is the renormalization constant of the combination $2c_6 + c_7$.
We denote $F_{1}^{p} \gamma_{\mu}+F_{2}^{p} \frac{i}{2 M} \sigma_{\mu \rho} q^{\rho}$ as $\Gamma^p_\mu$, the renormalized vertex function satisfies,
\begin{align}
    \Gamma^p_\mu=\Gamma_{\mu}^0+(Z_1-1)\gamma_\mu+(Z_1Z_c-1)\frac{i}{2M}(2c_6+c_7)M\sigma_{\mu\nu}q^\nu \ .
\end{align}
The first renormalization condition is that when $Q^2$ goes to $0$, $\Gamma^p_\mu$ defines the physical charge.
That is to say, $\Gamma^\mu(Q^2\to 0)=\gmu=F^{p, 0}_1(0)\gmu+(Z_1-1)\gmu$, which is,
\begin{align}\label{re vertex 1}
    1=F^{p,0}_1(0)+Z_1-1\ .
\end{align}
Another on shell renormalization condition requires $F_2^p(Q^2)$ returning to the proton magnetic moment at $Q^2 \to 0$~\footnote{Neglecting B$\chi$PT correction beyond NLO, the two LECs $c_6$ and $c_7$ can be related to the anomalous magnetic moments of the nucleon, $c_{6}=\frac{k_{p}+k_{n}}{4 M} \quad, c_{7}=\frac{k_{p}-k_{n}}{2 M}$, with $\kappa_p$ and $\kappa_n$ being anomalous magnetic moments of proton and neutron, respectively. 
The renormalization of $\kappa_p=(2c_6+c_7)/M$ is performed by identify with $1.793$ as an experimental input.},
\begin{align}\label{re vertex 2}
    0=F^{p, 0}_2(0)+(Z_1Z_c-1)(2c_6+c_7)M\ ,
\end{align}
using Eqs.~(\ref{re vertex 1}) and (\ref{re vertex 2}), the renormalization constants read,
\begin{align}\label{const Z}
    Z_1=&1-\frac{\alpha}{4\pi}\left\{\left[\frac{1}{\epsilon_{\mathrm{UV}}}-\gamma_{E}+\ln \left(\frac{4 \pi \nu^{2}}{M^{2}}\right)\right]+\left[\frac{1}{\epsilon_{\mathrm{IR}}}-\gamma_{E}+\ln \frac{4 \pi \nu^{2}}{Q^{2}}+\ln \frac{Q^{2}}{M^{2}}\right](2)+4\right\} \nonumber\\
    &-(2c_6+c_7)M\frac{\alpha}{4\pi}\left\{\left[\frac{1}{\epsilon_{\mathrm{UV}}}-\gamma_{E}+\ln \left(\frac{4 \pi \nu^{2}}{M^{2}}\right)\right]\left(\frac{3}{2}\right)+\frac{1}{2}\right\}\ , \\
    Z_1Z_c=&1-\frac{\alpha}{4\pi}\frac{2}{(2c_6+c_7)M}  \nonumber\\
    &+\frac{\alpha}{4\pi} \left\{\left[\frac{1}{\epsilon_{\mathrm{UV}}}-\gamma_{E}+\ln \left(\frac{4 \pi \nu^{2}}{M^{2}}\right)\right](4)-\left[\frac{1}{\epsilon_{\mathrm{IR}}}-\gamma_{E}+\ln \frac{4 \pi \nu^{2}}{Q^{2}}+\ln \frac{Q^{2}}{m^{2}}\right](2)-2\right\}\ ,
\end{align}
and the renormalized expressions for FFs of proton are ultimately given as,
\begin{align}
    &F_{1}^{p,(c)}\left(Q^{2}\right)=1+\left.\delta F_{1}^{\ell,(a),0}\left(Q^{2}\right)\right|_{v_{\ell} \rightarrow v_{N}, m\to M}-\left.\delta F_{1}^{\ell,(a),0}\left(0\right)\right|_{v_{\ell} \rightarrow v_{N}, m\to M}\ , \\
    &F_{2}^{p,(c)}\left(Q^{2}\right)=\left.F_{2}^{\ell,(a),0}\left(Q^{2}\right)\right|_{v_{\ell} \rightarrow v_{N}, m\to M}-\left.F_{2}^{\ell,(a),0}\left(0\right)\right|_{v_{\ell} \rightarrow v_{N}, m\to M}  \nonumber\\
    &=\frac{\alpha}{4 \pi} \frac{v_N^{2}-1}{v_N} \ln \frac{v_N+1}{v_N-1}-\frac{\alpha}{2\pi}\ ,\\
    &F_{1}^{p, (\mathrm{d,e,f})}\left(Q^{2}\right)=1+F_{1}^{p, (\mathrm{d,e,f}),0}\left(Q^{2}\right)-F_{1}^{p, (\mathrm{d,e,f}),0}\left(0\right)  \nonumber\\
    &=1+\left(2 c_{6}+c_{7}\right) M \frac{\alpha}{4 \pi}\left[\frac{3}{v_N} \ln \left(\frac{v_N-1}{v_N+1}\right)\right]\ ,  \\
    &F_{2}^{p, (\mathrm{d,e,f})}\left(Q^{2}\right)=F_{2}^{p, (\mathrm{d,e,f}),0}\left(Q^{2}\right)-F_{2}^{p, (\mathrm{d,e,f}),0}\left(0\right)  \nonumber\\
    &=(2c_6+c_7)M\frac{\alpha}{4\pi}\left\{\left[\frac{1}{\epsilon_\mathrm{IR}}-\gamma_{E}+\ln \frac{4 \pi \nu^{2}}{Q^{2}}+\ln \frac{Q^{2}}{M^{2}}\right] \left(\frac{v_{N}^{2}+1}{v_{N}} \ln \frac{v_{N}+1}{v_{N}-1}-2\right)\right. \nonumber\\
    &+\frac{v_{N}^{2}+1}{2 v_{N}} \ln \frac{v_{N}+1}{v_{N}-1} \ln \frac{v_{N}^{2}-1}{4 v_{N}^{2}}+\frac{5 v_{N}^{2}+3}{v_{N}} \ln \frac{v_{N}+1}{v_{N}-1}-10 \nonumber\\
    &\left.+\frac{v_{N}^{2}+1}{v_{N}}\left[\mathrm{Li}_2\left(\frac{v_{N}+1}{2 v_{N}}\right)-\mathrm{Li}_2\left(\frac{v_{N}-1}{2 v_{N}}\right)\right]\right\}\ .
\end{align}
Here, the process is essentially a renormalization of a EFT with $U(1)$ gauge symmetry. 
Therefore, we find that the Ward-Takahashi identity such as, $Z_1 = Z_2$, can be satisfied as verified by direct calculation.

\end{appendices}

\newpage

\bibliographystyle{hphysrev}
\bibliography{leptonprotonscattering}

\end{document}